# Creep Behavior of High-Entropy Alloys: A Critical Review

*Mingwei Zhang[1], Uwe Glatzel[2], Martin Heilmaier[3], Easo P. George[4,5]*

[1]Department of Materials Science and Engineering, University of California, One Shields Ave., Davis, CA, 95616, USA.

[2]Metals and Alloys, University of Bayreuth, Prof.-Rüdiger-Bormann-Str. 1, 95447, Bayreuth, Germany.

[3]Institute for Applied Materials (IAM-WK), Karlsruhe Institute of Technology (KIT), Engelbert-Arnold-Str. 4, 76131, Karlsruhe, Germany.

[4]Department of Materials Science and Engineering, University of Tennessee, Knoxville, TN 37996, USA.

[5]Institute for Materials, Ruhr University Bochum, 44801, Bochum, Germany.

Corresponding Author:

Mingwei Zhang, Assistant Professor

Department of Materials Science and Engineering

University of California, Davis

Email: mwwzhang@ucdavis.edu

**Abstract**

High-entropy alloys (HEAs) comprise a compositionally complex class of materials that in certain cases exhibit outstanding mechanical properties. While substantial progress has been made in understanding their phase stability, microstructure, and deformation mechanisms at room and cryogenic temperatures, the long-term creep behavior (>100 h) of HEAs at high temperatures (>0.6 $T_m$, where $T_m$ is the melting temperature) remains relatively underexplored. This knowledge gap is critical, as many engineering applications, including those in aerospace, power generation, aero engines and nuclear energy, require materials with good creep resistance to maintain structural integrity over extended service lifetimes. This review provides a focused and critical assessment of the current understanding of high-temperature deformation and creep behavior of HEAs, with particular attention paid to face-centered cubic HEAs and body-centered cubic refractory HEAs. The underlying deformation mechanisms governing their creep response and the influence of phase stability at elevated temperatures are examined in detail. Recent studies reveal mechanistic differences between HEAs and conventional dilute alloys that do not always lead to improved creep resistance belying their initial promise. Based on these findings, we discuss the challenges in designing HEAs for high-temperature structural applications and outline future research directions that may lead to creep-resistant HEAs.

## 1. Introduction

The pursuit of materials with better mechanical properties is a constant endeavor in materials research. Among these, resistance to creep deformation is one of the most important considerations for structural materials intended to operate at elevated temperatures over prolonged periods of time. Conventional superalloys perform well up to about 1100 °C in power generation and transportation roles due to their unrivaled combination of relevant properties related to oxidation [1-3], fatigue [4-6], creep [7-12] and fracture toughness [13, 14] . However, the need for materials that can operate at even higher temperatures while maintaining adequate creep properties has directed research into alternative alloy systems [15, 16].

Recently, high-entropy alloys (HEAs) comprising multiple constituent elements in large concentrations [17] have emerged as a new class of materials. In this review, we define HEAs rather broadly and include under its umbrella both single- and multi-phase alloys comprising 3 or more principal elements, as well as those referred to as multiple principal element alloys (MPEAs) or compositionally complex alloys (CCAs) [18-23]. To date, HEA research has mostly focused on two families. The first family is the face-centered cubic (FCC) HEAs consisting of $3d$-transition metals (V, Cr, Mn, Fe, Co, Ni, Cu) [17, 24-26]. These alloys exhibit exceptional mechanical properties at room and cryogenic temperatures; several exemplars have achieved excellent strength and ductility [27, 28] as well as record-breaking fracture toughness [29-31]. However, single-phase FCC HEAs tend to lose strength significantly at elevated temperatures [27, 32], which motivated the addition of (Al, Ti) to obtain multi-phase structures for improved strength (e.g., FCC-$L1_2$-B2 high-entropy alloys [33-36] and FCC-$L1_2$ high-entropy superalloys [37-41]).

The second family is the body-centered cubic (BCC) HEAs consisting of Ti, V, Cr, Zr, Nb, Mo, Hf, Ta, W [42-45]. Invented by Senkov et al. [45-47], these BCC HEAs are often termed refractory high-entropy alloys (RHEAs) due to their high melting points and were expected to perform at temperatures beyond those of current Ni-based superalloys. Some RHEAs, such as the equiatomic MoNbTaWV alloy, can retain a very high yield strength of ~500 MPa at 1600 °C [47]; however, it lacks tensile ductility [48, 49]. Others such as the equiatomic HfNbTaTiZr alloy are ductile and malleable [46] but have low yield strengths at elevated temperatures [50]. To increase strength, two approaches are followed: **1.** Al and Zr have been added to the latter family to promote the formation of a BCC–B2 dual-phase structure [51-53], analogous to the FCC–$L1_2$ structure in conventional superalloys. These two-phase RHEAs are sometimes referred to as refractory high-entropy superalloys (RHSAs); and **2.** Oxide dispersion strengthening (ODS) using $Y_2O_3$ particles has been employed to improve the creep resistance of both FCC [54-58] and BCC [59, 60] HEAs. In this review, the creep properties of the above single-phase and multi-phase FCC and BCC HEAs are covered in detail. Other types of HEAs, such as hexagonal close-packed (HCP) HEAs consisting of Cd, Mg, Os, Re, Ru, Tc, Zn, Co, Be, Sc, Ti, Zr, Hf [61, 62] and light-weight HEAs containing Li, Be, Mg, Al, Cr, Fe, Mn, Ti, V, Zr [63-65], have not demonstrated the desired combination of high-temperature capability and structural integrity and are therefore excluded from this review.

Creep deformation in pure metals and alloys typically progresses through three distinct stages: primary, secondary, and tertiary creep. Secondary creep is the most extensively studied stage because it typically accounts for the largest portion of total creep strain (and thus creep life). In this stage, the creep rate reaches a minimum value which can sometimes remain constant for an extended period due to a dynamic equilibrium between work hardening and recovery [66-70]; in the latter cases it is also referred to as steady-state creep. In other cases, no real steady state is reached during secondary creep, and it is better described by a minimum creep rate. The steady-state or minimum creep rate ($\dot{\varepsilon}_{SS\ or\ min}$) as a function of the applied stress ($\sigma$) is typically described by the following power-law relationship [67, 69-72]:

$$\dot{\varepsilon}_{SS\ or\ min} = A\sigma^{n}\, exp\left(-\frac{Q_C}{RT}\right), \tag{1}$$

where $A$ is a material constant that includes the dislocation density and elastic properties, $n$ is the stress exponent, $Q_C$ represents the apparent activation energy for creep, $R$ is the gas constant, and $T$ is the absolute temperature. The governing creep mechanisms are often characterized by the stress exponent $n$ and the activation energy $Q_C$, as summarized in **Table 1**.

**Table 1** Creep mechanisms and their characteristic stress exponent $n$ and apparent activation energy $Q_C$. In the table, $Q_{SD}$ refers to the activation energy for lattice self-diffusion, $Q_{GB}$ the activation energy for grain boundary diffusion, and $\tilde{Q}_D$ the activation energy for lattice interdiffusion.

| **Creep mechanism** | $\boldsymbol{n}$ | $\boldsymbol{Q_C}$ |
|---|---|---|
| Nabarro-Herring creep (through lattice diffusion) | 1 | ~ $Q_{SD}$ |
| Coble creep (through grain boundary diffusion) | 1 | ~$Q_{GB}$ |
| Grain boundary sliding | 2 | ~$Q_{GB}$ |
| Solute-drag (Viscous glide of dislocations through diffusion of solutes at dislocation cores) | 3 | $\tilde{Q}_D$ |
| Dislocation climb-controlled creep | 4-6 | ~ $Q_{SD}$ |
| Dislocation climb-glide-controlled creep (Power-law breakdown) | 8-30+ | >> $Q_{SD}$ |
| Dislocation-particle interaction | 8-30+ | >> $Q_{SD}$ |

Compared to the extensive investigations of the above creep mechanisms in conventional dilute alloys, the history of creep studies in HEAs is relatively short and much remains to be understood.

For example, the compositional complexity of HEAs can arguably be manifested as rugged potential energy landscapes that can influence phase stability [73-77], diffusion pathways [78-81], and defect energetics [82-86]. This raises fundamental scientific questions such as whether HEAs can exhibit superior creep resistance due to concentrated solid solution strengthening and severe lattice distortion, whether a concentrated solid solution with a rugged potential energy landscape and local composition fluctuations can slow down (or accelerate) dislocation kinetics and vacancy diffusion, and whether the high configuration entropy of HEAs can lead to enhanced phase stability during creep. It has been speculated that some of these factors can be beneficial for creep of HEAs [87-89], yet extensive creep data of single-phase HEAs have suggested that these hypotheses are either untrue or the purported effects are so insignificant that their creep resistance often falls short of existing commercial alloys [90-92]. In comparison, limited work on some multi-phase HEAs [57, 93] show intriguing hints of potentially surpassing their commercial counterparts by combining concentrated solid solution hardening and precipitate/dispersion strengthening. This review provides a detailed analysis of the deformation mechanisms and microstructural evolution of single-phase and multi-phase HEAs during creep, along with a critical evaluation of their suitability for elevated-temperature applications.

Numerous studies have investigated nanoindentation creep in HEAs [94-109]; however, the small deformation volumes involved lead to mechanisms that differ significantly from those in bulk samples. While the understanding gained from small-scale testing is scientifically important, nanoindentation creep is excluded from the scope of this review to keep the overall length manageable. Another exclusion is oxidation resistance, which is clearly a critical aspect of high-temperature materials; however, it has recently been comprehensively addressed elsewhere [110-114] and is therefore not covered here. We conclude our review with suggestions for future research directions aimed at advancing the fundamental understanding of high-temperature mechanical behavior and enhancing the creep resistance of HEAs.

## 2. Creep Properties of FCC High-Entropy Alloys

### *2.1. Single-Phase FCC High-Entropy Alloys*

Initial investigations into the creep behavior of HEAs primarily focused on the single-phase FCC, equiatomic CrMnFeCoNi (the so-called “Cantor alloy” [24]) and its derivatives [87, 88, 92, 115-125], which served as a model alloy system to evaluate the creep behavior in this novel class of materials. Important parameters for creep, including test temperatures, specimen microstructure, the stress exponent $n$, and the apparent activation energy for creep $Q_C$, are summarized in **Table 2**. For clarity, all alloy compositions in this paper are expressed in **atomic percent**, except in cases where equiatomic compositions are self-evident. This convention is adopted to eliminate ambiguity because several reported compositions in the literature are originally given in atomic ratios or weight percent. Secondary creep data as a function of applied stress, either directly from data tables in references or extracted using a plot digitizer, are provided in **Supplementary Table 1** in the **Appendix**. We note that in the current creep literature, the pre-exponential factor in Eqn. 1 is often omitted when reporting $n$ and $Q_C$. This poses a problem because, while $n$ and $Q_C$ are critical for understanding the creep mechanism, the actual creep rate can only be determined if the pre-exponential factor is known.

**Table 2 Creep parameters of single-phase FCC HEAs for different microstructural states and testing conditions. All compositions are expressed in atomic percent; equiatomic compositions are considered self-explanatory.**

*Acronyms: RX - recrystallized; SX - single crystal; AM - additively manufactured; hi - high; med - medium; lo - low.

| Alloy | Loading | Microstructure | T (ºC) | $n$ | $Q_C$ (kJ/mol) | Ref. |
|---|---|---|---|---|---|---|
| CrMnFeCoNi | Tension | FCC, RX | 600 | 6.2 | 394 | [115] |
| CrMnFeCoNi | Tension | FCC, RX | 500-600 | 5-6 (med-hi $\sigma$)<br>8.9-14 (hi $\sigma$) | 261-275<br>372-410 | [88] |
| CrMnFeCoNi | Tension | FCC, RX | 535-650 | 5.8-5.9 (med-lo $\sigma$)<br>2.6-3.2 (med $\sigma$) | 305-319<br>236-249 | [87] |
| CrMnFeCoNi | Tension | FCC, RX | 500-700 | 3.1-3.3 (med $\sigma$)<br>4.8-6.4 (med-hi $\sigma$) | -<br>302 | [116] |
| CrMnFeCoNi | Tension | FCC, RX | 650-700 | 4.1-5.9 | 303 | [117] |
| CrMnFeCoNi | Tension | FCC, RX | 700 | 6 | - | [118] |
| CrMnFeCoNi | Tension | FCC, RX | 750-900 | 3.6-3.8 | 219-236 | [92] |
| CrMnFeCoNi | Compression<br>Compression<br>Tension<br>Tension | FCC, RX<br>FCC, SX<br>FCC, RX<br>FCC, SX | 800<br>800<br>980<br>980 | 3.3<br>5.2<br>3.9<br>5.6 | - | [119] |
| CrMnFeCoNi | Tension | FCC, SX | 700-1100 | 5-6.8 | - | [120] |
| CrFeCoNi | Tension | FCC, RX | 600-662.5 | 6.3 | 351-368 | [121] |
| CrFeCoNi | Tension | FCC, RX | 700 | 5.4 | - | [118] |
| CrFeCoNi | Tension | FCC, RX | 650-725 | 5.4-6.8 | 295 | [117] |
| CrCoNi | Tension | FCC, RX | 550 | 3.4 | - | [122] |
| CrCoNi | Tension | FCC, RX | 700-800 | 4.8-5.5 | 370 | [123] |
| CrCoNi | Tension | FCC, RX<br>FCC, AM | 750-900 | 4.3-4.7<br>5.7-6 | 240-259<br>320-331 | [124] |
| $Ni_{42.5}Co_{21.3}Fe_{21.3}V_{10.6}Mo_{4.3}$ | Tension | FCC, RX | 700 | 3.1 | 456 | [125] |

Most of these studies investigated the creep behavior of cold worked and recrystallized polycrystalline HEAs at intermediate temperatures (500–700 °C, corresponding to approximately 0.5–0.6 $T_m$) and high temperatures (700–1100 °C, or 0.6–0.85 $T_m$). That is, the investigations started with comparable, well-annealed, low-dislocation-density microstructures. Generally, the activation energy for creep is in reasonable agreement with that obtained from bulk tracer diffusion of Cr, Mn, Fe, Co, Ni in a CrMnFeCoNi solid solution matrix, from which the arithmetic average activation energy for diffusion was determined to be 294-303 kJ/mol [89, 126] and that for Cr, Fe, Co, Ni in CrFeCoNi, which similarly yielded an activation energy of 280 kJ/mol [126].

At intermediate temperatures, stress exponents for CrMnFeCoNi exhibit a five-three-five power-law transition with increasing applied stress (**Fig. 1a,b**) before reaching power-law breakdown at still higher stresses (**Fig. 1c**) [87, 88, 116]. This phenomenon is widely observed in solid solutions when creep controlled by solute drag becomes dominant at a specific combination of temperature and stress at which the dislocation mobility aligns with the solute diffusion rate [127]. However, this power-law transition is not observed at higher temperatures, as **Fig. 1d** shows a uniform stress exponent of 3.7 ± 0.1 over the range 750-900 ºC [92]. The stress exponent falls between those associated with dislocation climb-controlled creep and solute drag creep, which indicates that both mechanisms are involved.

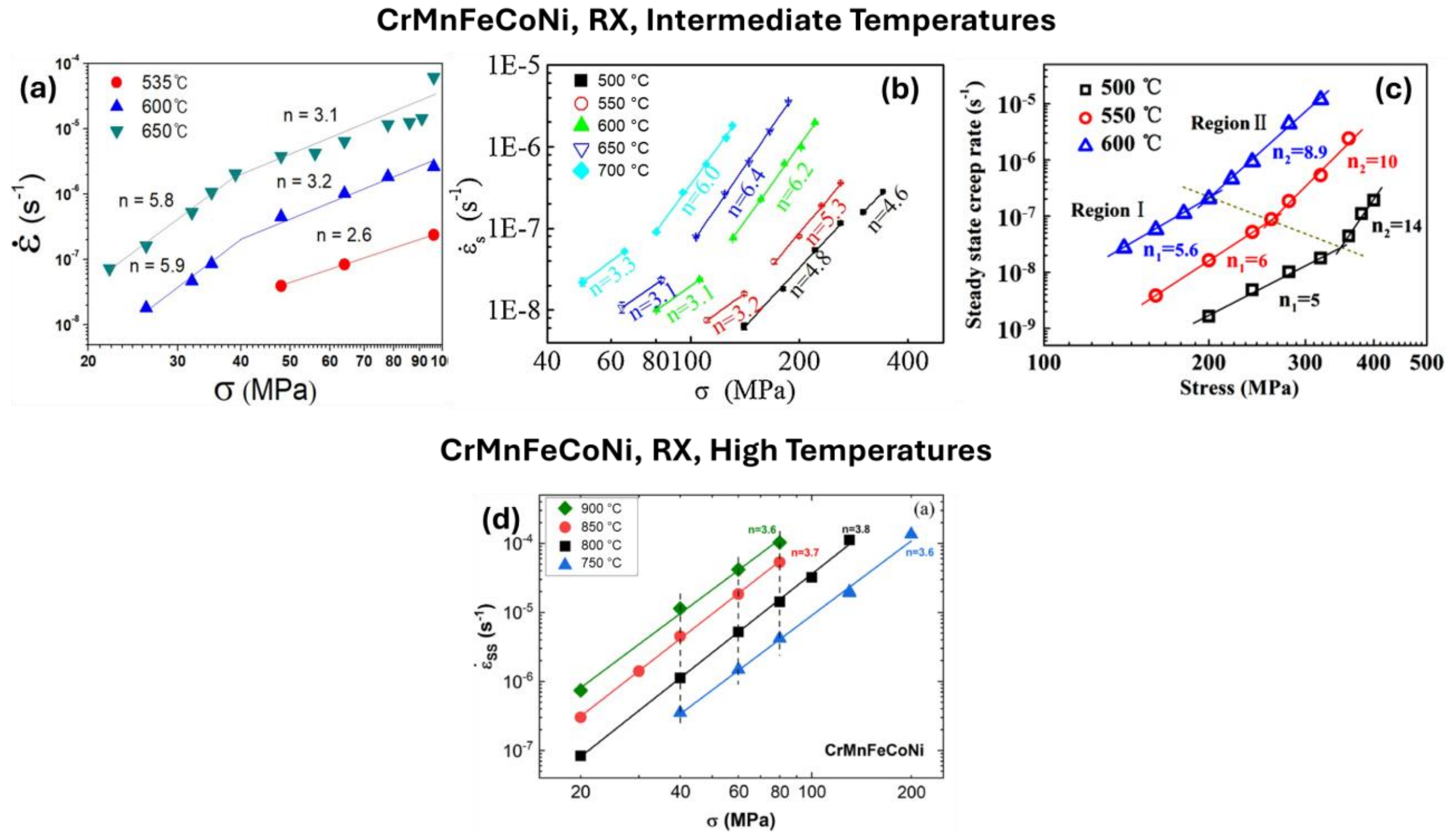

**Figure 1** Secondary creep strain rates of CrMnFeCoNi tested at different stresses at (a-c) intermediate temperatures and (d) high temperatures. The diagrams were taken from Refs. [87, 88, 92, 116].

It is important to examine the implications of a potential $n = 3$ solute-drag creep in HEAs, as the simple picture envisioned in conventional dilute alloys of solute atmospheres diffusing in a solvent matrix breaks down in equiatomic concentrated solid solutions, where "solutes" and "solvents" are ill-defined. One might initially hypothesize that the rate-controlling step for dislocation drag is either governed by the slowest diffusing element in the matrix, Ni, with an activation energy of 318 kJ/mol [89], or by the element with the largest atomic size misfit, Cr, which can segregate at dislocation cores with a diffusion activation energy of 293 kJ/mol [87]. However, the low measured activation energy for creep in this regime (< 250 kJ/mol) does not support this hypothesis. Záležák et al. [119] conducted an important experiment to address this discrepancy by comparing the creep properties of polycrystalline CrMnFeCoNi with those of its single-crystal counterpart.

As shown in **Fig. 2**, creep tests conducted at 800 °C and 980 °C under both compression and tension consistently yielded a stress exponent of approximately 3-4 for polycrystalline CrMnFeCoNi, and a stress exponent of around 5-6 for single crystals. The aforementioned hypotheses cannot explain the $n = 5$ power-law behavior observed in single crystals, suggesting that this discrepancy is likely linked to grain-boundary effects. It is reported that CrMnFeCoNi shows Class M behavior with normal primary creep and subgrain boundary formation during secondary creep [92, 119]. Therefore, it was argued that grain boundaries can enhance dynamic recovery by dislocation absorption at grain boundaries mediated by grain boundary diffusion, which accounts for the lower stress exponents and activation energies observed in CrMnFeCoNi compared to those associated with climb-controlled creep through the annihilation of dislocation dipoles (**Fig. 2c**). Interestingly, even if the grain boundary recovery mechanism appears widespread, it is not envisaged to significantly influence coarse-grained conventional alloys; rather, these alloys are affected only when they have ultrafine grain sizes, where creep is governed by grain boundary sliding with $n = 2$ and $Q_C = Q_{GB}$ [128, 129].

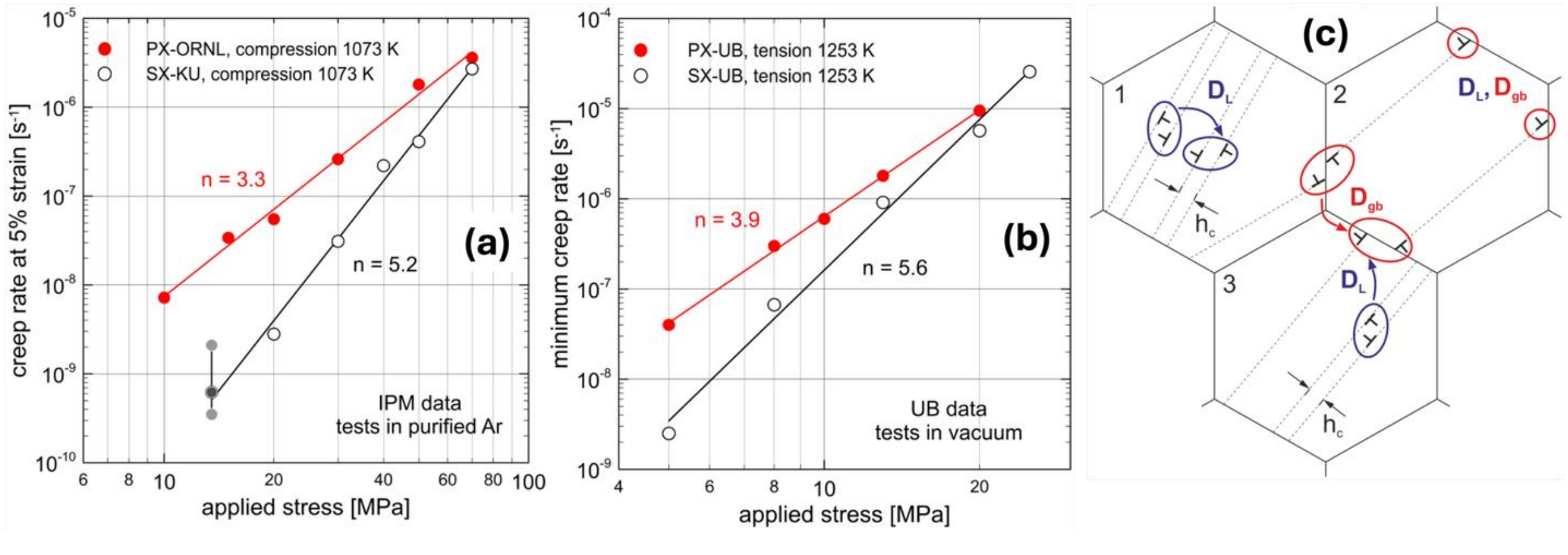

**Figure 2** Secondary creep data for single crystalline and polycrystalline CrMnFeCoNi tested (a) under compression at 800 ºC and (b) under compression at 980 ºC. A schematic illustrating the role of grain boundaries in creep is provided in (c). The images were taken from Ref. [119].

The above rationalization can account for most of the creep data in **Table 2**, except for some data for CrMnFeCoNi at intermediate temperatures associated with high stress exponents and apparent activation energies [88]. This behavior can be explained either by power-law breakdown or by the interaction of dislocations with secondary phase particles. Phase decomposition resulting in secondary phases is known to occur in originally single-phase FCC HEAs after prolonged intermediate-temperature thermal exposures both in the absence [130] and presence of stress [87, 117].

**Fig. 3a-f** shows the results of annealing experiments on CrMnFeCoNi at 700 ºC, which reveals the formation of a Cr-rich σ phase that can provide particle strengthening [130]. As shown in **Fig. 3g-l**, further annealing at 500 ºC reveals the decomposition of the FCC matrix into $L1_0$-NiMn, BCC-Cr, and B2-FeCo phases, which were observed in a long-term creep study at 600 ºC [115]. The FCC phase stability is better in CrFeCoNi and CrCoNi, where the formation temperatures for Cr-rich σ phase and other phases are lower [117] compared to those of CrMnFeCoNi. However, the formation of Cr-rich σ phase has been reported in a long-term creep study of CrFeCoNi from 600-650 ºC, along with $Cr_{23}C_6$ carbides at the grain boundaries [121]. The influence of these secondary phases on creep cannot be ignored, as they consistently lead to higher stress exponents and apparent activation energies.

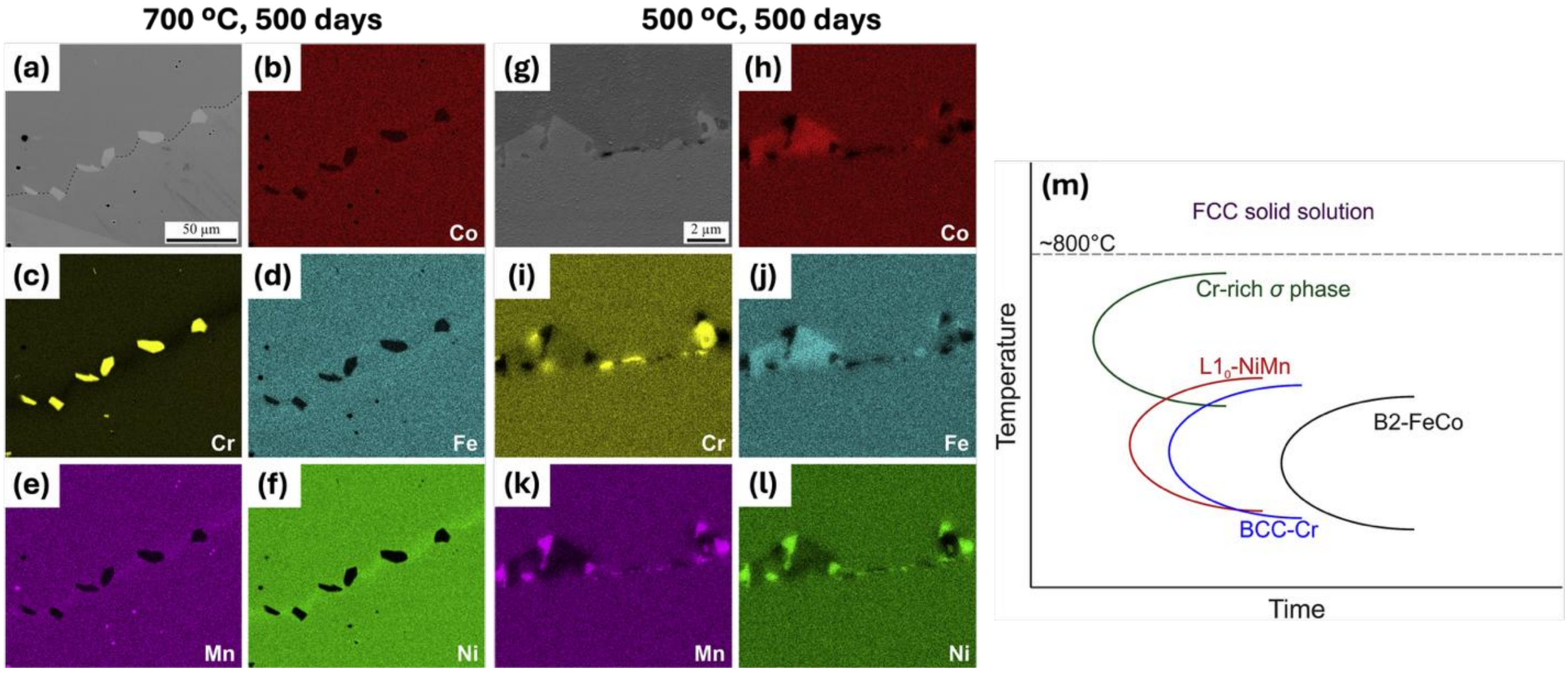

**Figure 3** Microstructure of CrMnFeCoNi HEA after annealing at (a-f) 700 °C for 500 days and (g-l) 500 ºC for 500 days; (m) presents a schematic TTT diagram, hypothesizing the kinetics of phase decomposition at temperatures below 800 ºC. The images were taken from Ref. [130].

The comparison of the creep properties of CrMnFeCoNi with its quaternary and ternary derivatives, CrFeCoNi and CrCoNi, is given in **Fig. 4**. **Fig. 4a-b** demonstrates that recrystallized CrFeCoNi is stronger than CrMnFeCoNi, while **Fig. 4c** indicates that recrystallized CrCoNi possesses superior creep resistance to CrMnFeCoNi. These results are in line with the observation that quasi-static mechanical properties do not necessarily improve with the number of constituent elements (i.e., entropy). For example, the temperature-dependent yield strengths of binary to quinary alloys in the CrMnFeCoNi system follow complicated trends that depend on the type rather than number of constituent elements [131], which has been explained on the basis of volume misfits [132].

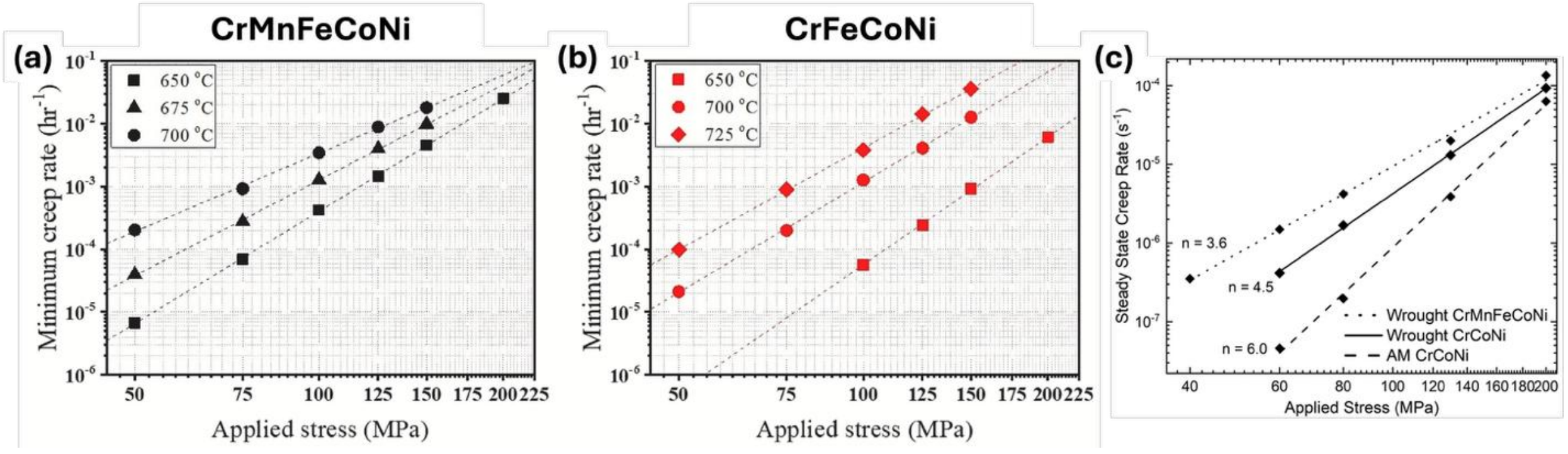

**Figure 4** Secondary creep data for recrystallized (a) CrMnFeCoNi and (b) CrFeCoNi tested between 650 and 725 ºC; and (c) shows a comparison among creep rates of recrystallized CrMnFeCoNi and CrCoNi as well as additively manufactured CrCoNi tested at 750 ºC. The images were taken from Refs. [117, 124].

There are limited, but intriguing, effects of additive manufacturing on the creep properties of FCC HEAs. Notably, AM CrCoNi has greater creep strength than its wrought counterpart, as shown in **Fig. 4c**. Laser powder bed fusion (LPBF)-AM utilized in this study [124] introduces numerous oxide dispersoids during processing (**Fig. 5**), which can interact with dislocations and increase creep resistance due to particle strengthening. This effect was used to rationalize the higher stress exponent and apparent activation energy for creep observed in the AM material [124]. These findings also motivate an oxide dispersion strengthening strategy, enabled by AM, to enhance creep performance, as discussed in the following section. Furthermore, AM followed by hot isostatic pressing (HIP) reduces the fraction of weaker low-angle grain boundaries (LAGBs) while increasing the proportion of high-angle grain boundaries (HAGBs) and twin boundaries (TB), thereby further elevating creep strength.

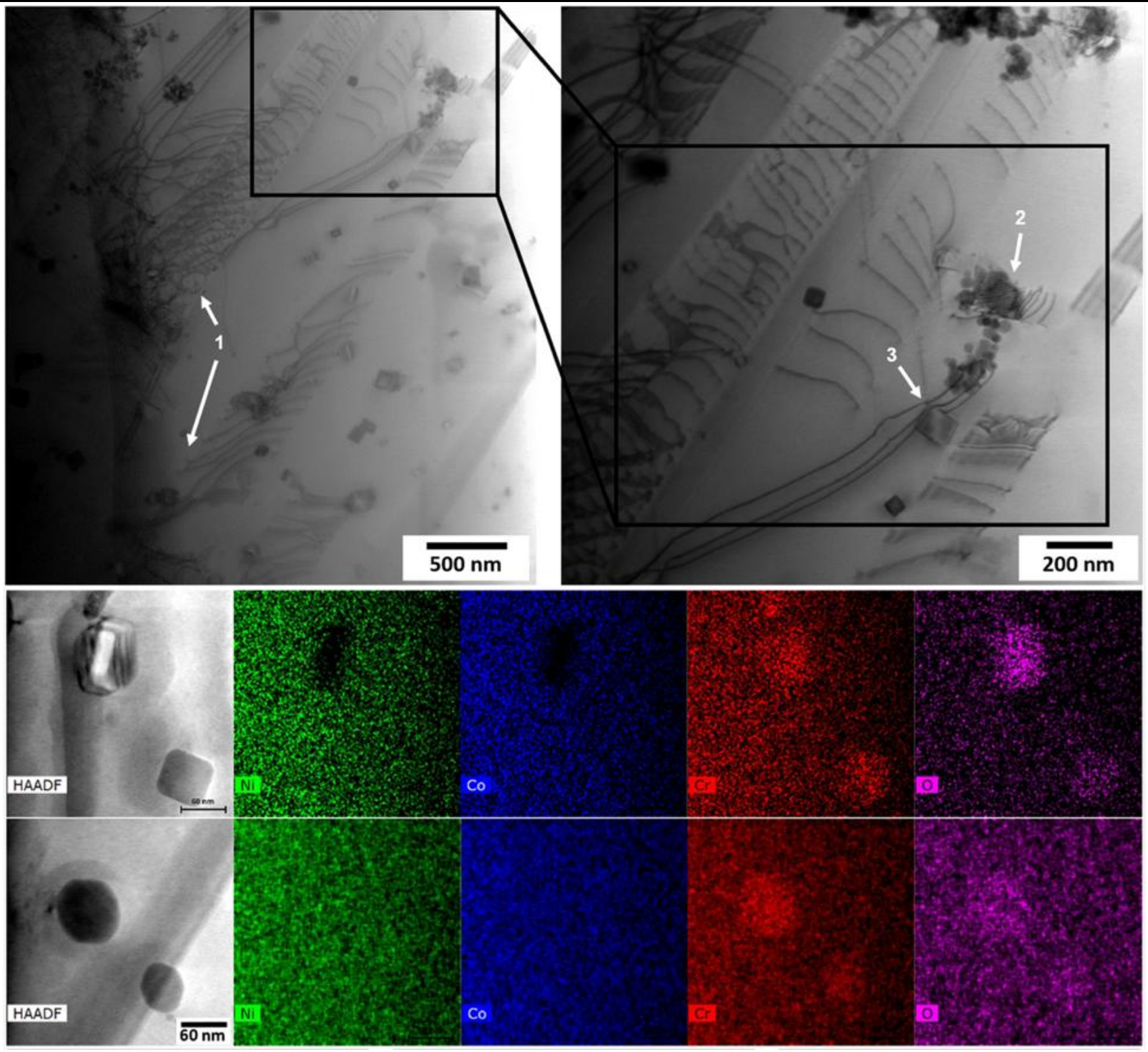

**Figure 5** Creep microstructure in AM CrCoNi crept to steady state at 1023 K and 80 MPa, showing planar slipping dislocations stopped by a grain boundary (arrows 1), generation of dislocations at an incoherent particle/matrix interface (arrow 2), and dislocation interactions with oxides such as bowing (arrow 3). The EDS image at the bottom right shows these particles are Cr-rich oxides. The images were taken from Ref. [124].

A final observation regarding the creep behavior of single-phase HEAs is the substantial improvement in creep resistance when derivatives of the CrMnFeCoNi system are alloyed with V and Mo, elements known to significantly increase lattice distortion and the activation energy for diffusion [125]. **Fig. 6** compiles available creep data for single-phase FCC HEAs for different temperatures and stresses that are normalized following the usual Mukherjee-Bird-Dorn approach [133]. The temperature-dependent elastic modulus, $E(T)$, for CrMnFeCoNi, CrFeCoNi, and CrCoNi are reported in Refs. [134, 135]. For all other HEAs with unreported $E(T)$ and $Q_C$, rough estimates of their values were obtained using a rule-of-mixture average. Caution is warranted here because the creep activation energy in HEAs can be dominated by a small number of elements, even when present only in minor amounts. As a result, simple rule-of-mixture estimates may substantially misrepresent the true activation energy, but they are performed here solely due to the limited availability of data. Therefore, to enable fair comparisons of creep resistance, it is recommended to use the original creep data. It can be observed that all data for CrMnFeCoNi, CrFeCoNi, and CrCoNi collapse to follow a similar trend line, whereas $Ni_{42.5}Co_{21.3}Fe_{21.3}V_{10.6}Mo_{4.3}$ stands out with a significantly lower creep rate by 1-2 orders of magnitude. Alloying with Mo can induce solute-drag creep with a stress exponent of 3, where viscous glide of dislocations is controlled by the diffusion of Mo, which results in a high measured activation energy of 456 kJ/mol.

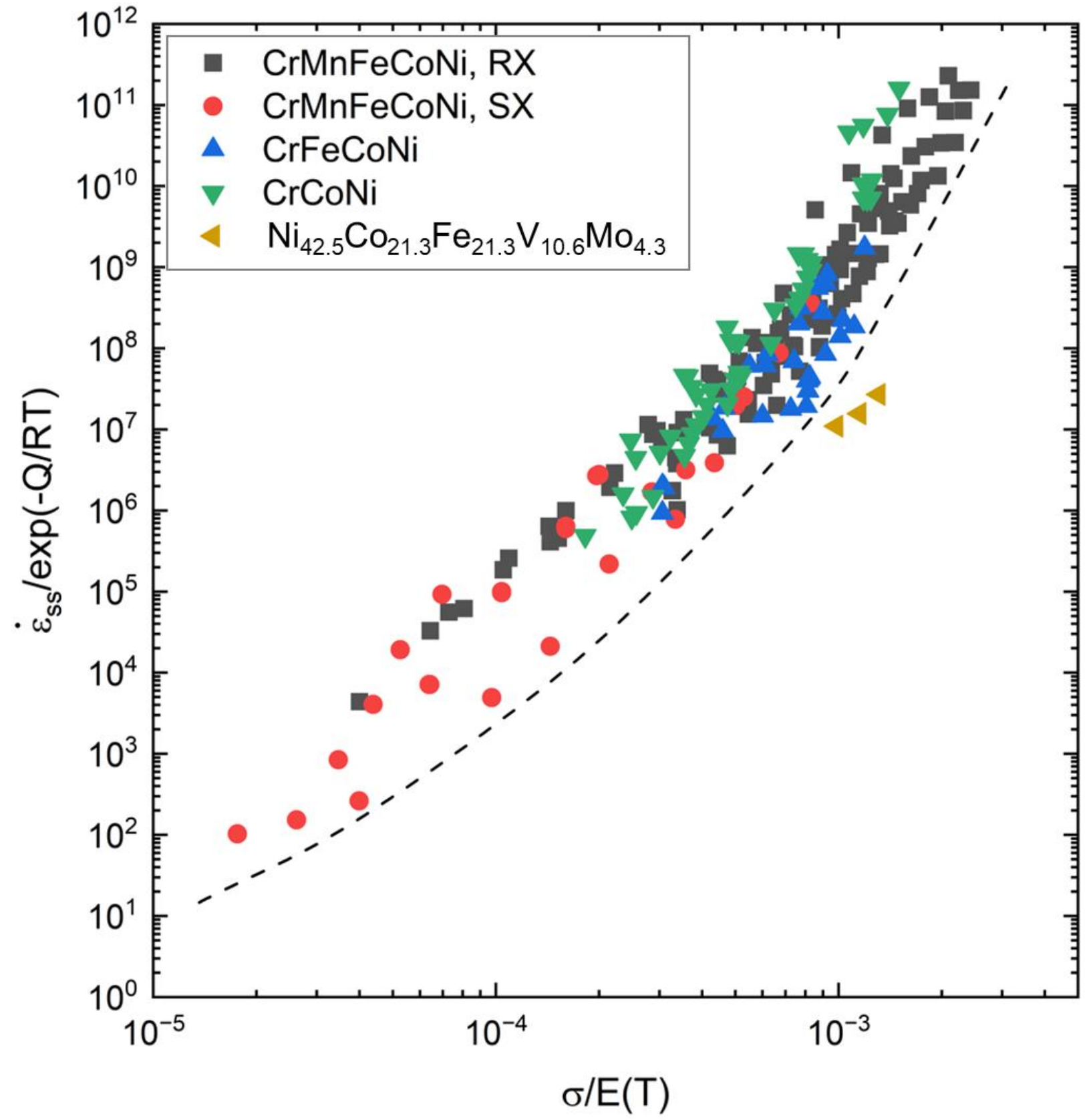

**Figure 6** Normalized secondary creep data for single-phase FCC HEAs. The dashed line separates the creep data for CrMnFeCoNi and its derivatives from those for $Ni_{42.5}Co_{21.3}Fe_{21.3}V_{10.6}Mo_{4.3}$. The creep data were adapted from Refs. [87, 88, 92, 115-125]. A comparison between the creep data of high entropy alloys and commercial alloys is given in the **Discussion** section.

*2.2. Multi-Phase FCC High-Entropy Alloys*

Although the creep properties of single-phase FCC HEAs have been extensively studied, there is broad consensus in the literature that their creep resistance remains insufficient for high-temperature applications. Therefore, secondary phase strengthening must be pursued to enhance their performance under such conditions. To date, there are mainly four strategies to introduce a multi-phase structure into FCC HEAs: **1.** Through the alloying of Al to achieve an FCC/BCC dual phase structure, sometimes with ordered phases such as B2+$L1_2$ (similar to eutectic HEAs) [136-139]; **2.** Through a similar alloying route as in Ni-based superalloys but with increased solute concentrations of Al, Cr, Co, and Fe and a reduced concentration of base Ni (i.e., high-entropy superalloys, HESAs) [40, 140]; **3.** Through the alloying of Mo and intermediate temperature exposure to form μ phase [141]; and **4.** Through oxide dispersion strengthening (i.e., ODS-HEAs) [142-144]. The creep parameters for multi-phase FCC HEAs are summarized in **Table 3**, and secondary creep data as a function of applied stress are given in **Supplementary Table 2**.

**Table 3 Creep parameters of multi-phase FCC HEAs for different microstructural states and testing conditions. All compositions are expressed in atomic percent; equiatomic compositions are considered self-explanatory.**

| Alloy | Loading | Microstructure | T (ºC) | $n$ | $Q_C$ (kJ/mol) | Ref. |
|---|---|---|---|---|---|---|
| $Al_{7.5}Cr_{18.5}Mn_{18.5}Fe_{18.5}Co_{18.5}Ni_{18.5}$ | Tension | FCC + BCC | 600-700 | 4.2-4.9 (hi $\sigma$)<br>1.6-2.2 (lo $\sigma$) | 346-351<br>254-262 | [136] |
| $Al_{10.2}Cr_{17.9}Mn_{17.9}Fe_{17.9}Co_{17.9}Ni_{17.9}$ | Tension | FCC + BCC | 600<br>650<br>700 | 5.8<br>3.5<br>2.8 | 182-193 | [136] |
| $Al_9Cr_{9.1}Mn_{27.3}Fe_{45.5}Co_{9.1}$ | Tension | FCC + BCC | 650 | 5 | 182 | [137] |
| $Al_{6.8}Cr_{23.3}Fe_{23.3}Co_{23.3}Ni_{23.3}$ | Tension | FCC + B2 + $L1_2$ | 700<br>730<br>760 | 3<br>4.5<br>6.5 | 391-548 | [138] |
| $Al_{16.4}Cr_{16.4}Fe_{16.4}Co_{16.4}Ni_{34.4}$ | Tension | FCC + B2 + $L1_2$ | 700-900 | 3.5-4.4 | 393 (700-800 ºC)<br>62 (800-900 ºC) | [139] |
| $Ni_{36}Al_{10}Co_{25}Cr_8Fe_{15}Ti_6$ | Tension | FCC + $L1_2$ | 750 | 10.8 | 486 | [140] |
| $Ni_{47.9}Al_{10.2}Co_{16.9}Cr_{7.4}Fe_{8.9}Ti_{5.8}Mo_{0.9}Nb_{1.2}W_{0.4}C_{0.4}$ | Tension | FCC + $L1_2$ | 750-982 | - | 290 | [40] |
| $Cr_{19}Fe_{19}Co_{19}Ni_{38}Mo_5$<br>$Cr_{18.5}Fe_{14.8}Co_{14.8}Ni_{44.4}Mo_{7.5}$<br>$Cr_{18.8}Fe_{12.5}Co_{12.5}Ni_{50.0}Mo_{6.3}$ | Tension | FCC + μ phase | 650 | - | - | [141] |
| $Y_2O_3$-$TiO_2$-CrMnFeCoNi | Compression | ODS-FCC | 700-800 | 1.8 (lo σ)<br>13.2 (hi σ) | 210<br>580 | [142] |
| $Y_2O_3$-CrCoNi | Tension | ODS-FCC | 750-900 | 6.4-6.7 | 335-367 | [143] |
| C-CrMnFeCoNi | Tension | Carbide-FCC | 600 | 3 (lo σ) 7 (hi σ) | - | [144] |

For Al-containing CrMnFeCoNi and derivatives, the stress exponents and activation energies for creep can vary drastically as a function of testing temperature and stress and are therefore less comprehensible than single-phase CrMnFeCoNi. At least part of the difficulty arises from the phase evolution during creep where the Al-rich BCC phase, NiAl B2 phase, and $Ni_3Al$ $L1_2$ phase can form, coarsen, or dissolve depending on testing temperature and time [138]. Furthermore, the addition of Al can also increase the tendency for the decomposition of the FCC matrix to form Cr-rich σ phase [136, 138]; for example, in the case of $Al_{10.2}Cr_{17.9}Mn_{17.9}Fe_{17.9}Co_{17.9}Ni_{17.9}$, the amount of σ in the FCC phase can reach 43% by volume when creep tested at 700 ºC. These precipitates can improve creep strength, but they decrease the time to rupture due to local stress concentration and easier crack and void formation [136, 137]. Chen et al. [138] performed dedicated creep tests at different temperatures interrupted at various times to reveal the microstructural evolution in an $Al_{6.8}Cr_{23.3}Fe_{23.3}Co_{23.3}Ni_{23.3}$ HEA. They illustrated the coarsening of B2 and $L1_2$ phases by comparing specimens deformed at 700 ºC under 75 MPa for 50 h and 250 h, as shown in **Fig. 7a-b**. At 760 ºC, the volume fraction of the $L1_2$ and the B2 phase was determined to be low for the specimen creep tested at 75 MPa for 20 h (**Fig. 7c**). However, **Fig. 7d** shows that complementary annealing tests at 760 ºC for 50 h reveal a large increase in volume fraction. These results highlight that it is not straightforward to interpret the stress exponents and activation energies for creep of $Al_{6.8}Cr_{23.3}Fe_{23.3}Co_{23.3}Ni_{23.3}$ (**Fig. 7e-f**); at high temperatures and high stresses, the minimum creep rate is elevated because rupture occurs quickly, preventing precipitates from reaching their equilibrium volume fraction. In contrast, the creep rate at lower temperatures and stresses for longer testing durations reflects a competition between increasing precipitate volume fraction and precipitate coarsening. Therefore, the microstructure corresponding to the minimum creep rate at a given stress and temperature can vary significantly, rendering the extracted apparent stress exponents and activation energies less meaningful.

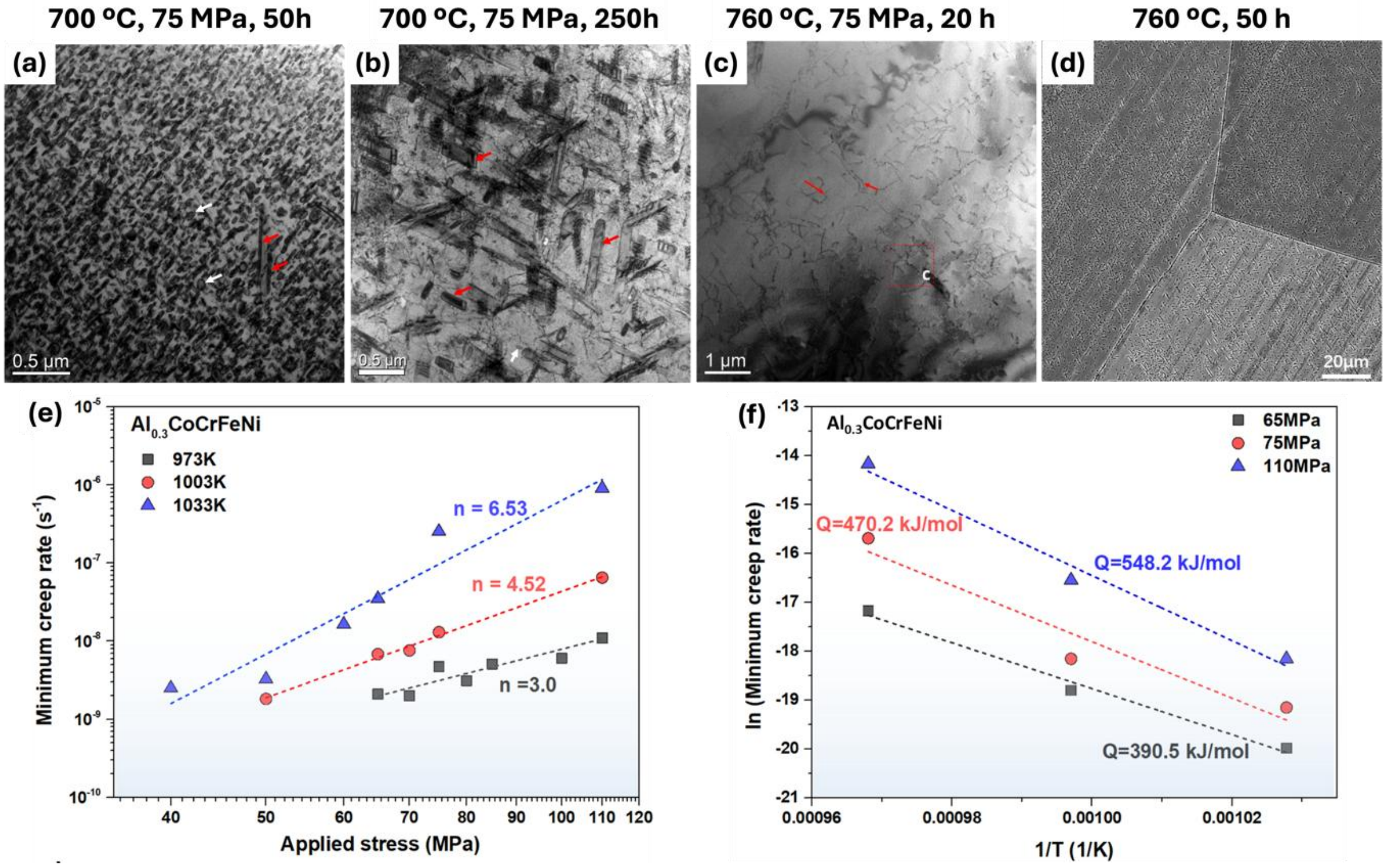

**Figure 7** Microstructure of $Al_{6.8}Cr_{23.3}Fe_{23.3}Co_{23.3}Ni_{23.3}$ (written as $Al_{0.3}CrFeCoNi$ in the paper) after creep testing: (a) 75 MPa at 700 ºC for 50 h, (b) 75 MPa at 700 ºC for 250 h, (c) 75 MPa at 760 ºC for 20 h, and (d) after just annealing at 760 ºC for 50 h. The red and white arrows indicate $L1_2$ and B2 phases, respectively. The minimum creep rate is plotted against applied stress in (e); and (f) shows an Arrhenius plot of the minimum creep rate against inverse temperature to extract the apparent activation energy. The images were taken from Ref. [138].

An ideal mechanistic study would pre-age samples to achieve equilibrium microstructures at the creep testing temperature prior to loading. However, in practice, most studies begin with metastable, supersaturated solid solutions. Nevertheless, it may be concluded that multi-phase FCC HEAs strengthened through Al alloying can exhibit higher creep resistance than CrMnFeCoNi, primarily due to precipitation hardening, where the Orowan bowing mechanism is frequently observed [138, 139].

The multi-phase processing strategy that yields an FCC + $L1_2$ (or γ + γ') structure in HEAs [40, 140] is not reviewed in detail here, given its close resemblance to that of conventional Ni-based superalloys. As a result, the creep resistance of these HESAs is comparable to commercial superalloys, such as RENE' 80 and NX-188 [40]. In addition, the tensile yield strength and creep resistance of HESAs remain lower than those of the CMSX series, which contain substantial amounts of refractory elements [145-147]. Increasing the ideal configurational entropy of the FCC matrix in HESA design appears to have little impact on creep resistance; however, this strategy can help stabilize the $\gamma + \gamma'$ microstructure by suppressing the formation of detrimental topologically close-packed (TCP) phases at the expense of a slightly lower solvus temperature for the $\gamma'$ phase at 1199 ºC [40]. Further compositional design of HESAs aimed at optimizing $\gamma'$ phase volume fraction and morphology, anti-phase boundary (APB) energy, and particle–matrix misfit must take full advantage of the vast literature available on superalloys instead of focusing on, and comparing with, only what is available in the HEA literature. It is also worth remembering that existing superalloys have been optimized over many decades, and it seems highly unlikely that the same set of alloying elements would produce HESAs (or other variants) that have better combinations of characteristics (including creep strength, rupture life, ductility, fracture toughness, oxidation resistance, fatigue properties, processability, and cost).

It has been shown above that the creep properties of single-phase FCC HEAs can significantly benefit from alloying with Mo despite Rozman et al. [141] showing that intermediate temperature aging or creep testing can result in the formation of the Mo-rich μ phase (**Fig. 8**). In addition, Mo tends to react with carbon present as an impurity in the HEA to form $Mo_6C$ carbides at grain boundaries. The formation of these secondary phases increased the time to rupture by ~100 times compared to their single-phase counterparts without Mo. Out of the three Mo-containing compositions they designed, $Cr_{18.5}Fe_{14.8}Co_{14.8}Ni_{44.4}Mo_{7.5}$ shows favorable intragranular precipitation of μ phase with an area fraction of 0.057. This precipitation hardening effect offsets the reduction of solid solution hardening due to loss of Mo from the matrix. In addition, the formation of $Mo_6C$ carbides can strengthen the grain boundaries and prevent the μ phase from forming at grain boundaries, which causes grain boundary embrittlement as in superalloys [148]. Overall, the addition of Mo can enhance the creep resistance of CrFeCoNi-based HEAs to levels comparable to Inconel 625, whereas the base HEA exhibits creep behavior similar to that of austenitic stainless steels [141].

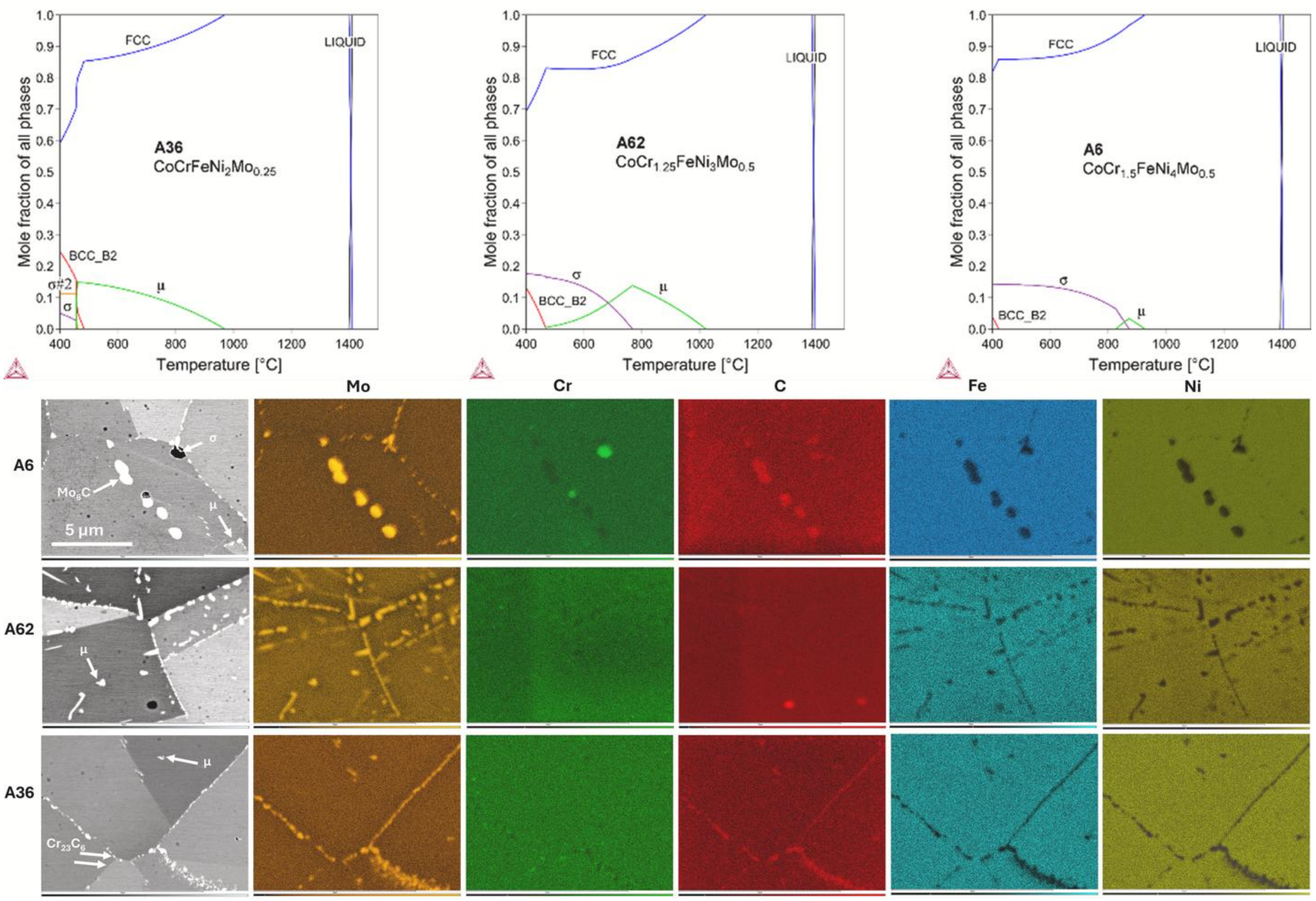

**Figure 8** CALPHAD results for $Cr_{19}Fe_{19}Co_{19}Ni_{38}Mo_5$, $Cr_{18.5}Fe_{14.8}Co_{14.8}Ni_{44.4}Mo_{7.5}$, $Cr_{18.8}Fe_{12.5}Co_{12.5}Ni_{50.0}Mo_{6.3}$ (written as $CrFeCoNi_2Mo_{0.25}$, $Cr_{1.25}FeCoNi_3Mo_{0.5}$, and $Cr_{1.5}FeCoNi_4Mo_{0.5}$ in the paper, respectively) showing the formation of Mo-rich μ phase at intermediate temperatures. SEM-EDS confirms the existence of μ phase along with $Mo_6C$ carbides, Cr-rich σ phase, as well as $Cr_{23}C_6$ carbides. The images were taken from Ref. [141].

Dispersion strengthening offers an alternative approach to producing multi-phase alloys that leaves the HEA matrix largely unaltered by incorporating oxide or carbide dispersoids into the base alloy through rapid solidification processing [149-151] or additive manufacturing techniques [57, 58, 152]. Compared to precipitation hardening, dispersion strengthening can remain effective at even higher temperatures, as it avoids issues related to precipitate coarsening and dissolution, though this comes at the cost of a more modest strengthening effect at intermediate temperatures. Dobeš et al. [142] fabricated ODS-CrMnFeCoNi by blending powders of pure elements and adding $O_2$ gas, Y, and Ti to form ~0.3 wt% oxides during ball milling. The mechanically alloyed powder with oxides was then compacted by spark plasma sintering (SPS), which yielded finely dispersed oxides in CrMnFeCoNi with an ultrafine grain size of ~800 nm. After creep, the grain size remained stable due to oxide pinning of the grain boundaries. The creep data presented in **Fig. 9** demonstrate that ODS significantly increases the creep resistance of CrMnFeCoNi, with improvements ranging from one to three orders of magnitude. The strengthening effect is more pronounced at higher temperatures. Additionally, a transition in the creep mechanism is observed between low and high stress regimes, as evidenced by the evolution of the stress exponent from 1.8 to 13.2. The corresponding activation energies in the two regimes were reported to transition from 210 to 580 kJ/mol.

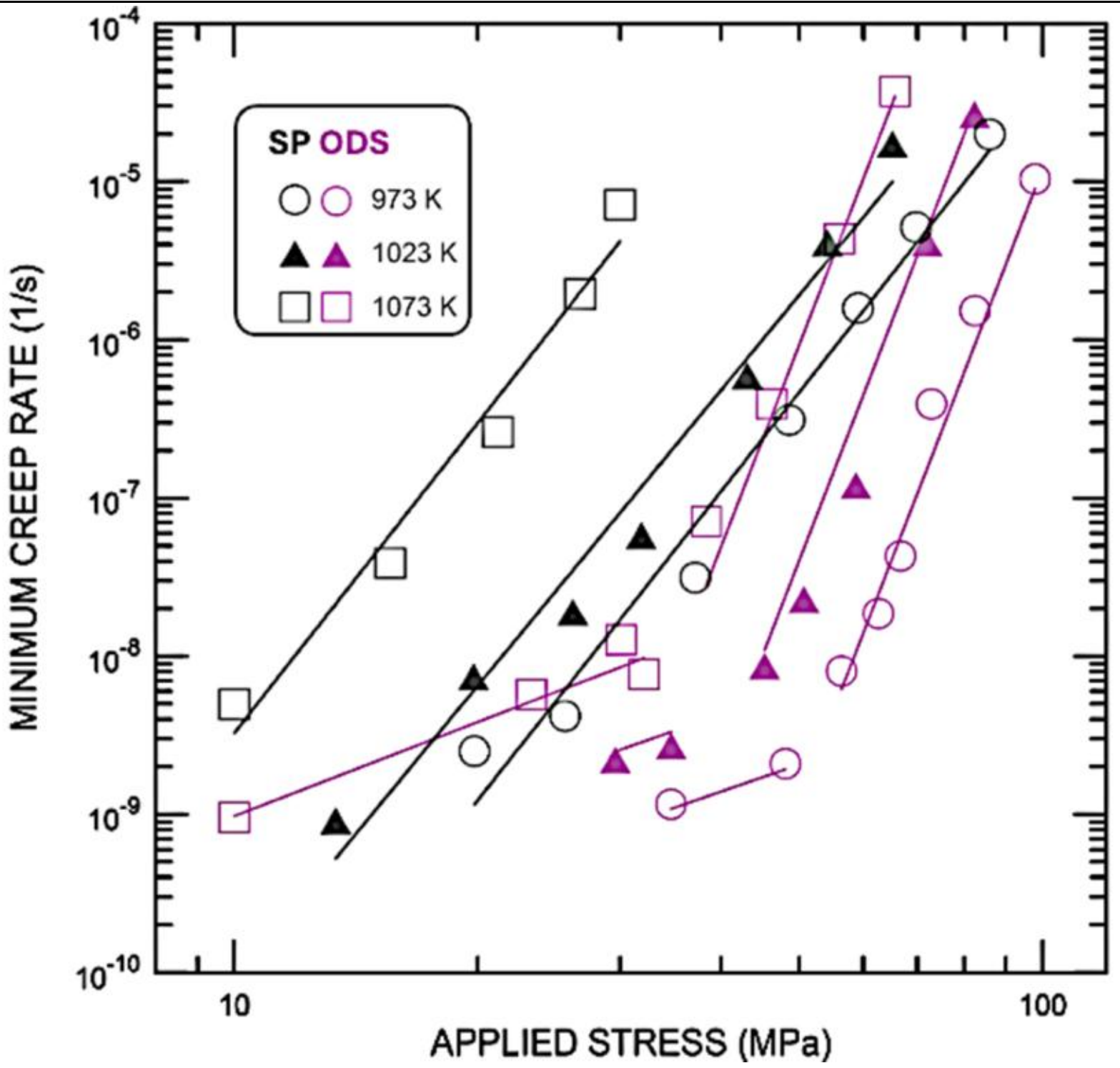

**Figure 9** Minimum creep rate as a function of applied stress for single-phase (SP) CrMnFeCoNi and ODS-CrMnFeCoNi. The image was adapted from Ref. [142].

The high stress exponents and activation energies observed in ODS alloys are often attributed to the presence of a back stress on dislocations induced by particles [153-155]. The creep equation can be modified as:

$$\dot{\varepsilon}_{SS} = A'(\sigma - \sigma_p)^n \exp\left(-\frac{Q_C}{RT}\right), \tag{2}$$

where $\sigma_p$ is the back stress, and $n$ is approximately that for the single-phase CrMnFeCoNi, which is 6.3 in this study. The threshold stress for ODS-CrMnFeCoNi was determined to be in the range of 25–35 MPa, which is quite substantial at 800 °C according to the creep data in **Fig. 9**. Creep deformation below this stress did not vanish entirely but instead transitioned to a different mechanism, characterized by a stress exponent of 1.8 and an activation energy of 210 kJ/mol. Although the stress exponent is similar to that for grain boundary sliding, Dobeš et al. [142] attributed this regime to Coble creep since the grain boundaries are pinned by oxide particles, and the modest tensile ductility reported for the ODS alloy in a separate study [55] does not support the occurrence of grain boundary sliding typically associated with superplasticity.

In addition to the conventional powder metallurgy route for producing ODS-HEAs, Smith et al. [58] demonstrated that LPBF-AM can also achieve a uniform dispersion of oxides within the HEA matrix. This is enabled by resonant acoustic mixing to coat a prealloyed CrCoNi powder with nanoscale $Y_2O_3$ particles (**Fig. 10a-e**). The AM material features elongated grains along the build direction as well as tensile residual stresses at the surface, as shown in **Fig. 10f**. HIP was employed to relieve the residual stress and further consolidate the samples, while preserving the elongated grain structure due to the pinning effect of the oxide particles on grain boundaries. This behavior contrasts with the oxide-free AM-CrCoNi described above, where recrystallization led to the formation of a high density of annealing twins.

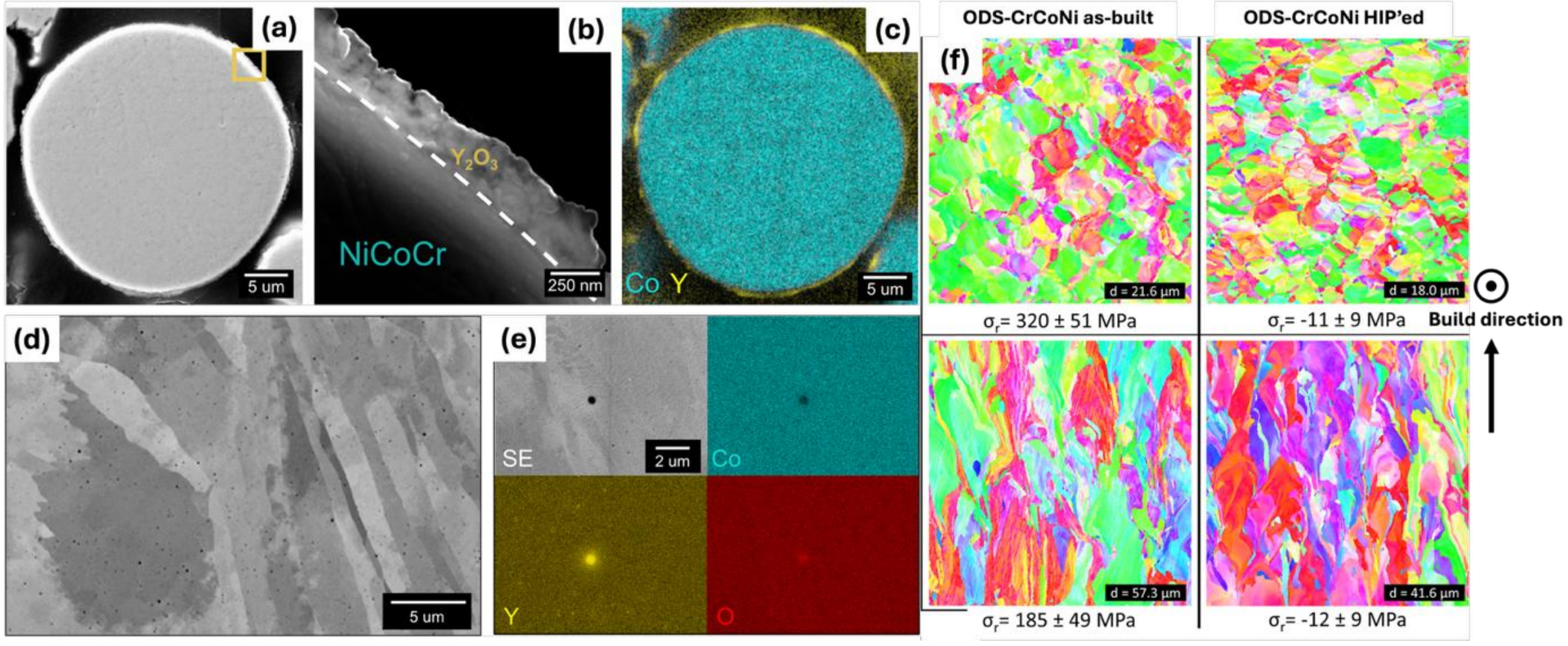

**Figure 10** (a-c) SEM images of prealloyed CrCoNi powder coated with $Y_2O_3$, which resulted in (d, e) an ODS-CrCoNi alloy with uniformly dispersed $Y_2O_3$ particles after LPBF-AM; (f) shows that the alloy exhibits an equiaxed structure perpendicular to the build direction and an elongated structure along the build direction. Tensile residual stresses were measured in the as-built AM material, which were then eliminated by HIP'ing. The images were adapted from Ref. [58].

Sahragard-Monfared et al. [143] studied the creep properties of this AM ODS-CrCoNi alloy. As shown in **Fig. 11**, the AM ODS-CrCoNi alloy exhibits a creep rate approximately one order of magnitude lower than that of AM CrCoNi and two orders of magnitude lower than wrought CrCoNi. Furthermore, it demonstrates superior creep resistance compared to the SPS ODS-CrMnFeCoNi alloy. This improvement is attributed to the inherently higher creep strength of the CrCoNi matrix compared to CrMnFeCoNi, as well as the retention of a high dislocation density during additive manufacturing.

**Fig. 12** presents the TEM dislocation microstructure of HIP'ed and annealed AM ODS-CrCoNi, revealing a high dislocation density and extensive dislocation–particle interactions. Particle pinning effectively inhibits dislocation annihilation during annealing. After creep, distinct slip traces and continued dislocation–particle interactions were observed.

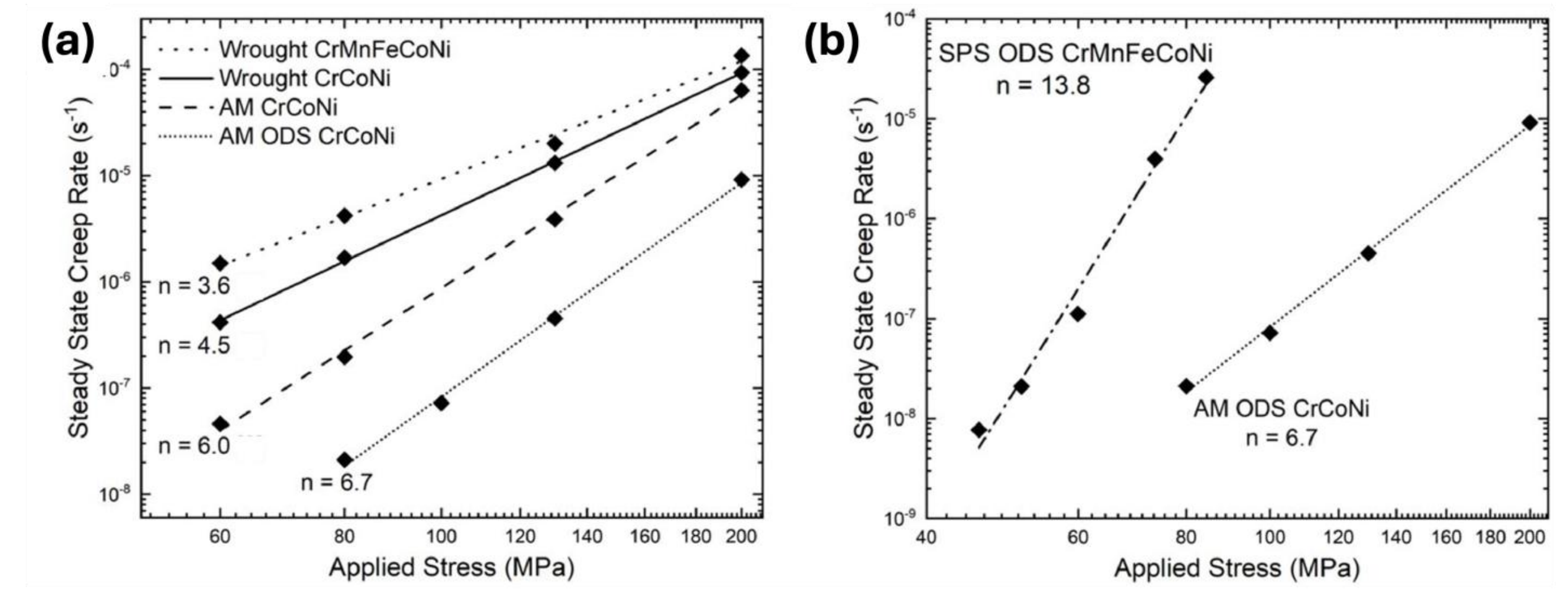

**Figure 11** (a) Steady-state creep rate as a function of applied stress for wrought CrMnFeCoNi, wrought CrCoNi, AM CrCoNi, and AM ODS-CrCoNi; (b) Comparison between the creep rates of the SPS ODS-CrMnFeCoNi reported by Dobeš et al. [142] and AM ODS-CrCoNi. The creep tests were carried out at 750 °C. The images were taken from Ref. [143].

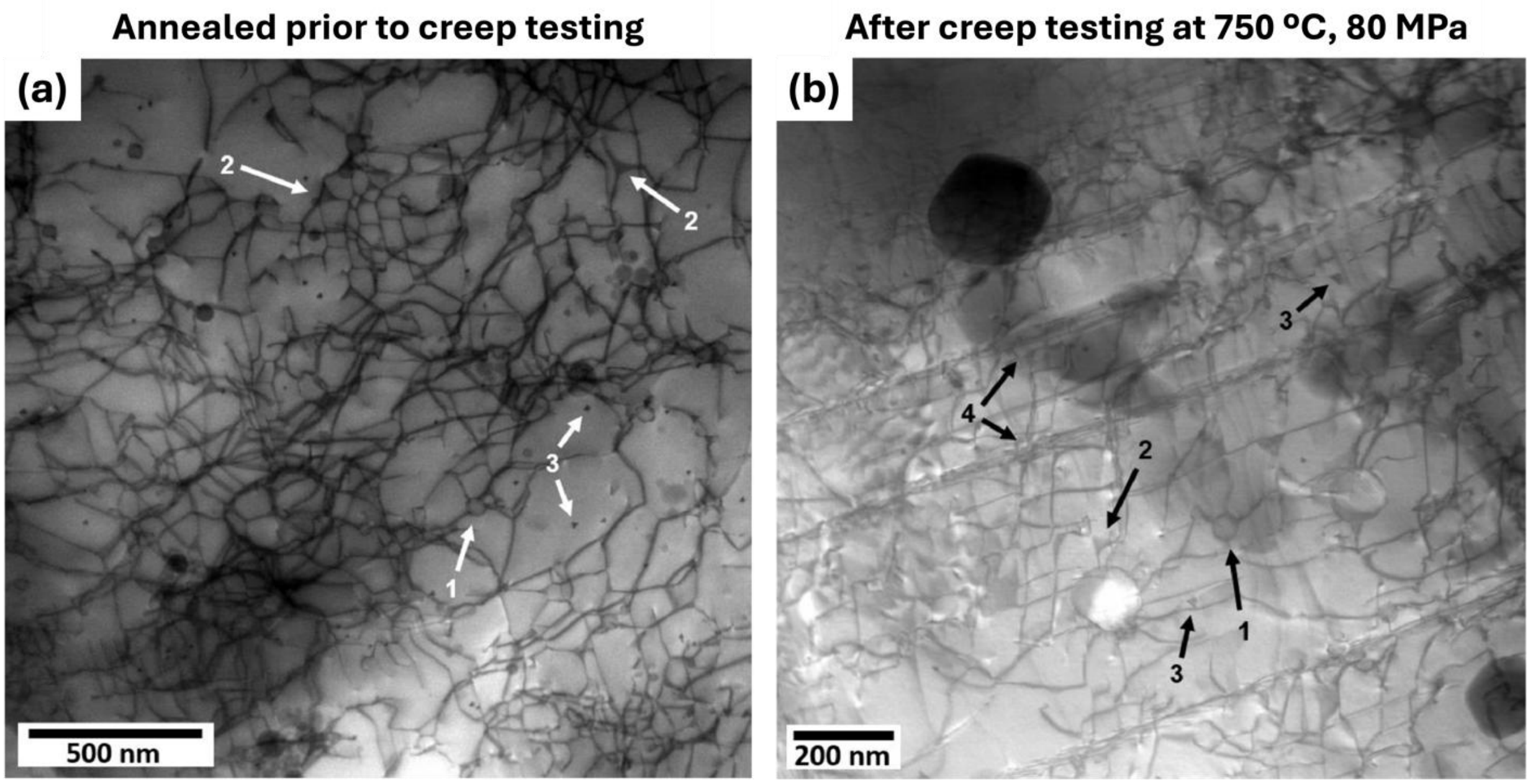

**Figure 12** TEM microstructure of AM ODS-CrCoNi samples (a) annealed at 750 ºC for 3.5h prior to creep testing and (b) creep tested at 750 ºC under 80 MPa. The images were taken from Ref. [143].

Smith et al. [57] further engineered the composition of AM ODS-CrCoNi and developed NASA GRX-810, an $Y_2O_3$ dispersion-strengthened $Ni_{31.9}Co_{32.6}Cr_{32.5}Re_{0.5}Al_{0.6}Ti_{0.3}Nb_{0.5}W_{0.9}C_{0.2}$ (in at.%) alloy with minor additions of Al, Ti, and refractory elements for further solid solution strengthening. The matrix of the alloy remains mostly single-phase FCC with some (Nb,Ti)-carbides at grain boundaries that can increase grain boundary strength [156, 157]. GRX-810 exhibits a remarkable enhancement in creep performance, as shown in **Fig. 13**, with a 2000-fold increase in creep rupture life compared to AM CrCoNi and a 200-fold increase compared to AM ODS-CrCoNi at 1093 °C under 30 MPa. Furthermore, the creep rupture life at 1093 °C significantly surpasses that of many commercial 3D-printable Ni- and Co-based superalloys, including Haynes 230, Inconel 625, and Hastelloy X. Although GRX-810 does not yet match the creep strength of other superalloys (e.g., CMSX-10), whose creep rupture time can exceed 2000 hours at stresses above 200 MPa at 1100 ºC [147], its 3D printability presents advantages for rapid near-net-shape manufacturing, highlighting its potential for commercialization.

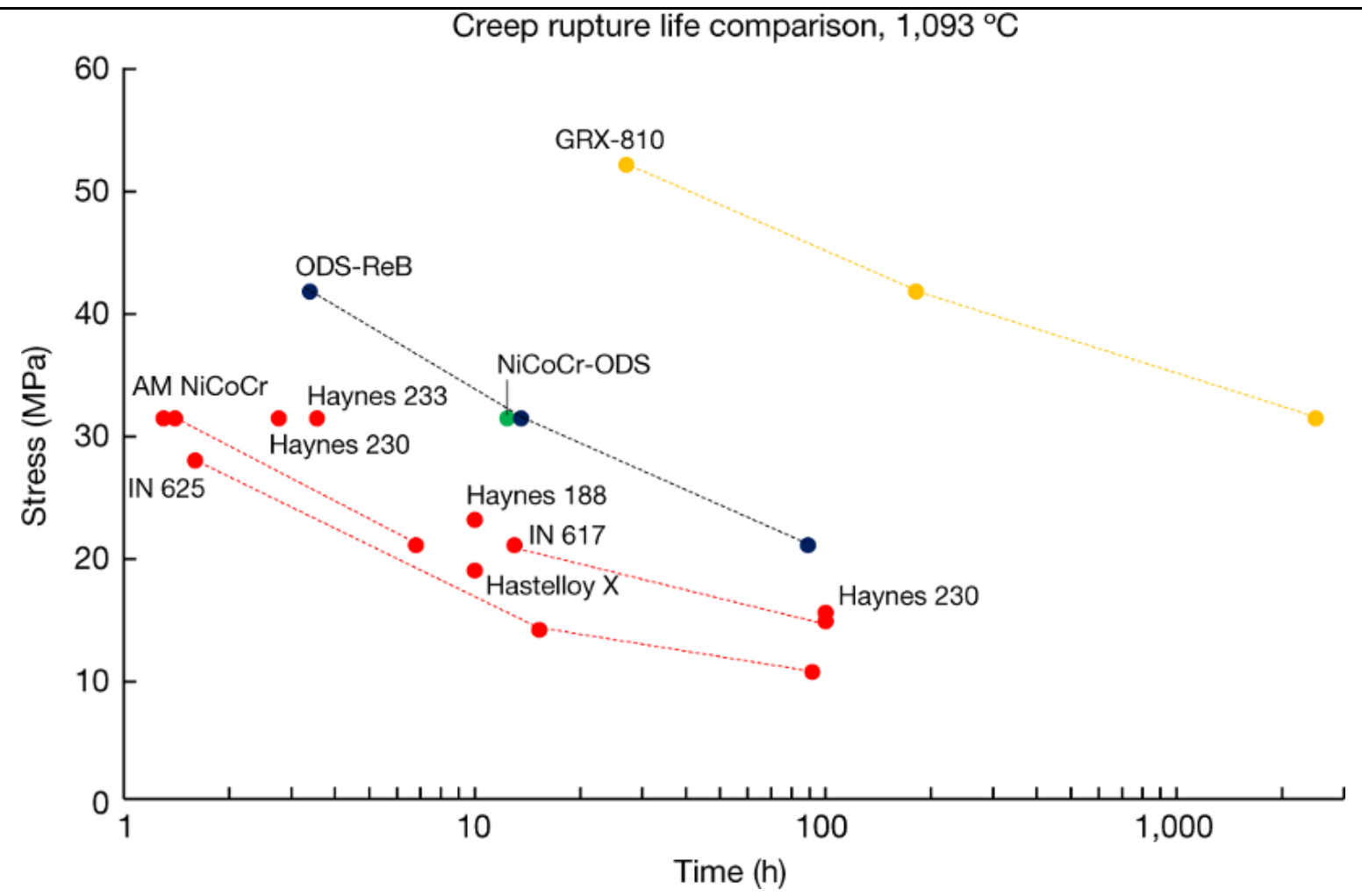

**Figure 13** Comparison of the creep rupture life at 1093 ºC of various AM high-temperature alloys. The image was taken from Ref. [57]. Note here that all mentioned alloys do not contain the g' phase for particle strengthening.

A comparative overview of the creep properties of multi-phase FCC HEAs is presented in **Fig. 14**. Admittedly, this compilation is less reliable than the single-phase FCC HEAs shown in **Fig. 6**, due to the ambiguity of the correct activation energy needed to normalize the creep rates for multi-phase materials. For example, use of the abnormally high apparent activation energies reported for ODS-HEAs or HESAs can lead to artificially high normalized creep rates. Conversely, use of the activation energy of the matrix alone overlooks the temperature dependence resulting from dislocation–particle interactions. While the threshold stress model can partially address this issue by providing a more appropriate activation energy, it has not been applied in most of the existing studies. As a compromise, we adopt a rule-of-mixtures approach for the activation energies for diffusion and the temperature-dependent elastic moduli across all datasets. While this introduces some uncertainty, it does not obscure the clear trend that multi-phase alloys have enhanced creep resistance relative to single-phase HEAs. Notably, $\gamma + \gamma'$ HESAs exhibit fundamentally superior creep strength compared to all the other multi-phase FCC HEAs. However, given the comparable melting points of superalloys and FCC HEAs, research on HEAs has yet to yield novel alloy design and processing strategies that can outperform $\gamma + \gamma'$ superalloys at operating temperatures below 1100 °C.

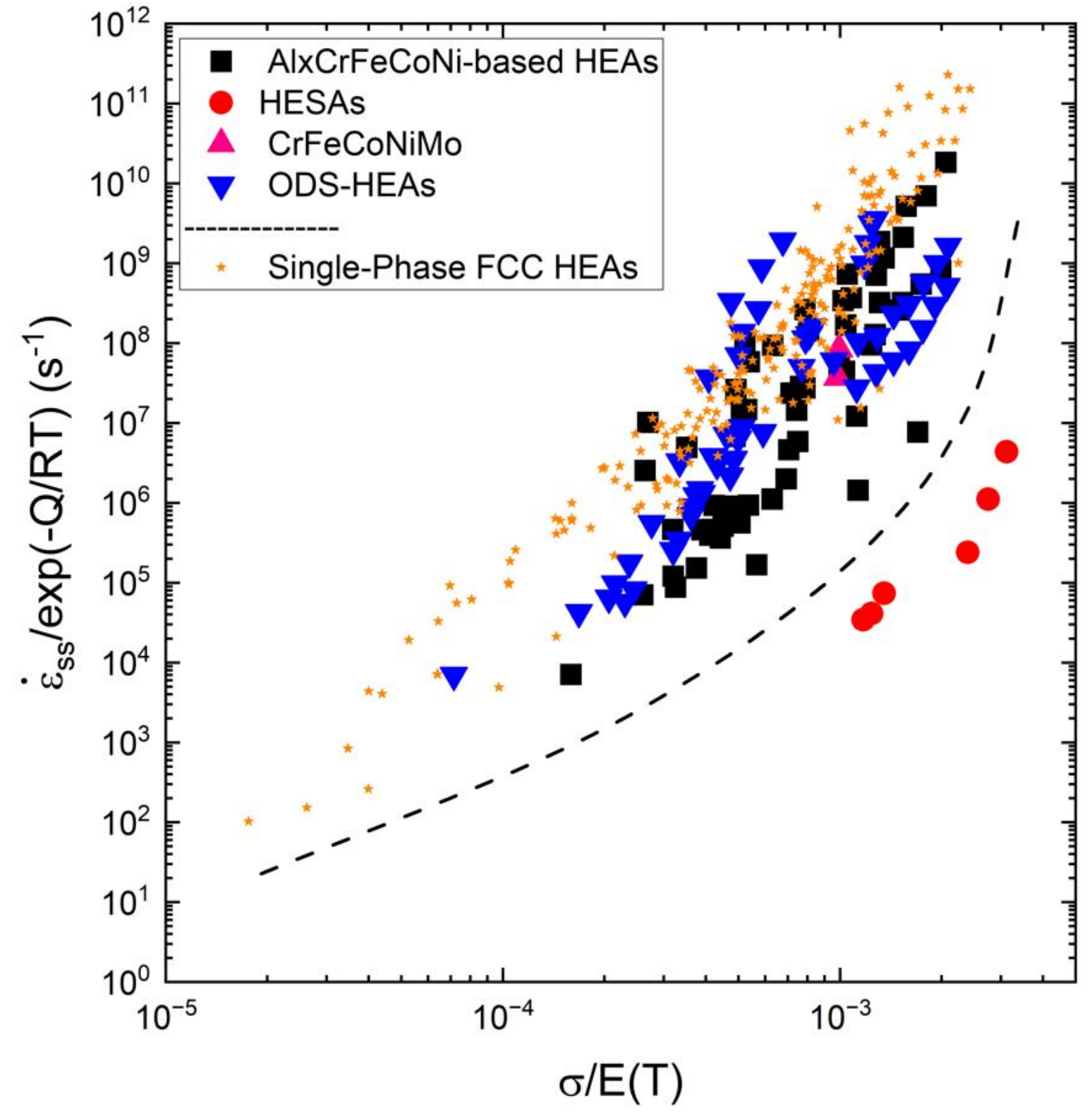

**Figure 14** Normalized secondary creep data for multi-phase FCC HEAs compared to their single-phase counterparts. The dashed line separates the creep data for HESAs from all the other multi-phase FCC HEAs. The data to create this plot were taken from [40, 136-144]. A comparison between the creep data of high entropy alloys and commercial alloys is given in the **Discussion** section.

### *2.3. Elevated-Temperature Deformation Mechanisms in FCC High-Entropy Alloys*

Previous high-temperature tensile tests on CrMnFeCoNi revealed a steep decline in strength with increasing temperatures [32]. As shown in **Fig. 15**, the yield strength drops to approximately 100 MPa at 800 ºC and an engineering strain rate of $10^{-3}$ $s^{-1}$, regardless of grain size. The pronounced temperature dependence of strength is also evident in creep experiments. Gadelmeier et al. [120] utilized single crystals to demonstrate how the temperature dependence of solid solution strengthening influences the creep strength of CrMnFeCoNi, isolating the effects from grain boundaries. As shown in **Fig. 16**, a creep strength advantage of 80 MPa over single-crystal Ni at 700 ºC was observed, which diminishes to just 2 MPa at 980 ºC. By 1100 ºC, the solid solution hardening effect is entirely lost. These observations suggest that deformation in CrMnFeCoNi is strongly thermally activated, with creep strength primarily governed by concentrated solid solution strengthening at intermediate temperatures. However, this strengthening effect diminishes significantly by 980 °C, partially due to the lowered solidus temperature of the alloy.

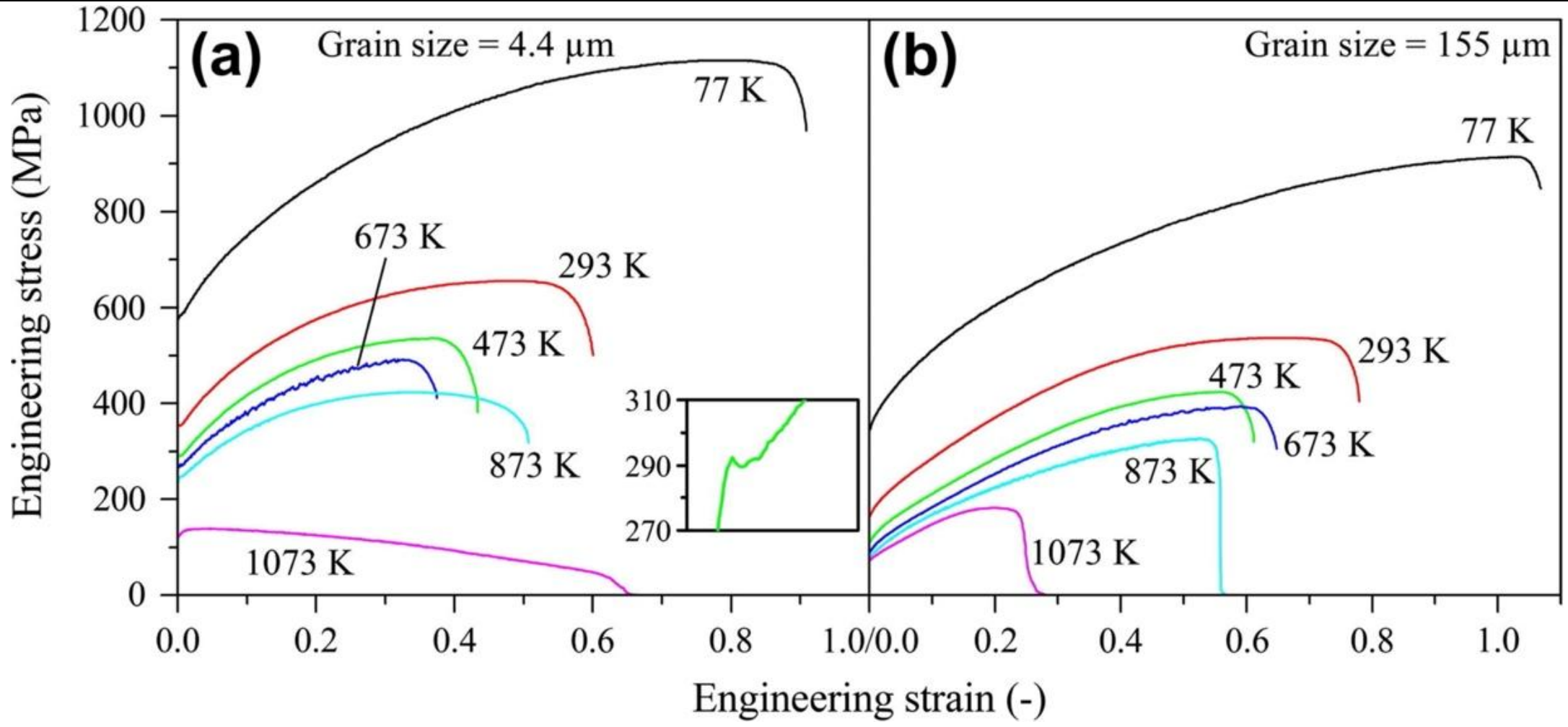

**Figure 15** Tensile stress-strain curves for CrMnFeCoNi from 77 K to 1073 K with a grain size of (a) 4.4 μm and (b) 155 μm. The images were taken from Ref [32].

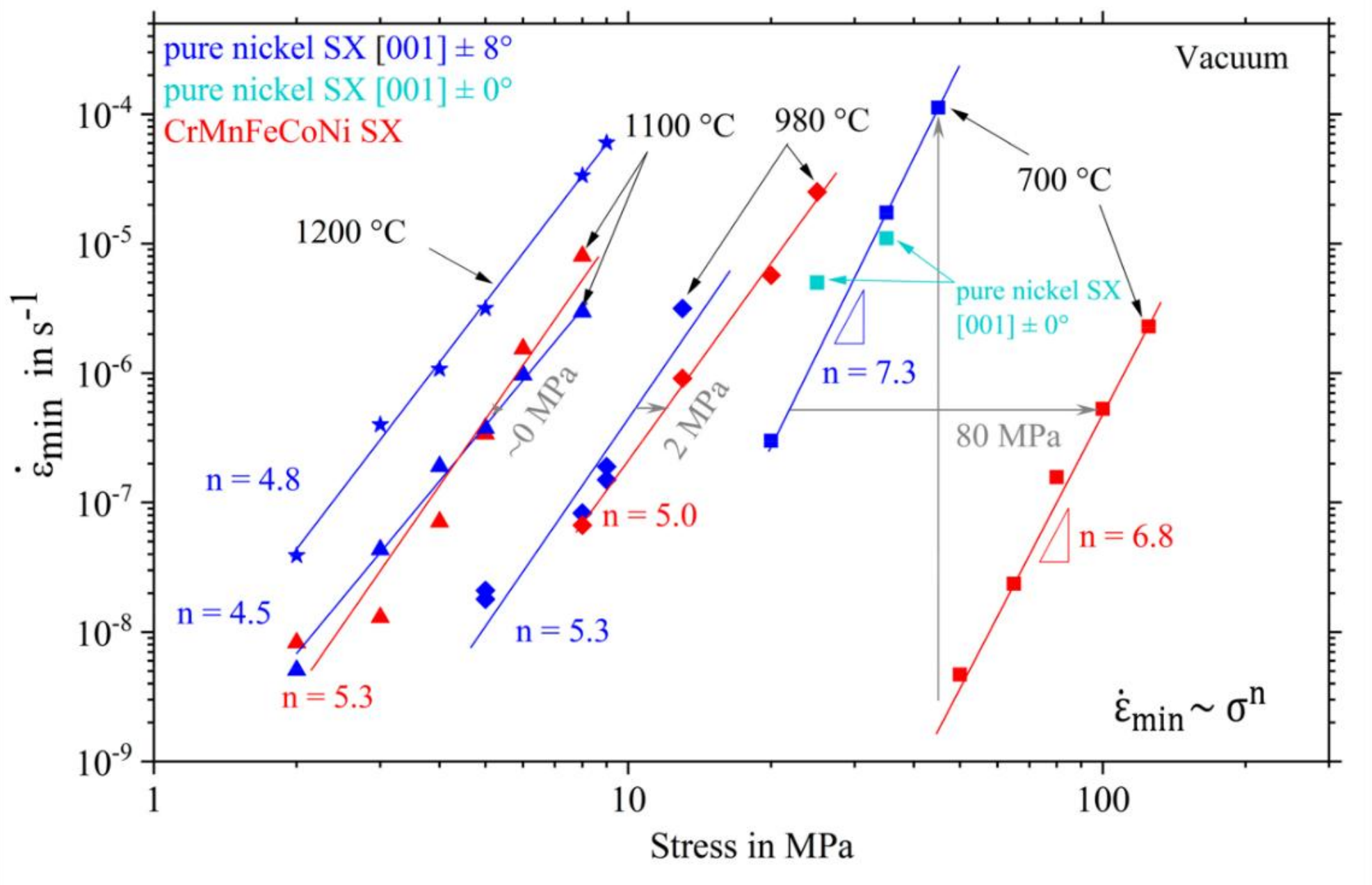

**Figure 16** Minimum creep rates as a function of applied stress for single-crystal CrMnFeCoNi and single-crystal Ni at 700, 980, and 1100 ºC. The image was taken from Ref [120].

Zhang et al. [158] carried out a detailed mechanistic study on creep of CrMnFeCoNi at 800 ºC using stress change creep experiments to determine key parameters that can characterize the rate-controlling deformation mechanisms, including the activation area and the Gibbs free energy of activation. This analysis is based on the thermally activated dislocation glide description of plasticity by Kocks, Argon, and Ashby [159], where the strain rate is governed by the thermal activation of dislocations as they overcome rate-controlling obstacles under a given applied stress. This framework is consistent with classical creep mechanisms such as climb-controlled or solute-drag creep. In all cases, plastic strain arises from dislocation glide, while the overall creep rate is governed by the slowest step, whether it is dislocation climb, solute diffusion, or another process, that controls how quickly dislocations can bypass obstacles and continue to glide. Specifically, the strain rate can be defined as:

$$\dot{\varepsilon}_{SS} = \dot{\varepsilon}_0 \, exp\left(-\frac{\Delta G}{k_B T}\right) = \dot{\varepsilon}_0 \, exp\left(-\frac{\Delta F - \sigma b \Delta a/M}{k_B T}\right), \quad (3)$$

where $\dot{\varepsilon}_0$ is a pre-exponential factor, $\Delta G$ is the Gibbs free energy of activation, $k_B$ is the Boltzmann constant, $\Delta F$ is the Helmholtz free energy of activation, $M$ is the Taylor factor (3.06 for FCC), $b$ is the magnitude of the Burgers vector, and $\Delta a$ is the activation area. The quantity $b\Delta a$ is often referred to as the activation volume, $\Delta V$ [159, 160]. It can be observed from Eqn. 3 that the thermally activated dislocation glide model treats plastic deformation as a stress assisted, thermally activated process, where the total energy of activation, $\Delta F$, is the sum of the thermal contribution, $\Delta G$, and the contribution from the work done by the applied shear stress, $\sigma b \Delta a/M$. In addition, $\Delta a \approx b\Delta l$ contains information about the length scale of inter-obstacle spacing $\Delta l$, since it can be physically interpreted as the area swept by the dislocation during the activation event.

As shown in **Fig. 17**, curvy dislocations were observed with copious evidence of solute pinning and inter-dislocation interactions. The activation areas for creep at 800°C in CrMnFeCoNi were determined to exhibit a magnitude of approximately 100 $b^2$ and scale inversely with the applied stress. These findings are consistent with activation area measurements obtained at room and cryogenic temperatures [161-163] for the same material, indicating similar types of obstacles. This suggests that the primary deformation mechanisms for creep in CrMnFeCoNi are dislocation-solute interactions and forest dislocation interactions, which can be pictured schematically as shown in **Fig. 17c**.

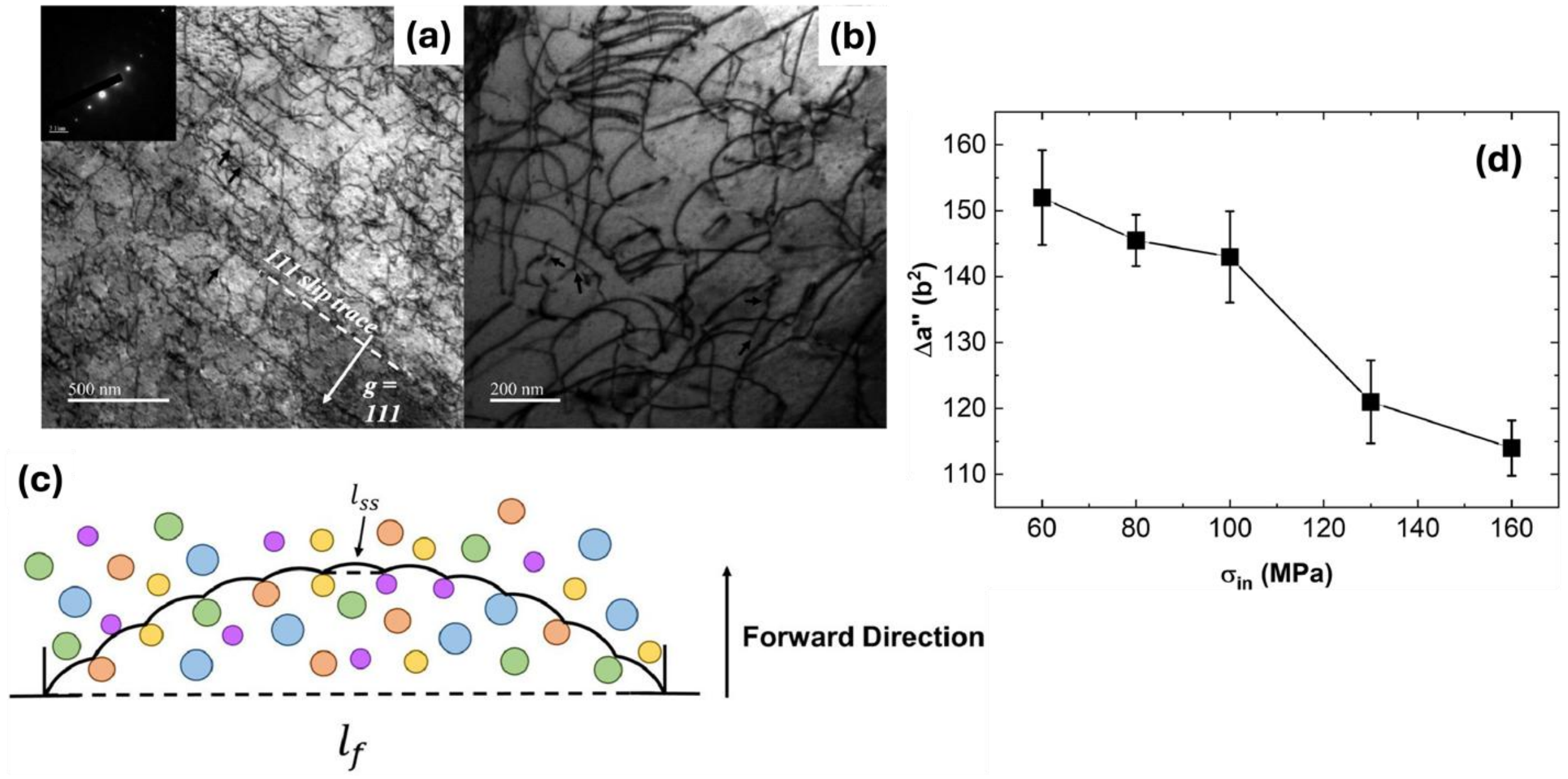

**Figure 17** (a-b) TEM micrographs of the dislocation substructure during creep of CrMnFeCoNi at 800 ºC; (c) Schematic diagram of the elevated-temperature deformation mechanism in CrMnFeCoNi, where dislocation-solute interactions and forest dislocation interactions co-determine the deformation rate; and (d) Operational activation areas of creep in CrMnFeCoNi at different applied stresses. The images were taken from Ref. [158].

**Table 4 Thermal activation parameters for dislocation-solute interactions and forest dislocation interactions in CrMnFeCoNi deformed at 800 ºC [158].**

*Terms: $\sigma$ is the applied stress. $\sigma_{ss}$ is the partitioned flow stress from solid solution hardening mechanism. $\sigma_f$ is the partitioned flow stress from forest dislocation hardening mechanism. $\hat{\sigma}_{ss}$ is the theoretical strength of solid solution hardening at 0 K. $\hat{\sigma}_f$ is the theoretical strength of forest dislocation hardening at 0 K. $\Delta a_{ss}$, $l_{ss}$, $\Delta a_f$ , and $l_f$ are the activation areas and inter-obstacle spacings of solid solution hardening mechanism and forest dislocation hardening mechanism, respectively.

| $\boldsymbol{\sigma}$ (MPa) | $\boldsymbol{\sigma_{ss}}$ (MPa) | $\boldsymbol{\sigma_f}$ (MPa) | $\boldsymbol{\hat{\sigma}_{ss}}$ (MPa) | $\boldsymbol{\hat{\sigma}_f}$ (MPa) | $\boldsymbol{\Delta a_{ss}}$ ($b^2$) | $\boldsymbol{l_{ss}}$ (nm) | $\boldsymbol{\Delta a_f}$ ($b^2$) | $\boldsymbol{l_f}$ (nm) |
|---|---|---|---|---|---|---|---|---|
| 160 | 64 | 96 | | 140 | | | 373 | 95 |
| 130 | 60 | 70 | | 114 | | | 460 | 117 |
| 100 | 55 | 45 | 318 | 47 | 164 | 42 | 1111 | 283 |
| 80 | 52 | 28 | | 41 | | | 1284 | 327 |
| 60 | 48 | 12 | | 25 | | | 2145 | 547 |

Recently, Varvenne et al. [164] developed a solid solution strengthening theory for FCC HEAs by combining the conventional solid solution hardening mechanism with additional strengthening contributions arising from the statistical distribution of local concentration fluctuations. The theory results in a theoretical strength of 318 MPa for CrMnFeCoNi from concentrated solid solution hardening at 0 K. Zhang et al. [158] performed mechanical separation of the flow stresses and activation areas resulting from solid solution hardening and forest dislocation hardening, which are shown in **Table 4**. The experimental value for the characteristic length scale of dislocation-solute interaction is ~ 40 nm, which is consistent with that obtained by Varvenne et al. in atomistic simulations [165]. It is important to note that the experimentally determined flow stress due to solid solution hardening at 800 ºC is significantly lower than the theoretical strength (i.e., a low $\sigma_{ss}/\hat{\sigma}_{ss}$), indicating substantial strength degradation at elevated temperatures. This observation is consistent with the relatively low stress-dependent $\Delta G$ of 111-141 kJ/mol, which reflects the thermal assistance required for dislocations to overcome resistance from solutes and forest dislocations. These findings are well aligned with the single-crystal study from Gadelmeier et al. mentioned above [120], which suggests that the pinning effect from concentrated solid solution hardening in CrMnFeCoNi weakens considerably at high temperatures. The detailed understanding of the elevated-temperature deformation mechanisms in CrMnFeCoNi can explain the limited creep resistance of $3d$-transition metal-based single-phase FCC HEAs, and the need for secondary phase strengthening. The high-temperature deformation mechanisms of multi-phase FCC HEAs that have been uncovered so far are consistent with those observed in conventional precipitation-hardened alloys and ODS alloys that are reviewed elsewhere [39, 153, 166-172]. To date, no novel elevated-temperature deformation mechanisms have been discovered in either single-phase or multi-phase FCC HEAs.

## 3. Creep Properties of BCC Refractory High-Entropy Alloys

### *3.1 Single-Phase BCC Refractory High-Entropy Alloys*

The potential for high-temperature performance offered by BCC HEAs has attracted widespread research interests since the invention of refractory HEAs (RHEAs) by Senkov et al. [45]. The available creep parameters for BCC RHEAs are compiled in **Table 5**, and the detailed secondary creep data are provided in **Supplementary Table 3** [90, 91, 173].

One of the first RHEAs developed was MoNbTaWV, which retains a compressive yield strength above 400 MPa at 1600 °C [47]. Whether this will translate to high creep strength remains unknown because creep tests have been performed only to 900 °C [174]. Despite their impressive compressive strength at elevated temperatures, Kumar et al. [48] demonstrated the poor tensile properties in a MoNbTaW RHEA: no tensile ductility is observed from room temperature up to 1200 °C, and the tensile strength remains below 70 MPa. If the *intrinsic* ductile-to-brittle transition temperature (DBTT) is not as high as 1200 °C, then *extrinsic* embrittlement mechanisms may be responsible for the lack of tensile ductility up to this temperature (although this remains an open question). Brittle oxide formation along grain boundaries was observed in this RHEA, which could lead to grain boundary embrittlement and degradation of tensile properties. To address this issue, Wang et al. [175] applied grain boundary engineering by doping the MoNbTaW RHEA with 400–8000 ppm of boron. Atom probe tomography (APT) revealed that boron preferentially segregates to the grain boundaries, effectively replacing brittle oxides. This modification led to a significant improvement in the alloy's deformability in compression. However, tensile and creep data are not yet available to reveal the effectiveness of this strategy. In addition, this system is predicted to be intrinsically brittle at room temperature through first-principle electronic calculations [176], which adds further difficulty to their direct application.

**Table 5 Creep parameters of single-phase BCC RHEAs for different microstructural states and testing conditions. All compositions are expressed in atomic percent; equiatomic compositions are considered self-explanatory.**

| Alloy | Loading | Microstructure | T (ºC) | $n$ | $Q_C$ (kJ/mol) | Ref. |
|---|---|---|---|---|---|---|
| HfNbTaTiZr | Tension | BCC RX | 980-1100 | 4.2 (hi $\sigma$) 2 (lo $\sigma$) | 273 | [90] |
| HfNbTaTiZr | Tension | BCC RX | 1100-1250 | 2.5-2.8 | 258-288 | [91] |
| $Nb_{45}Ta_{25}Ti_{15}Hf_{15}$ | Tension | BCC RX | 900 | 4.1 | 253 | [173] |

The second RHEA system developed by Senkov et al. [46], the HfNbTaTiZr system, exhibited extensive tensile ductility, a striking contrast to the brittle behavior of the MoNbTaWV system. The creep properties of HfNbTaTiZr were investigated by Gadelmeier et al. [90] and Liu et al. [91]. Their results are presented in **Fig. 18**, which reveal the inadequate creep resistance of HfNbTaTiZr at elevated temperatures. At 980 °C, CMSX-4 exhibits a creep strength approximately 25 times greater than that of HfNbTaTiZr to achieve a minimum creep rate of $10^{-7}$ $s^{-1}$; this disparity increases to nearly 70 times at 1100 °C. In addition, phase decomposition at 980 °C (**Fig. 18c**) driven by the segregation of Zr and Hf from the matrix led to further cracking, which contributed to deterioration of creep ductility.

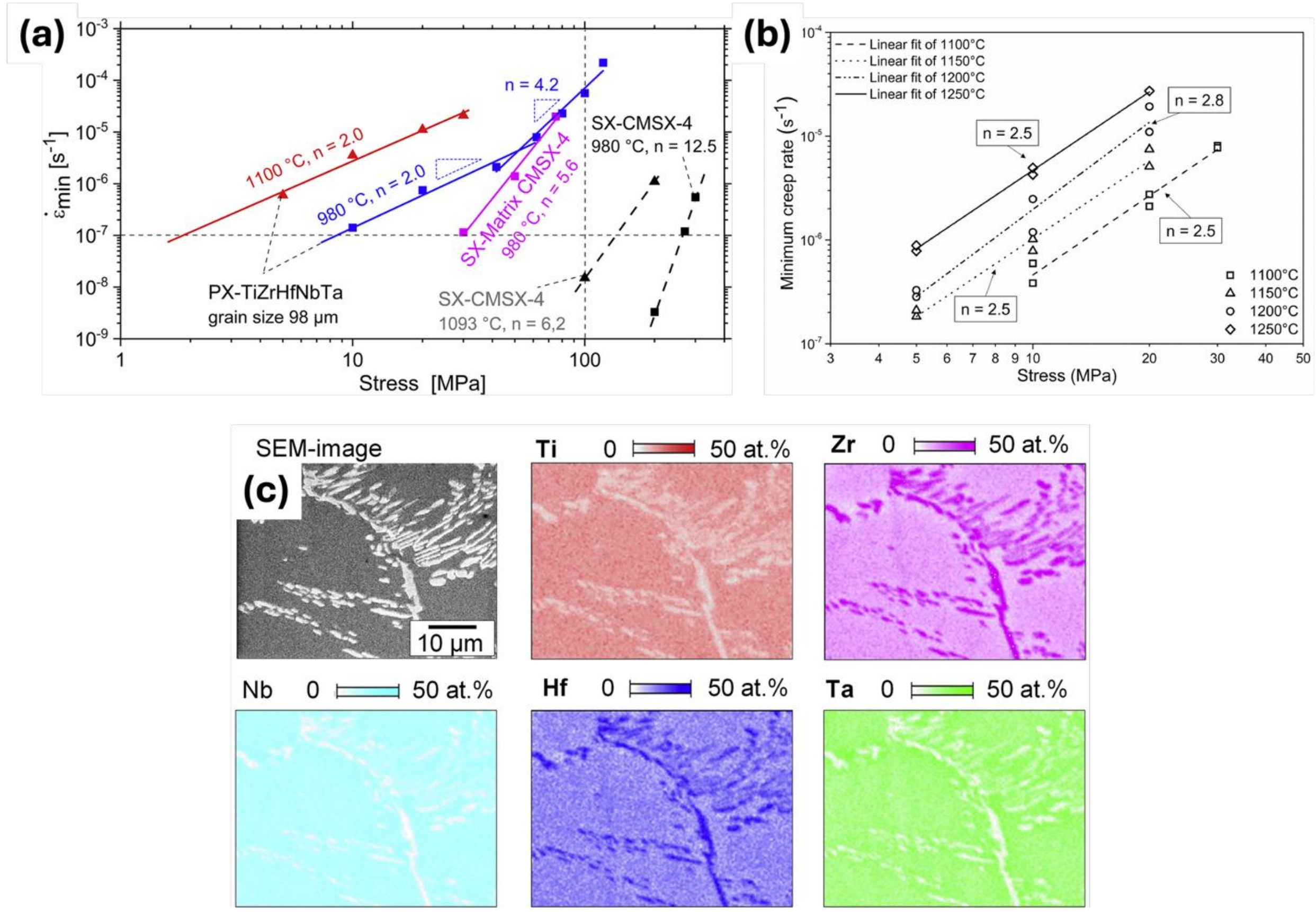

**Figure 18** Minimum creep rates for HfNbTaTiZr (a) at 980-1100 ºC, showing a comparison with CMSX-4, a single-crystal Ni-based superalloy; (b) at 1100-1250 ºC; and (c) phase decomposition during creep at 980 ºC using SEM-EDS. Images (a) and (c) were taken from Ref. [90]and (b) from [91].

A more equitable comparison can be made between the creep strength of the single-phase BCC HfNbTaTiZr RHEA and the single-phase FCC matrix of CMSX-4, excluding the contribution of the ~65 vol.% $L1_2$ precipitates. Even after this adjustment, the RHEA still lags behind by a factor of three (**Fig. 18a**). Furthermore, the stress exponent and activation energy for HfNbTaTiZr near a minimum creep rate of $10^{-7}$ $s^{-1}$ is approximately 2 and 250 kJ/mol, respectively. Gadelmeier et al. [90] extrapolated the creep behavior of the recrystallized polycrystalline HfNbTaTiZr to a single-crystal configuration, yet the alloy still exhibits a factor of two lower creep strength compared to the single-phase matrix of CMSX-4. At temperatures exceeding 1100 °C, the stress required to sustain a creep rate of $10^{-7}$ $s^{-1}$ falls below 5 MPa. This highlights a critical limitation of single-phase BCC HfNbTaTiZr, which mirrors the challenge faced by single-phase FCC CrMnFeCoNi: a rapid degradation of strength at high temperatures and the insufficiency of solid solution strengthening alone to resist creep.

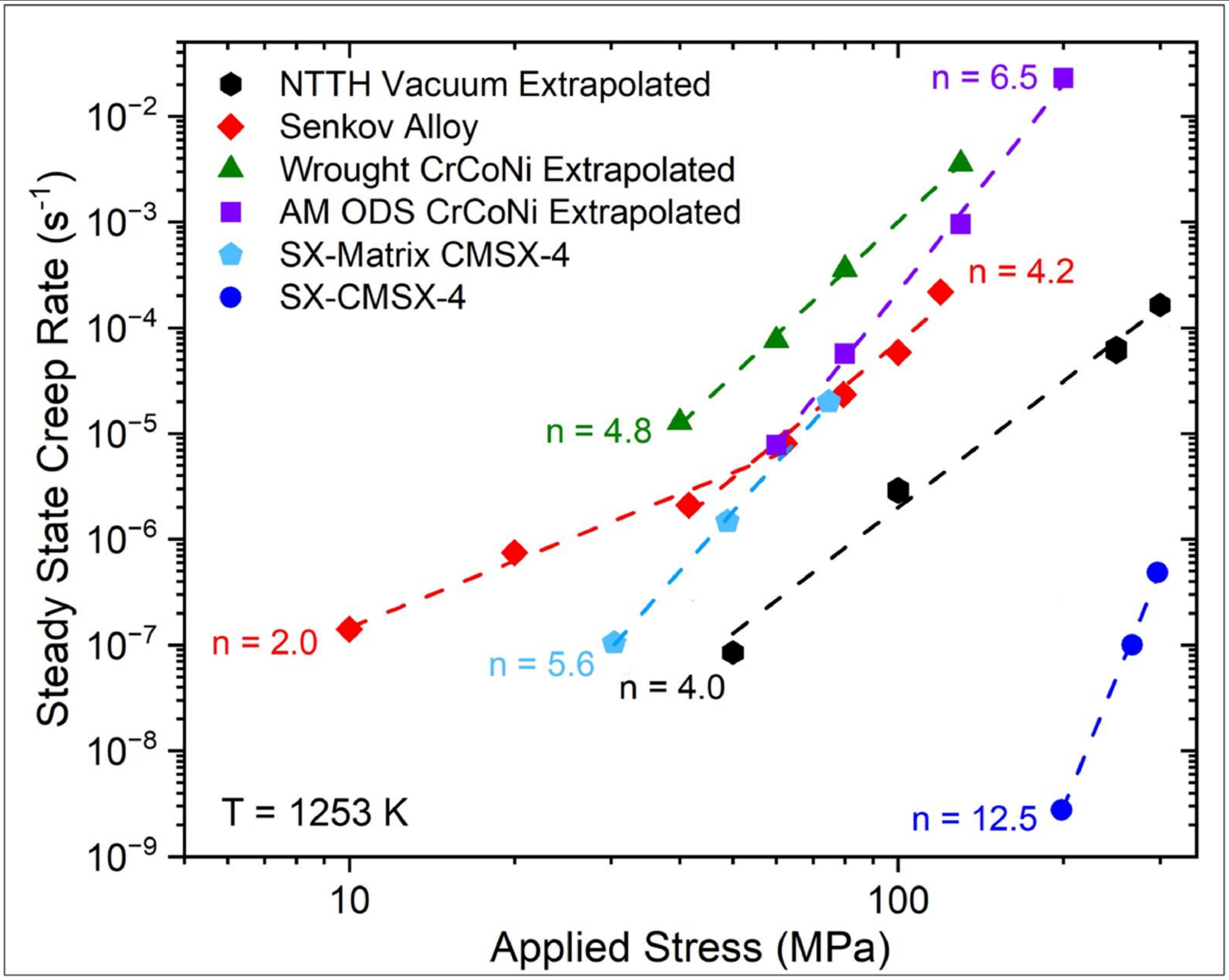

**Figure 19** Minimum or steady-state creep rates for $Nb_{45}Ta_{25}Ti_{15}Hf_{15}$ (NTTH) compared to HfNbTaTiZr (Senkov alloy), wrought CrCoNi, AM ODS-CrCoNi, CMSX-4, and the single-phase matrix of CMSX-4. All creep data were either taken at 980 ºC or extrapolated to 980 ºC using the reported activation energies. The image was taken from Ref. [173].

Sahragard-Monfared et al. [173] studied the creep behavior of a related RHEA, $Nb_{45}Ta_{25}Ti_{15}Hf_{15}$ (in at.%), at a lower temperature of 900 °C. The corresponding creep data were compared with several studies discussed above and the results are shown in **Fig. 19**. $Nb_{45}Ta_{25}Ti_{15}Hf_{15}$ is projected to exhibit superior creep strength compared to HfNbTaTiZr (equiatomic) at 980 °C, as well as outperform the single-phase matrix of CMSX-4 and additively manufactured ODS-CrCoNi. Nevertheless, a significant performance gap remains between the creep resistance of single-phase BCC RHEAs and $\gamma + \gamma'$ CMSX-4, highlighting the need for secondary phase strengthening strategies in single-phase RHEAs.

Furthermore, the study revealed another critical challenge for RHEAs: their pronounced environmental sensitivity, as even trace amounts of oxygen can significantly degrade creep performance. As shown in **Fig. 20**, samples tested in ultrahigh-purity argon exhibited considerable variability in creep rates at lower applied stresses. This variability leads to a misinterpretation of the deformation mechanism as diffusional creep, due to an apparent stress exponent near unity. However, premature creep rupture was observed in the sample tested in argon, where the steady-state creep rate appeared artificially elevated due to the early onset of tertiary creep.

Detailed microstructural analysis (**Fig. 20c–e**) revealed substantial oxygen ingress even in ultrahigh-purity argon environments, with the formation of $HfO_2$ along grain boundaries contributing to embrittlement and intergranular fracture. Under a stress of 100 MPa for 29h, oxygen penetrated the entire sample, resulting in intergranular fracture. In contrast, at 250 MPa for 4.5h, intergranular fracture was confined to the surface region, while the interior exhibited ductile fracture features. Notably, such degradation was absent in tests conducted under high vacuum, where the significantly lower oxygen partial pressure minimized environmental effects.

These findings underscore the inability of this RHEA to resist creep in oxidizing environments such as air, where oxidation becomes a dominant degradation mechanism. Given the generally poor oxidation resistance of many RHEAs [111, 112, 114], protective coatings are essential to shield RHEAs from environmental attack during high-temperature service. In parallel, alloy design that enhances the intrinsic oxidation resistance of the base RHEA is essential to ensure the material's integrity in the event of coating failure. Alternatively, recent observations on a ternary refractory Cr-Mo-Si alloy show potential to overcome this barrier [177]. It demonstrates both pesting resistance at intermediate temperatures of up to 800°C as well as oxidation resistance at elevated temperatures up to 1100°C, while simultaneously maintaining substantial compressive deformability at room temperature.

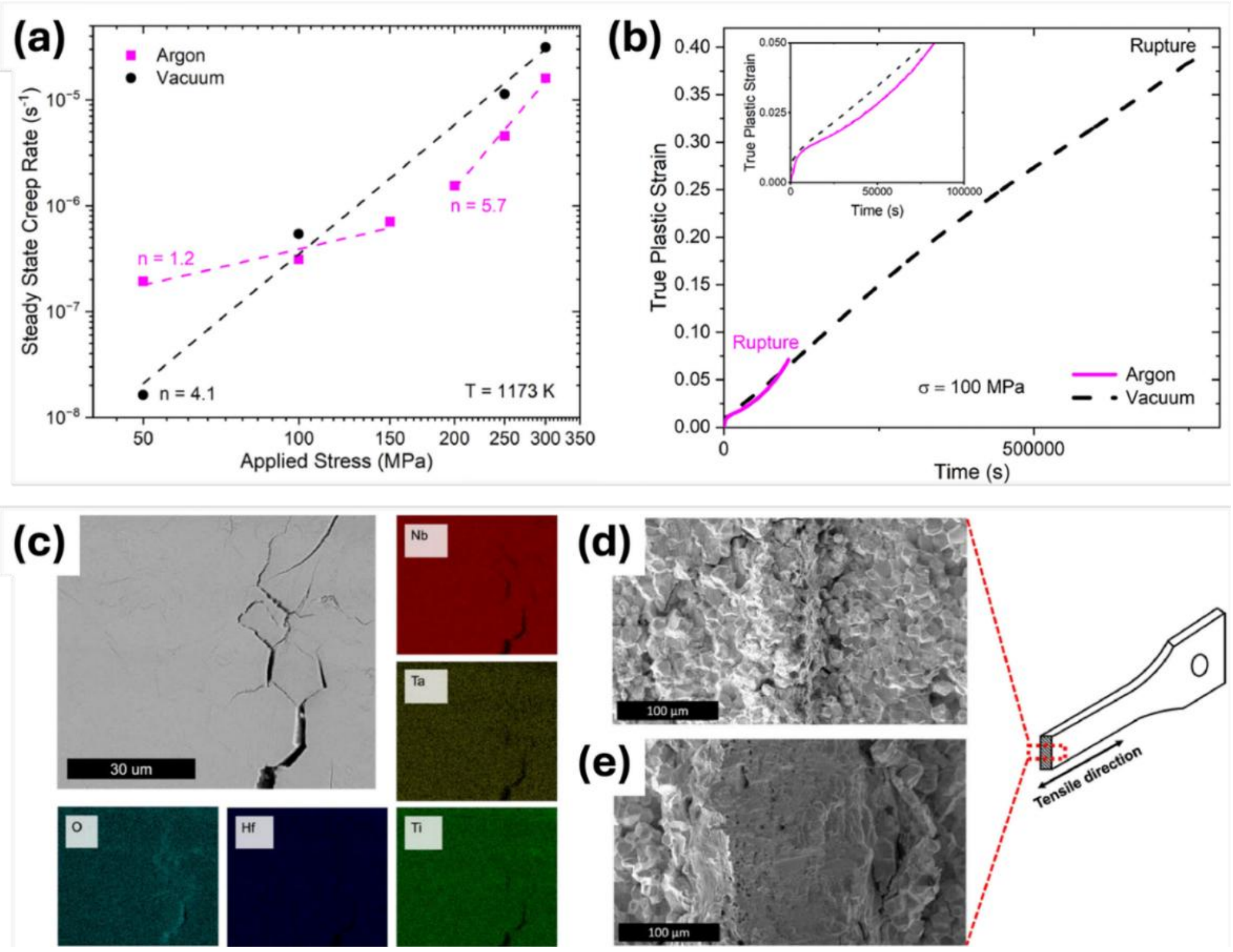

**Figure 20** (a) Steady-state creep data and (b) creep curves for $Nb_{45}Ta_{25}Ti_{15}Hf_{15}$ tested in high-purity argon and high vacuum; (c) The SEM-EDS scan for a crept sample after rupture in argon; and (d-e) fractographs of crept samples tested in argon at 100 MPa for 29h and 250 MPa for 4.5h, respectively. The images were taken from Ref. [173].

The overall normalized creep rates for BCC single-phase RHEAs are summarized in **Fig. 21**. The limited availability of data restricts the comparison to only HfNbTaTiZr and $Nb_{45}Ta_{25}Ti_{15}Hf_{15}$ RHEAs, where $Nb_{45}Ta_{25}Ti_{15}Hf_{15}$ appears stronger than its Senkov alloy counterpart.

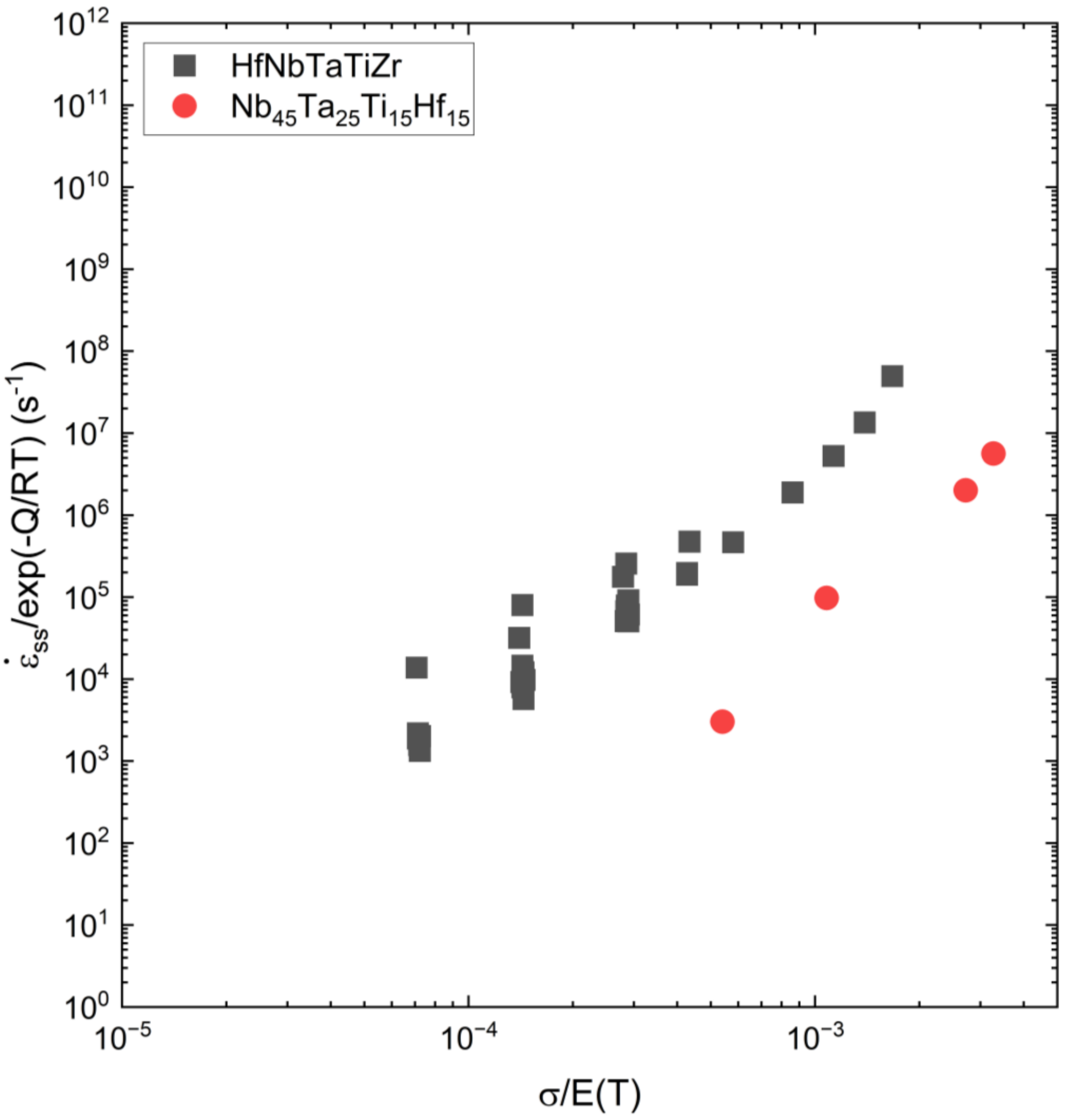

**Figure 21** Normalized secondary creep data for single-phase BCC RHEAs. The creep data were adapted from Refs. [90, 91, 173]. A comparison between the creep data of high entropy alloys and commercial alloys is given in the **Discussion** section.

*3.2 Multi-Phase BCC Refractory High-Entropy Alloys*

Since ductile single-phase BCC RHEAs experience a loss in high-temperature strength, similar to their FCC counterparts, it is necessary to seek multi-phase strengthening mechanisms. The creep parameters for multi-phase BCC RHEAs are compiled in **Table 6**, and the reported secondary creep data are given in **Supplementary Table 4** [93, 178, 179].

**Table 6 Creep parameters of multi-phase BCC RHEAs for different microstructural states and testing conditions. All compositions are expressed in atomic percent; equiatomic compositions are considered self-explanatory.**

| Alloy | Loading | Microstructure | T (ºC) | $n$ | $Q_C$ (kJ/mol) | Ref. |
|---|---|---|---|---|---|---|
| $Ta_{27.3}Mo_{27.3}Ti_{27.3}Cr_8Al_{10}$ | Compression | BCC + B2 | 1030 | 18.7 (hi $\sigma$) 4.3 (lo $\sigma$) | 390 | [93] |
| $Al_{23.5}Ti_{23.5}V_{23.5}Nb_{23.5}Zr_6$ | Compression | B2 + $Zr_5Al_3$ + $Nb_2Al$ | 800 | 3 | 256 | [178] |
| $Ti_{33.3}Ta_{11.1}Nb_{22.2}Zr_{22.2}Mo_{11.2}$ | Tension | BCC + Zr-rich precipitates | 650-750 | 3.2-3.4 | 242-281 | [179] |

The most widely adopted strategy involves designing a BCC-B2 dual-phase structure that imitates the FCC–$L1_2$ structure found in Ni-based superalloys. As a result, these alloys are often referred to as refractory high-entropy superalloys (RHSAs). **Fig. 22** shows the microstructure of $Al_{10}Nb_{20}Ta_{16}Ti_{30}V_4Zr_{20}$ (in atomic proportions) RHSA after isothermal annealing at 600 ºC for 0.5-120 h [180]. The annealing induces spinodal decomposition of the high-temperature BCC phase, leading to the development of a continuous B2 matrix with discrete cuboidal BCC precipitates aligned along the <001> directions. The spinodal decomposition reflects an inherently unstable microstructure at this early stage, resulting in a continuous brittle phase and discrete ductile phase, which is an unfavorable configuration that limits overall ductility. However. prolonged subsequent annealing at 600 °C can lead to the formation of necking constrictions along the B2 channels, ultimately pinching off these channels and rendering the BCC phase continuous, with discrete B2 precipitates. This phase inversion process improves the compressibility of the material, as the ductile BCC phase becomes continuous to allow long-range dislocation glide.

**Fig. 22d-e** demonstrates that the 120-hour annealing at 600 °C enhances compressive ductility while maintaining high strength at 600 °C. While a dedicated creep study for the $Al_{10}Nb_{20}Ta_{16}Ti_{30}V_4Zr_{20}$ RHSA does not yet exist, Yang et al. [93] investigated the compressive creep properties of a BCC-B2 $Ta_{27.3}Mo_{27.3}Ti_{27.3}Cr_8Al_{10}$ (in at.%, abbreviated as TMT-8Cr-10Al) RHSA (**Fig. 23**). It has been shown that TMT-8Cr-10Al exhibits comparable creep resistance to CMSX-4 at approximately 1030 °C under stresses below 125 MPa. Considering that the solvus temperature of the B2 precipitates in TMT-8Cr-10Al, which ranges between 1060 and 1070 °C, is significantly lower than the solvus temperature of the $L1_2$ phase in CMSX-4 at 1280 °C, and the B2 phase fraction of around 20% is much smaller than the $L1_2$ fraction in CMSX-4, TMT-8Cr-10Al demonstrates remarkable creep strength as a polycrystalline material, outperforming the single-phase HfNbTaTiZr by orders of magnitude. However, the relatively high stress exponents of $n > 10$ measured at stresses above 125 MPa lead to a significant increase in the creep rate, highlighting the need for a more detailed mechanistic study of stress dependence on dislocation interactions with matrix and precipitates. Furthermore, to surpass the creep properties of Ni-based superalloys, further optimization is required to enhance the B2 volume fraction and raise the solvus temperature. Currently, it is not known whether either can be accomplished in the Al-containing RHEAs.

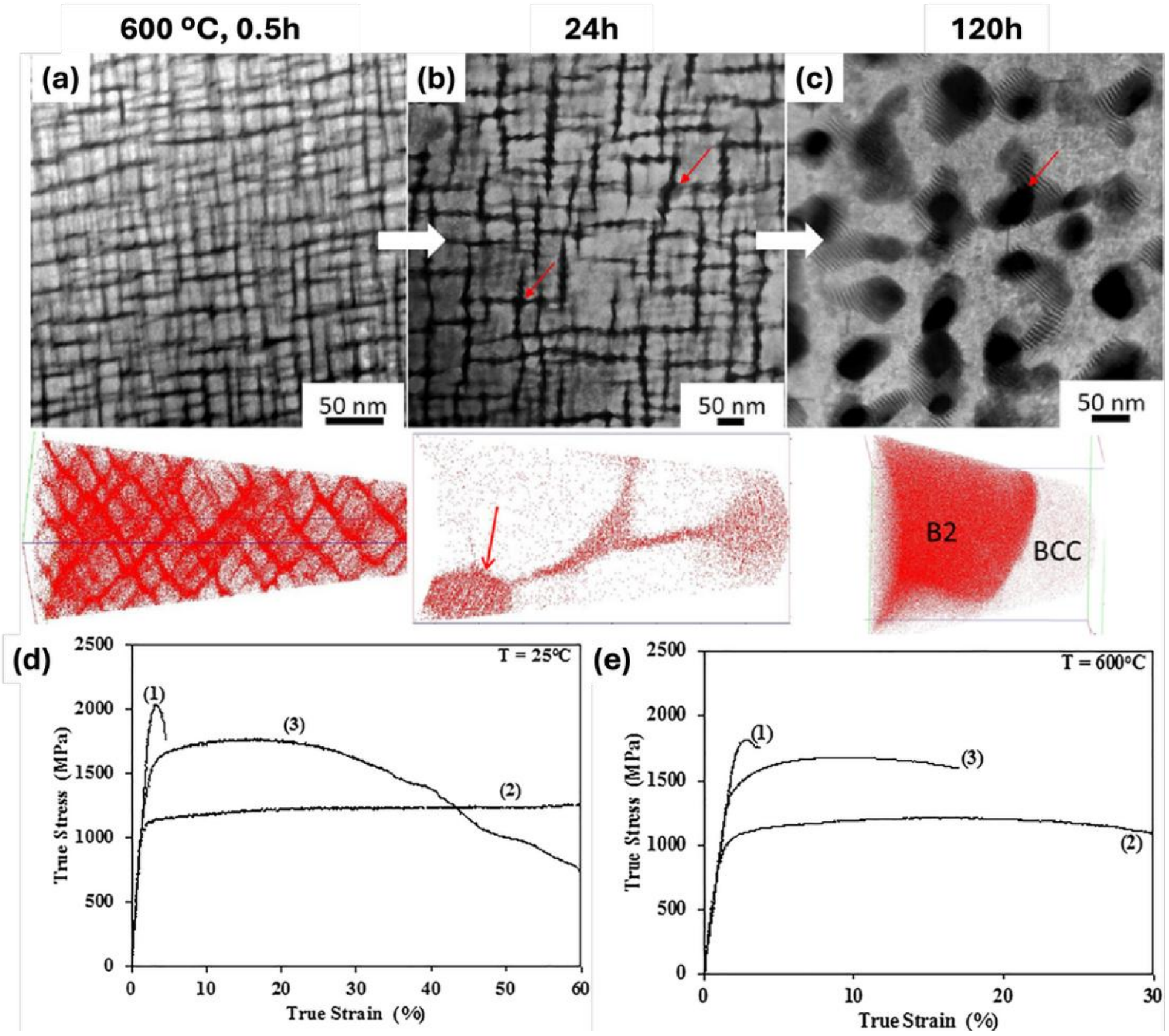

**Figure 22** Phase microstructure of 600 ºC annealed $Al_{10}Nb_{20}Ta_{16}Ti_{30}V_4Zr_{20}$ for (a) 0.5 h, (b) 24 h, and (c) 120 h. The dark phase in the images in (a-c) is B2 and the light phase is BCC. In a related study, the compressive stress-strain curves for three processing conditions are shown: (1) as-cast and HIP'ed alloy, followed by homogenization at 1200 °C for 24 h with slow cooling; (2) the same process with additional annealing at 1400 °C for 20 min, followed by water quenching; and (3) the same process with further annealing at 600 °C for 120 h, also followed by water quenching. The results for room temperature compressive tests are shown in (d), and those for 600 °C tests are shown in (e). The images were taken from Refs. [180, 181].

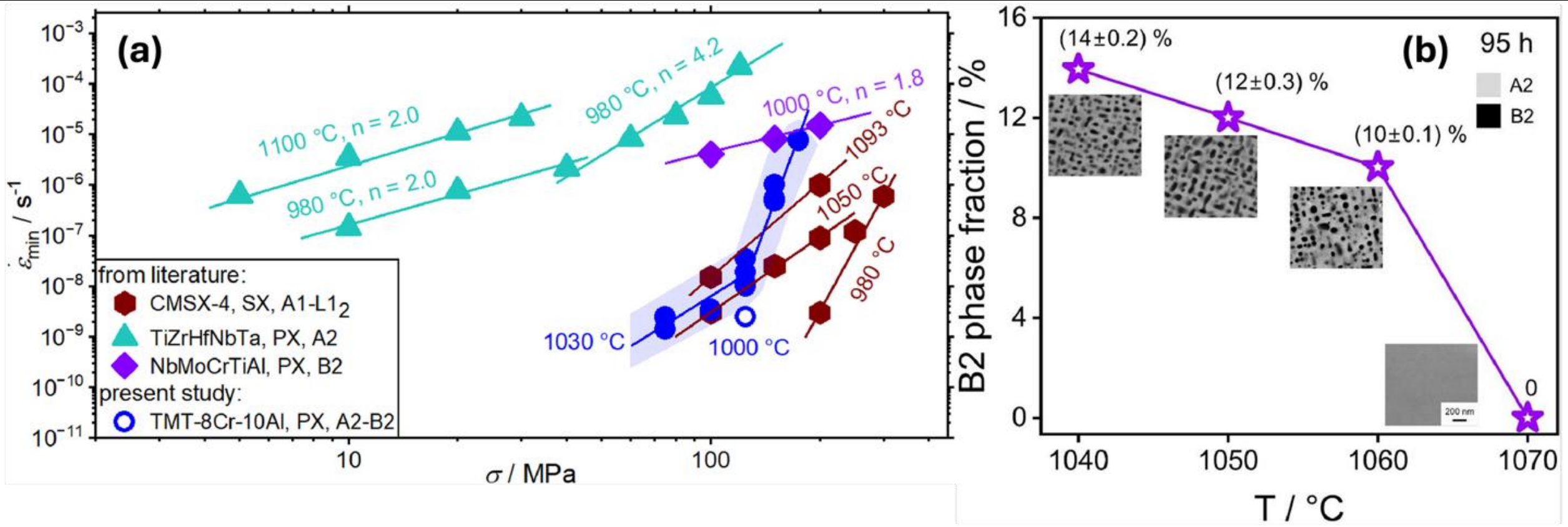

**Figure 23** (a) Minimum creep rates of $Ta_{27.3}Mo_{27.3}Ti_{27.3}Cr_8Al_{10}$ RHSA compared with CMSX-4, HfNbTaTiZr, and NbMoCrTiAl; (b) the B2 phase fraction as a function of annealing temperature for 95h of annealing. The images were taken from Refs. [93].

It is worth noting that most mechanical tests on RHSAs were conducted under compression. Similar to the single-phase MoNbTaWV system discussed earlier, RHSAs face the challenge of grain boundary embrittlement that significantly impairs their tensile properties. Kumar et al. [182] compared the compressive and tensile properties of a $Al_{10}Nb_{25}Ta_{25}Ti_{20}Zr_{20}$ RHSA, revealing a strong contrast between the two. **Fig. 24** shows that no tensile ductility was observed up to 1200 °C, with fracture strength falling below 50 MPa at elevated temperatures. The observed tensile brittleness was attributed to the formation of Al-Zr-rich precipitates at the grain boundaries, as well as local shearing in the precipitation-free zones near these boundaries, leading to brittle intergranular fracture. Future research efforts must address this issue, either through the fabrication of single-crystal BCC-B2 RHSAs or by employing grain boundary engineering techniques, such as doping with boron or carbon, to occupy grain boundary free volumes and prevent grain boundary segregation and precipitation.

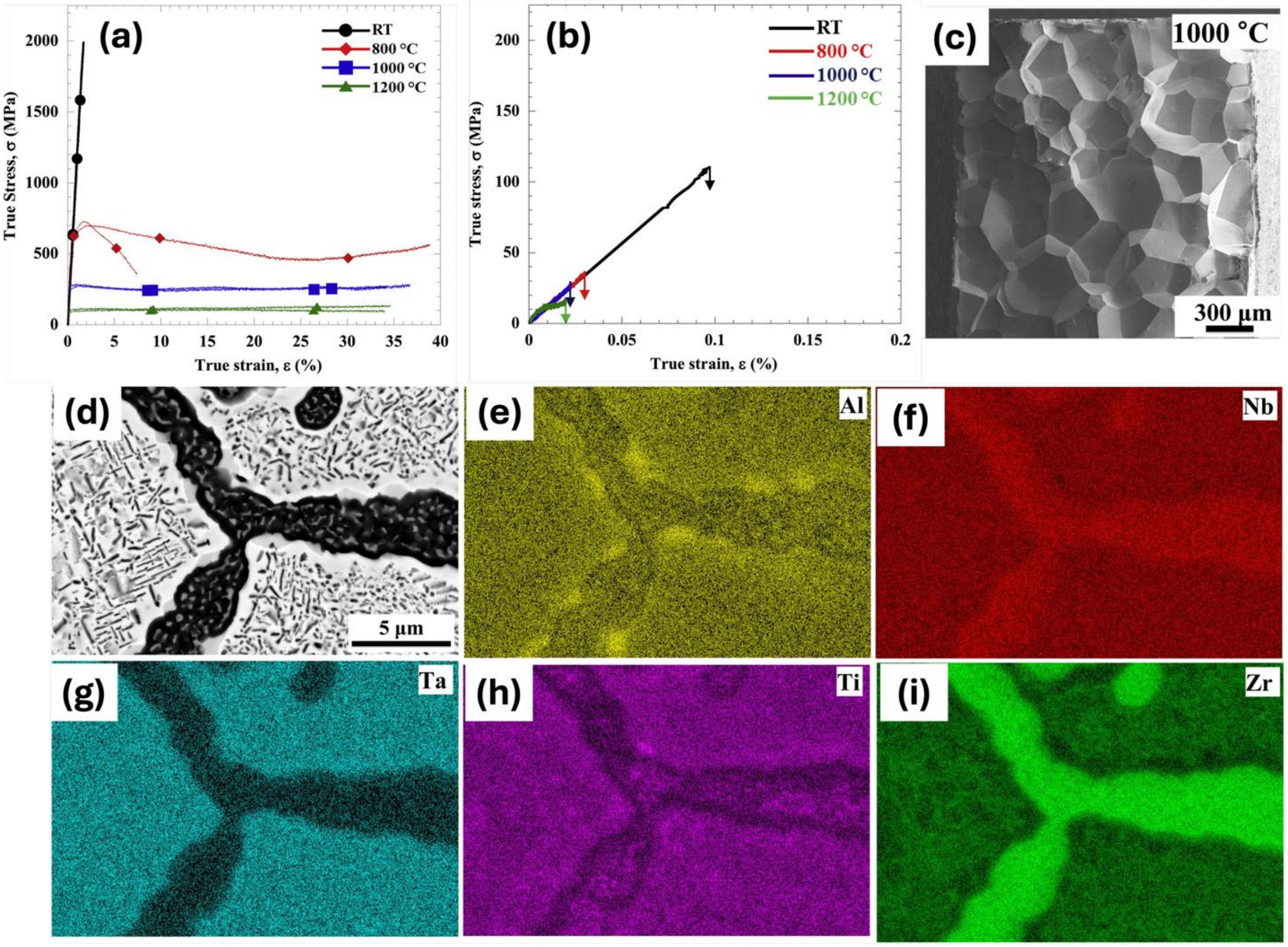

**Figure 24** (a) Compressive and (b) tensile properties of the $Al_{10}Nb_{25}Ta_{25}Ti_{20}Zr_{20}$ RHSA; (c) brittle intergranular fracture in a sample tensile tested at 1000 ºC; and (d-i) SEM-EDS showing grain boundary segregation and a precipitate free zone immediately adjacent to the Zr-rich precipitate. The images were taken from Ref. [182].

Other multi-phase strategies to improve the creep resistance of BCC RHEAs are sparse, although precipitation hardening strategies for single principal component BCC systems (e.g., Cr-based superalloys [183]) have been widely applied with promising creep properties. Kral et al. [178] investigated the compressive creep behavior of an $Al_{23.5}Ti_{23.5}V_{23.5}Nb_{23.5}Zr_6$ RHEA with a B2 matrix phase, coarse $Zr_5Al_3$ precipitates, and $Nb_2Al$ sigma phase. However, the material is unlikely to exhibit tensile ductility due to the brittle B2 matrix and grain boundary precipitates. Feng et al. [179] fabricated a $Ti_{33.3}Ta_{11.1}Nb_{22.2}Zr_{22.2}Mo_{11.2}$ RHEA by LPBF-AM, where the BCC matrix is strengthened by two types of Zr-rich continuous precipitates in the grain interior and at the grain boundaries. However, as shown in **Fig. 25**, neither of the two RHEAs exhibits a clear advantage in creep resistance compared to single-phase BCC RHEAs at elevated temperatures. This may be attributed to the low melting points, elastic moduli, and activation energies of the constituent elements in $Al_{23.5}Ti_{23.5}V_{23.5}Nb_{23.5}Zr_6$, as well as the insufficient strengthening provided by the small volume fraction (lower than 6%) of Zr-rich precipitates in $Ti_{33.3}Ta_{11.1}Nb_{22.2}Zr_{22.2}Mo_{11.2}$.

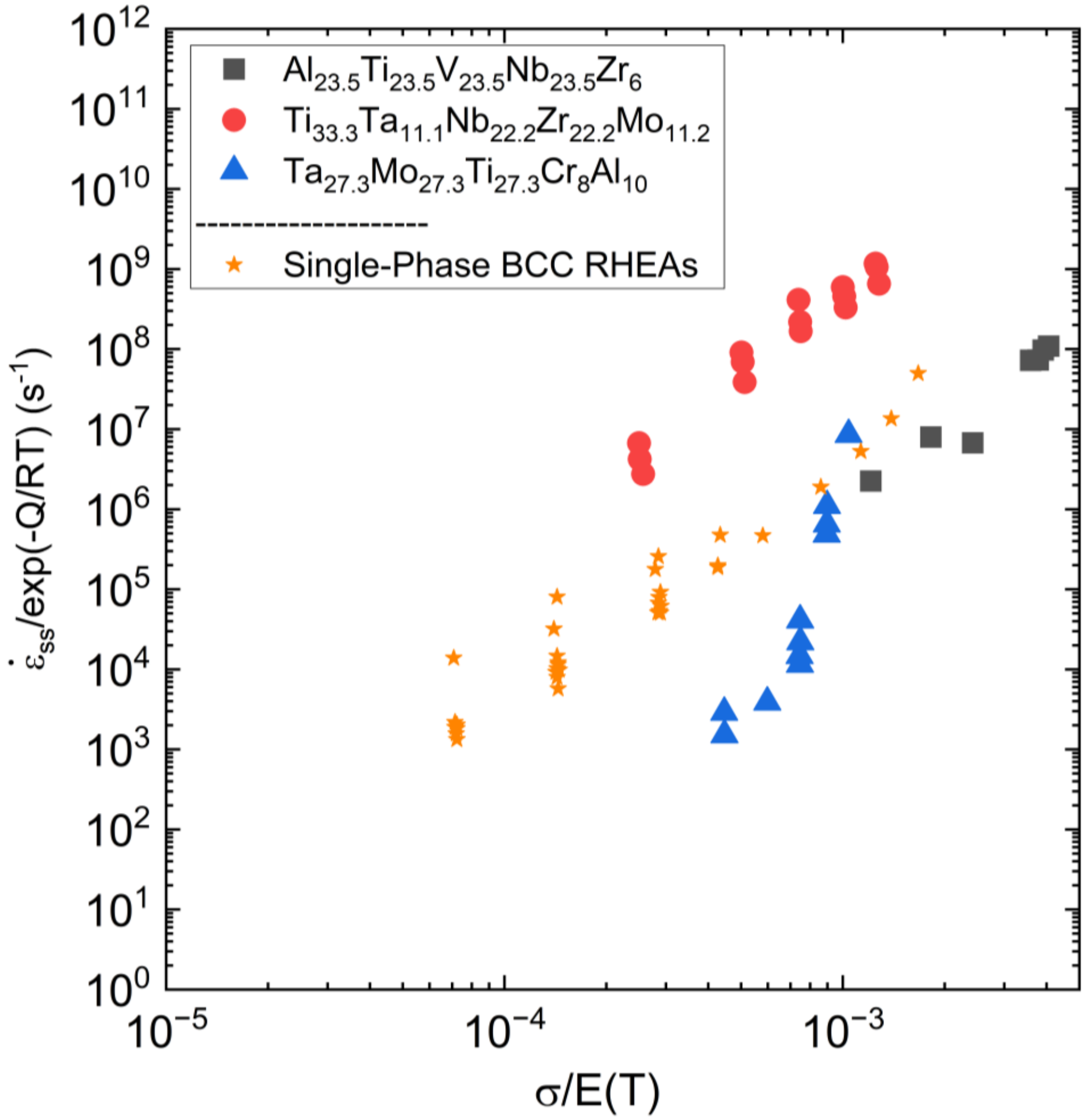

**Figure 25** Normalized secondary creep data for multi-phase BCC RHEAs compared to their single-phase counterparts. The creep data were adapted from Refs. [93, 178, 179]. A comparison between the creep data of high entropy alloys and commercial alloys is given in the **Discussion** section.

*3.3 Elevated-Temperature Deformation Mechanisms in BCC Refractory High-Entropy Alloys*

Unlike FCC HEAs, deformation at lower temperatures in BCC RHEAs is mainly controlled by the glide of screw dislocations [160, 184], although edge dislocations can sometimes be important [185-187]. Maresca and Curtin [188] developed a theory of screw dislocation strengthening in concentrated BCC solid solutions, attributing plastic deformation to mechanisms such as Peierls advancement, lateral kink glide, and the unpinning of cross-kinks, all of which are related to screw dislocations (**Fig. 26**). At elevated temperatures, the increased thermal activation can improve screw dislocation mobility and enable cross-slip, thereby making the cross-kink/dipole/jog unpinning mechanism dominant. Rao et al. [189] modified the Suzuki model [184, 190, 191] that describes dislocation-solute interactions at BCC screw dislocation cores. They proposed that plastic deformation is controlled by either cross-kink pinch-off under stress and thermal activation, resulting in the formation of debris in the form of vacancy or interstitial loops, or by jog dragging. In the jog-dragging mechanism, impinging cross-kinks drag the motion of the screw segment through the non-conservative motion of the edge components within the cross-kinks, which is facilitated by vacancy diffusion. Both mechanisms were directly observed in creep deformed $Nb_{45}Ta_{25}Ti_{15}Hf_{15}$ by Sahragard-Monfared et al. [173] using TEM, as shown in **Fig. 27**. Long, straight screw dislocations were observed on multiple {110} glide planes, accompanied by a high density of debris. Additionally, bowed segments were frequently observed, indicating jog pinning and suggesting that creep deformation is governed by jog dragging. It is noted that the dislocation substructure during creep in the BCC RHEA differs from that in FCC HEAs illustrated in **Fig. 17**. Based on these findings, Rao et al. [192] recently rationalized the creep deformation of $Nb_{45}Ta_{25}Ti_{15}Hf_{15}$ between 850 ºC and 950 ºC using the Rao-Suzuki screw dislocation glide model [189], achieving good agreement with experimental results.

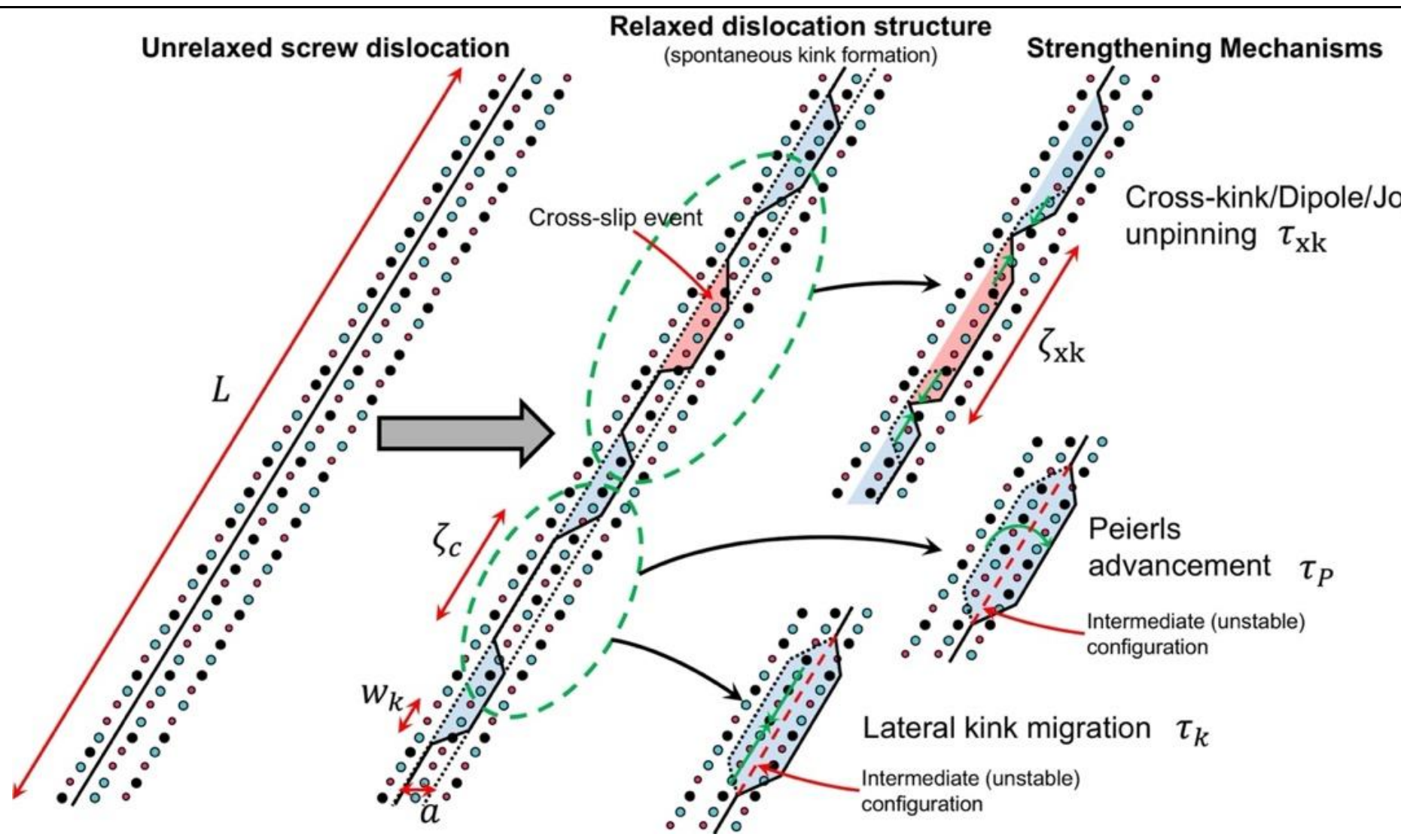

**Figure 26** Schematic of screw dislocation glide-controlled deformation in single-phase BCC RHEAs. The image was taken from Ref [188].

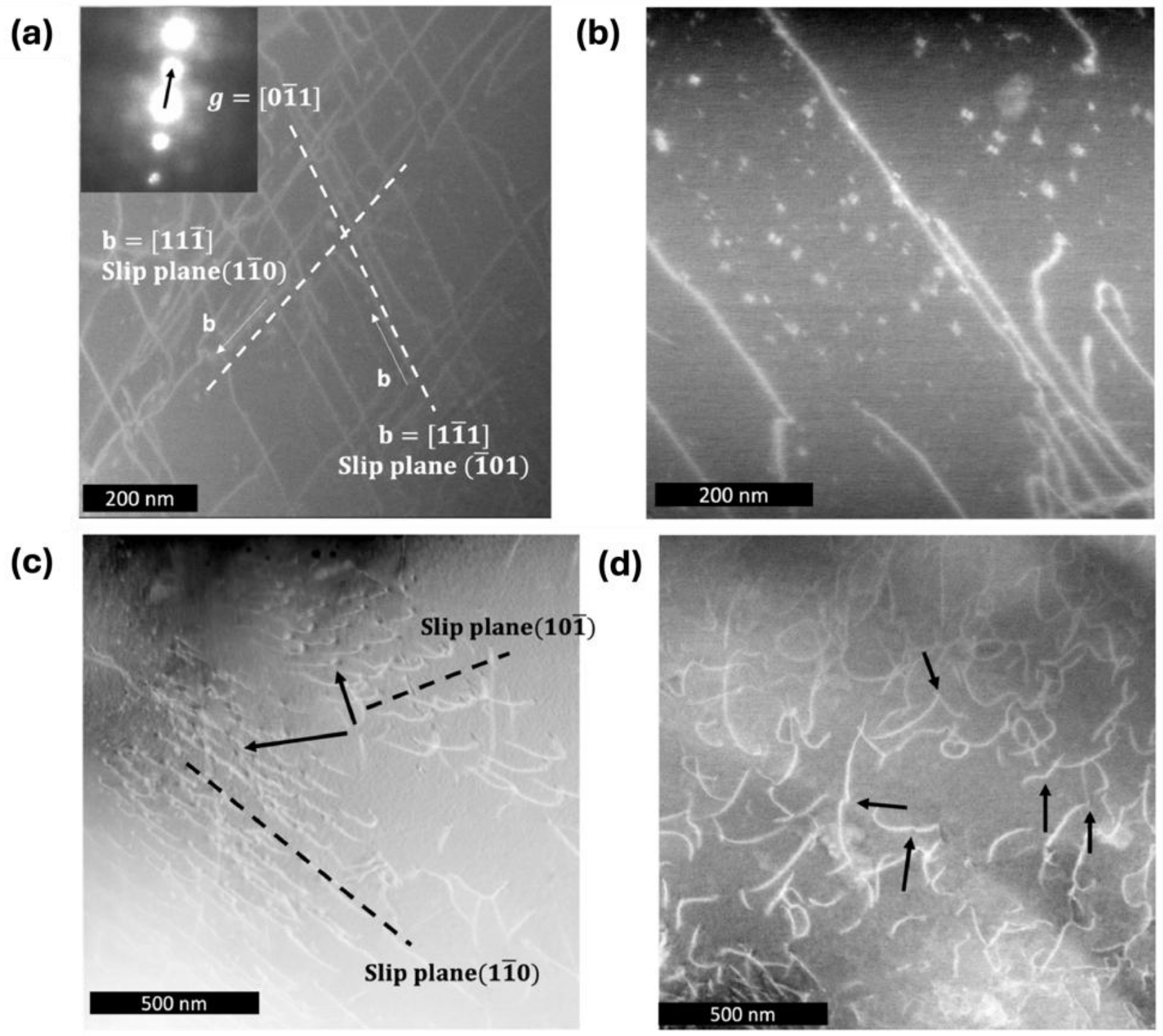

**Figure 27** STEM images of the dislocation substructure during creep at 900 ºC in $Nb_{45}Ta_{25}Ti_{15}Hf_{15}$. (a-b) show long straight screw segments on intersecting glide planes with a large number of debris; and (c-d) show screw dislocations pinned by jogs (indicated by black arrows). The image was taken from Ref. [173].

Combined creep data and mechanistic analysis reveal that NbTaTiHf-based RHEAs exhibit insufficient creep resistance at elevated temperatures compared to Ni-based superalloy counterparts. This limitation arises from significantly reduced glide barriers for screw dislocations and lower activation energies for creep through cross-kink/jog unpinning/dragging. Notably, recent theoretical and experimental studies have demonstrated that edge dislocation glide can govern plastic deformation in some BCC RHEAs at elevated temperatures [187, 193]. This behavior is attributed to the trapping of dislocations in statistically favorable random solute environments, which imposes intrinsically high energy barriers to edge dislocation motion, as manifested in the MoNbTaWV system [193] and the NbTaTiV system [187]. Detailed mechanistic studies are still required to understand how edge dislocation glide control can affect the creep behavior in BCC RHEAs.

The elevated-temperature deformation mechanism of a BCC-B2 $Al_{10}Nb_{20}Ta_{16}Ti_{30}V_4Zr_{20}$ RHSA at 600 ºC was studied in detail by Couzinié et al [194]. Dislocations with a/2<111> Burgers vectors are highly localized in slip bands, as illustrated in **Fig. 28**. A closer look using high-resolution STEM discloses the discrete nature of the bands, which form a zigzag morphology along {110} and {112} planes intermittently, suggesting frequent cross-slip. As a result, the overall band trajectory follows a direction in between the {110} trace and the {112} trace. The B2 precipitates strongly pin dislocations before being subsequently sheared. This can potentially explain the power-law breakdown behavior in the creep study on $Ta_{27.3}Mo_{27.3}Ti_{27.3}Cr_8Al_{10}$ by Yang et al. [93], where a high applied stress can lead to the buildup of back stress that allows the leading super-dislocation pairs to cut through the particles. This behavior is frequently observed in Ni-based superalloys [195-197]. Back stress can be applied toward the threshold stress to rationalize the high creep stress exponent.

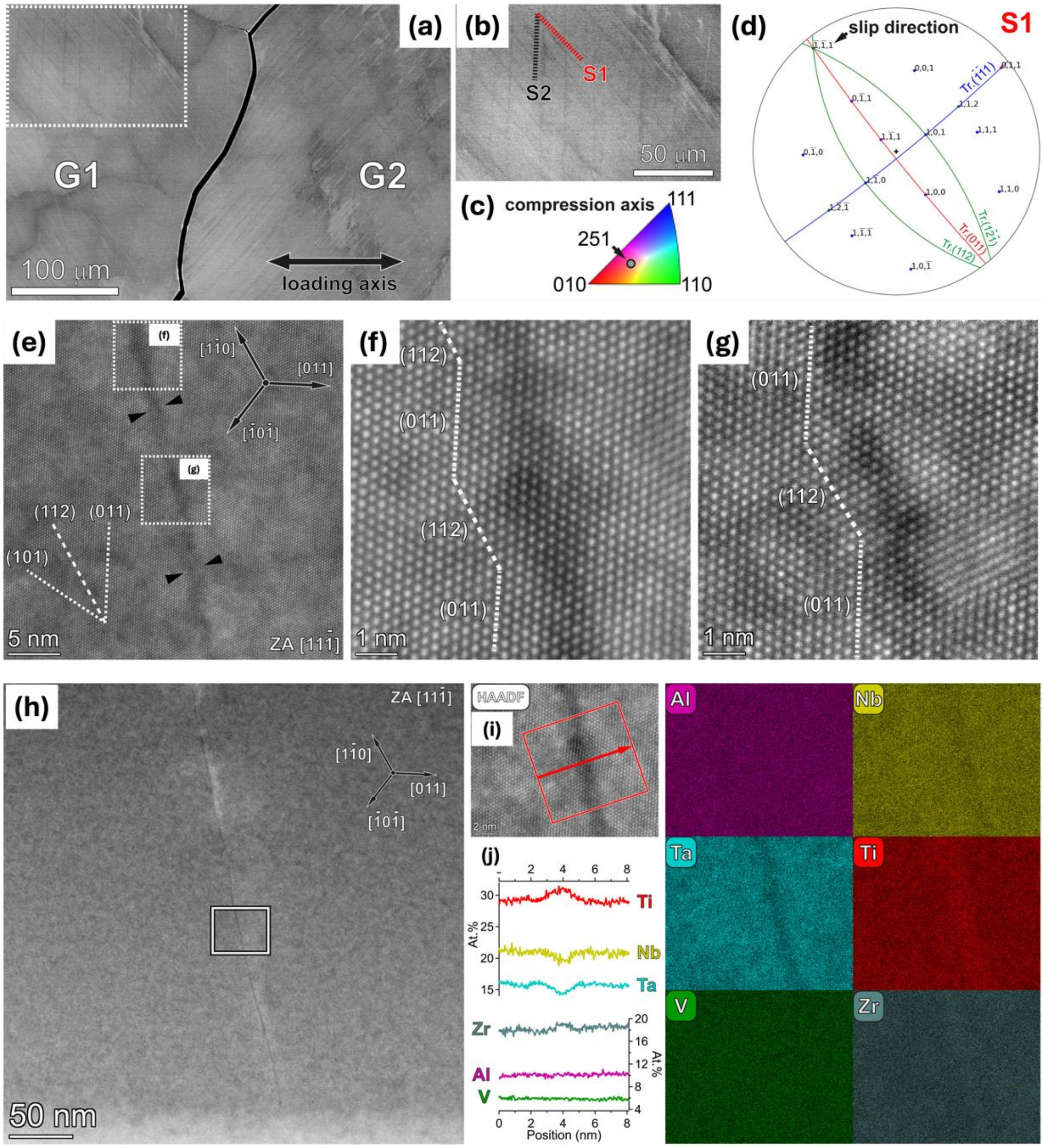

**Figure 28** (a-d) SEM images of the dislocation substructure after compression testing at 600 ºC in $Al_{10}Nb_{20}Ta_{16}Ti_{30}V_4Zr_{20}$ and the corresponding crystallography; (e-g) high-resolution STEM images showing the zigzag nature of the deformation band; (h) STEM image of a deformation band; and (i-j) gives the STEM-EDS analysis that shows Ti segregation at the band core. The images were taken from Ref. [194].

It is important to understand the evolution of the B2 phase under both temperature and stress that can strongly affect the creep behavior. Yang et al. [93] reported stress-induced precipitate coarsening and rafting behavior in $Ta_{27.3}Mo_{27.3}Ti_{27.3}Cr_8Al_{10}$, as shown in **Fig. 29**. Creep deformation led to a greater increase in the B2 phase fraction compared to simple annealing at the same temperature and duration, indicating faster kinetics caused by higher defect densities in the deforming material. Although rafting in polycrystalline materials can be complex, N-type rafting (characterized by precipitate elongation perpendicular to the applied stress) was observed during compression. This behavior is attributed to the positive lattice mismatch between the B2 precipitates and the BCC matrix and can be translated to P-type rafting (precipitate elongation parallel to the applied stress) under tension.

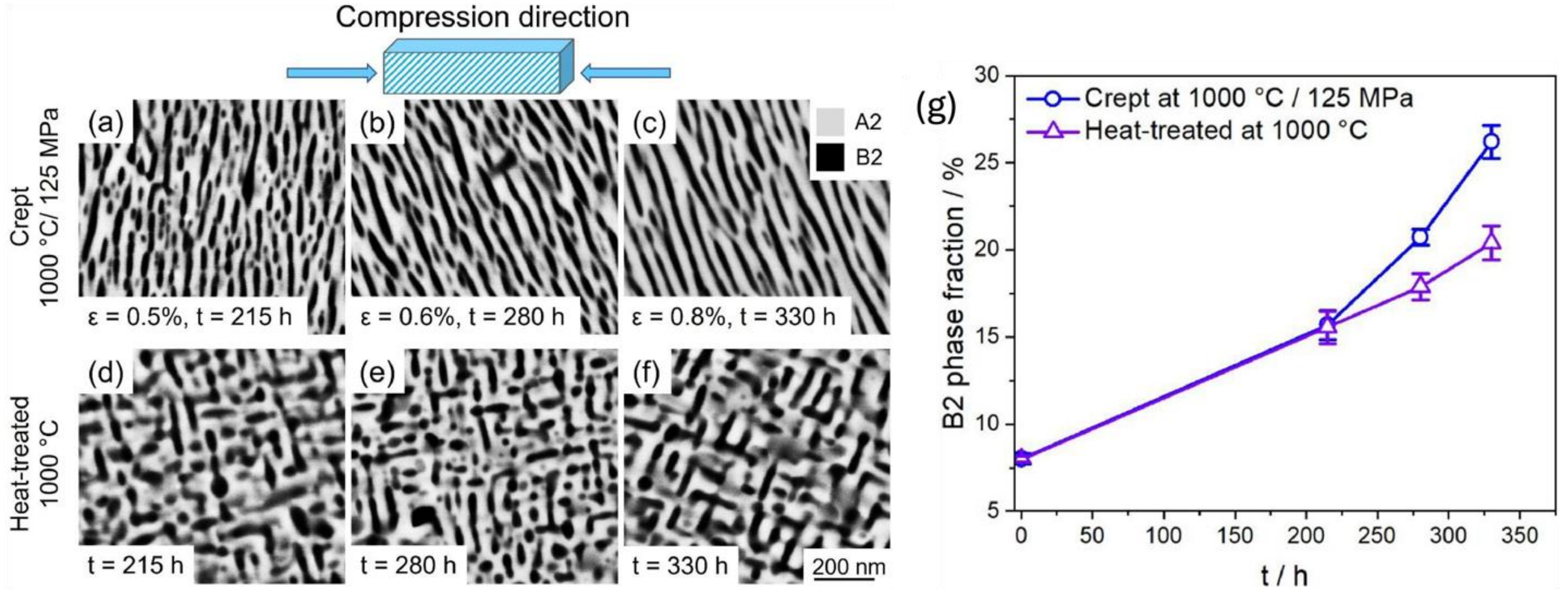

**Figure 29** SEM images of B2 precipitate morphology in $Ta_{27.3}Mo_{27.3}Ti_{27.3}Cr_8Al_{10}$ after compression creep testing at 1000 ºC for (a) 215 h, (b) 280 h, and (c) 330 h; as well as after heat treatment at (d) 215 h, (e) 280 h, and (f) 330 h; the B2 phase fraction as a function of creep testing or heat treatment time is shown in (g). The images were taken in Ref. [93].

## 4 Discussion and Future Prospects

In terms of creep resistance, no current HEA system has demonstrated performance superior to that of Ni-based single-crystal superalloys. The comparison of FCC HEAs with FCC-based commercial alloys is given in **Fig. 30.** Many single-phase FCC HEAs exhibit creep resistance comparable to that of 304 stainless steel, which is expected given their similar compositions and microstructures. This trend aligns with the fact that CrMnFeCoNi-based HEAs possess high-temperature tensile properties similar to those of austenitic stainless steels [27, 28, 32]. Some variants, such as CrCoNi or Mo-containing FCC HEAs, along with multi-phase Al-containing or ODS-FCC HEAs (excluding HESAs), demonstrate creep performance comparable to Ni-based alloys strengthened by $M_{23}C_6$ carbides, γ″, and δ phases, such as Inconel 625. Nevertheless, a significant performance gap remains when compared to high-volume-fraction γ′-strengthened superalloys, particularly single-crystal variants like CMSX-4. Among HEA-based systems, only γ + γ′ HESAs approach the creep performance of such superalloys, and as such, are arguably better categorized as superalloys rather than conventional HEAs under the original definition.

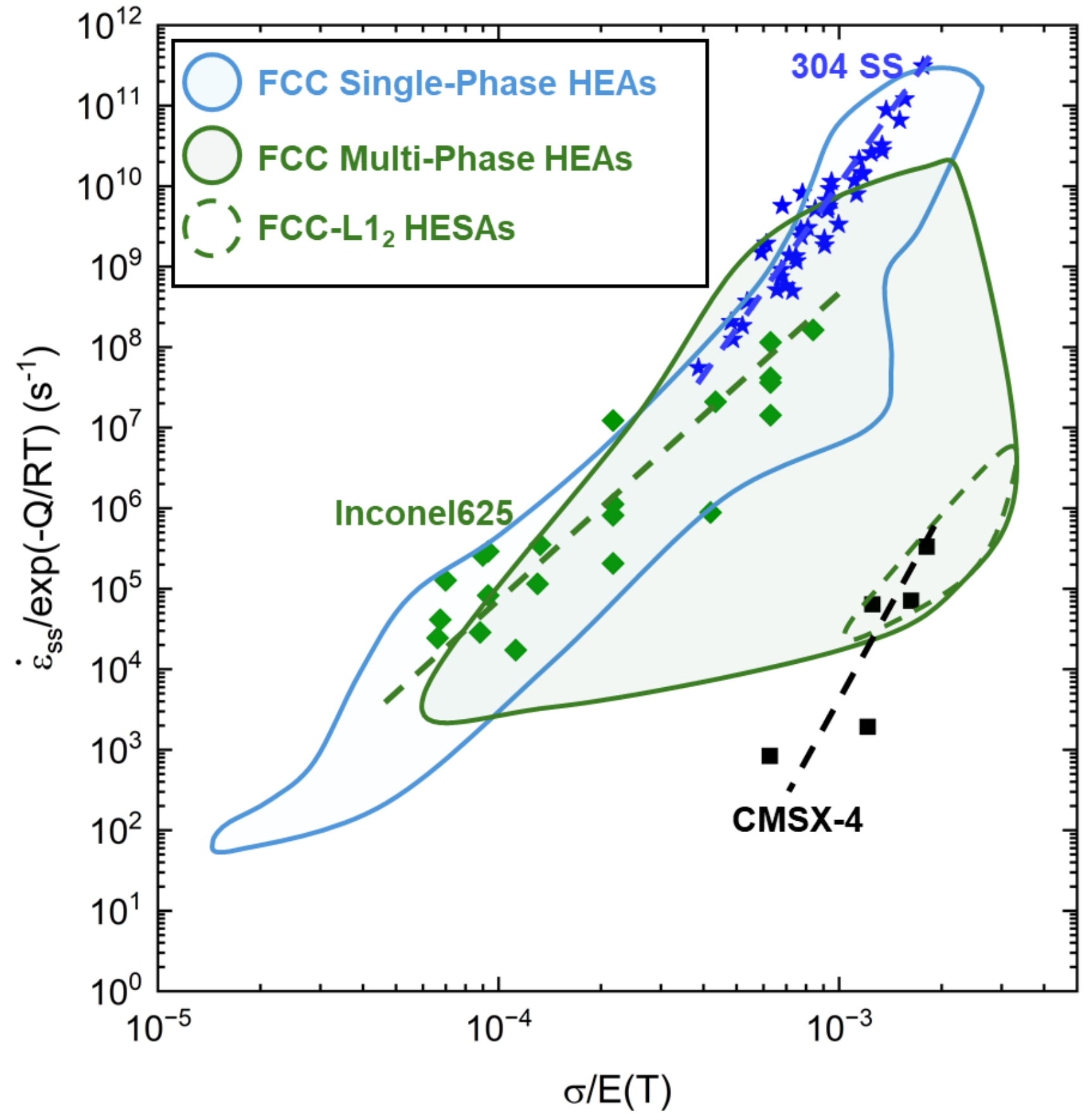

**Figure 30** Normalized secondary creep data for FCC HEAs considered in this review compared to select commercial alloy counterparts. The creep data for commercial alloys were adapted from Refs. [8, 198, 199].

Notably, these $\gamma + \gamma'$ superalloys continue to outperform most BCC single-phase and multi-phase RHEAs in terms of creep strength, benefiting from a unique and highly effective strengthening mechanism retained below their solvus temperature (~1100 °C). To assess the potential of RHEAs for higher-temperature applications, they must be creep tested at temperatures well above 1100 °C and benchmarked against conventional refractory alloys such as TZM ($Mo_{99.8}Ti_{0.11}Zr_{0.02}C_{0.07}$, in at. %), W-Re, and WC3009 ($Nb_{72}Hf_{22.4}W_{5.6}$, in at. %). It should be noted, however, that none of these refractory alloys are suitable for use in air or combustion environments due to their poor oxidation resistance. Given the lack of creep data for RHEAs beyond 1100 ºC, we have taken the approach of normalizing the existing creep data for all HEAs studied to date [8, 198-201], and compared them with a few commercial alloys, as shown in **Fig. 31**. Caveats are identified for such an approach, however, as normalization of stress by $E(T)$ and creep rate by $\exp{(Q_c/RT)}$ can obscure actual creep rates at specific temperatures and stresses that are directly relevant to engineering applications. For example, at 1200 ºC, minimum creep rates for TZM alloy range from approximately $10^{-7}$ to $10^{-6}$ $s^{-1}$ under an applied stress of 300 MPa. In contrast, achieving a similar creep rate in the single-phase BCC HfNbTaTiZr RHEA at the same temperature requires only 5–10 MPa of stress, representing a 30- to 60-fold reduction. However, the normalized creep rates for TZM appear higher than single-phase BCC RHEAs due to the significantly higher elastic modulus and activation energy for creep of Mo (330 GPa at room temperature, and 405 kJ/mol, respectively [202]) compared to those of HfNbTaTiZr (90.8 GPa [203] at room temperature and ~250 kJ/mol). Normalization, while useful from a scientific perspective, can give a false impression because what an engineer or designer needs to know is for how long a given material can withstand a certain stress at a certain temperature before reaching unacceptably high strains (of order 1%). To address this limitation, **Fig. 32** compares the actual creep rates of BCC RHEAs with those of commercial

BCC alloys at a variety of temperatures. It is evident that TZM exhibits lower creep rates at 1200 °C than all currently tested BCC RHEAs, even at lower temperatures. This comparison highlights the barrier that BCC RHEAs have to overcome for creep-resistant applications.

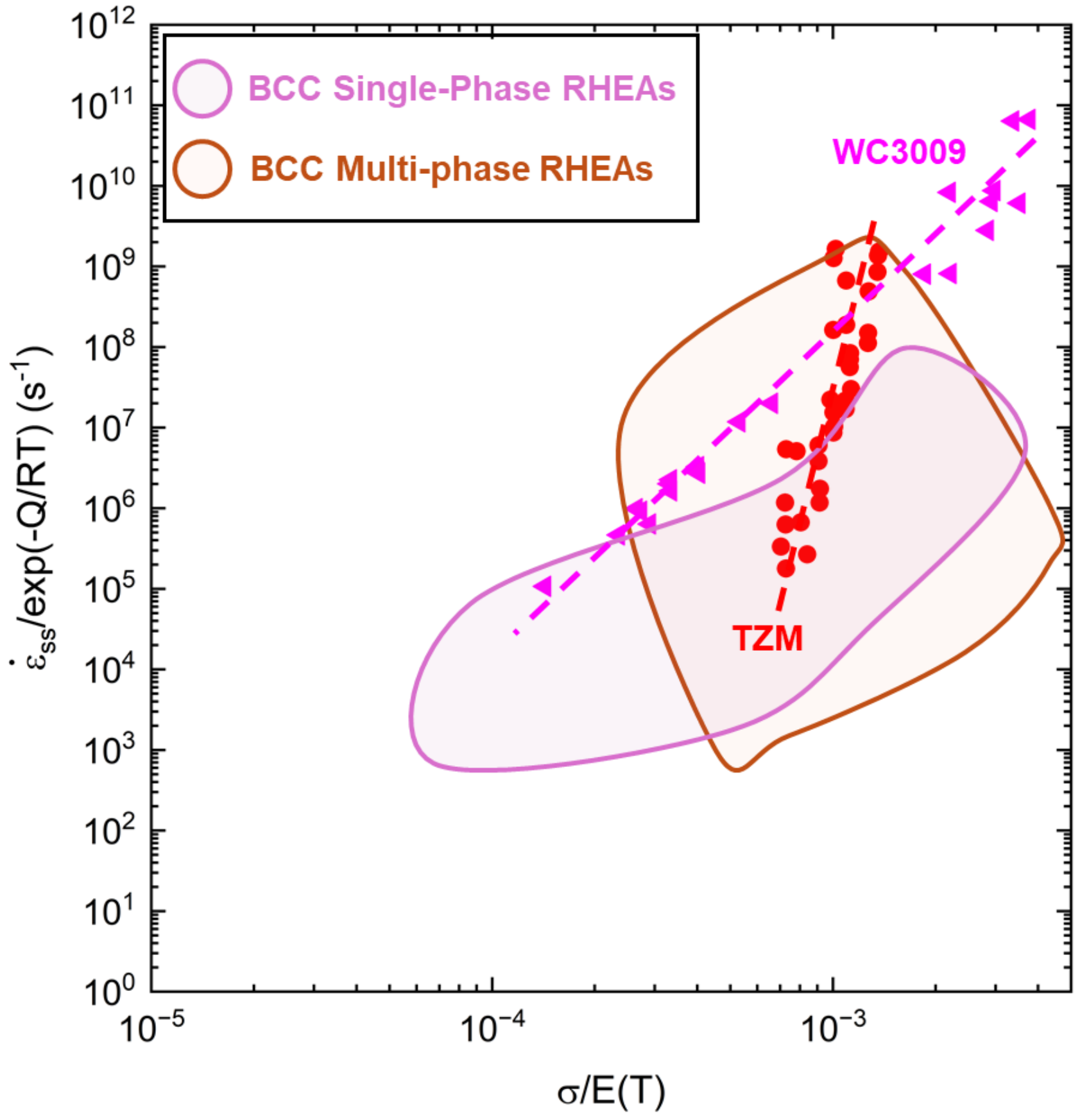

**Figure 31** Normalized secondary creep data for BCC HEAs considered in this review compared to select commercial alloy counterparts. The creep data for commercial alloys were adapted from Refs. [200, 201].

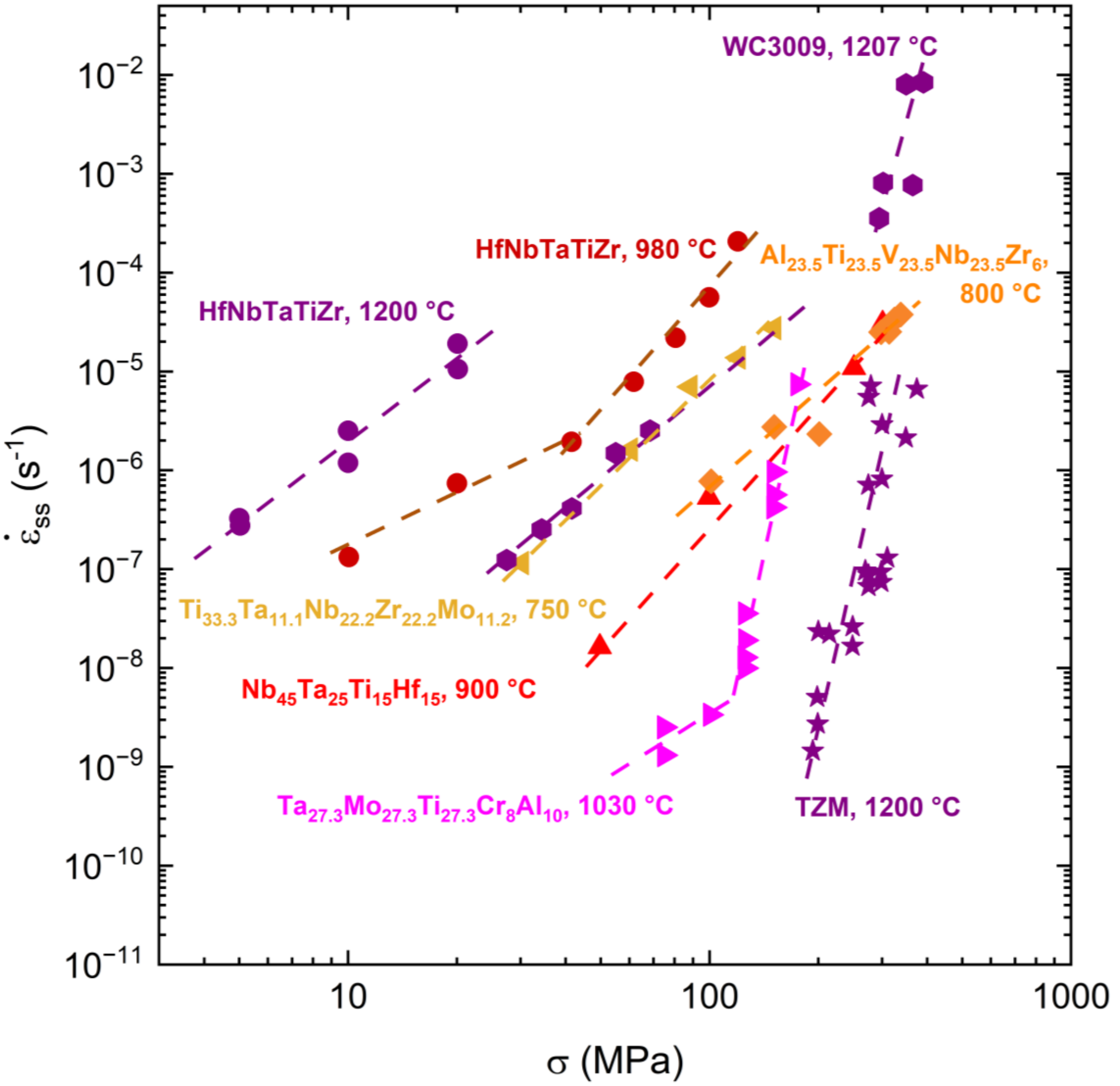

**Figure 32** Secondary creep data for single and multi-phase BCC RHEAs compared to commercial BCC alloys. The creep data were adapted from Refs. [90, 91, 93, 173, 174, 178, 179, 200, 201, 204].

Therefore, we foresee the future development of HEAs for creep-resistant applications to follow two primary directions.

1. For applications below 1100 °C, research should prioritize the design of multi-phase FCC-based HEAs that offer a balance of high toughness and processibility with sufficient creep resistance at a reduced cost and lead time. That said, conventional Ni-base superalloys have been highly optimized over decades and there is a wealth of data available about their processing capabilities and properties. Significant cost savings will have to be demonstrated (or some other equally compelling reason) before multi-phase FCC HEAs attract industrial interest beyond the current academic interest.
2. For temperatures exceeding 1100 °C, where conventional superalloys typically lose effectiveness, the focus should shift toward developing multi-phase, lightweight RHEAs with both high-temperature strength and intrinsic oxidation resistance. These alloys must be designed in conjunction with robust thermal- and environmental-barrier coatings to ensure adequate protection during service, as sufficient oxidation resistance is essential for any prospective aeroengine application. Currently, creep data above 1100 °C are extremely limited for BCC RHEAs, leaving their performance in the most relevant application temperature range largely unexplored. In particular, tensile creep data are scarce because many RHEAs exhibit brittleness and lack the damage tolerance required for sustained tensile loading at high temperatures. Therefore, priority should be given to conducting tensile creep tests on promising RHEA prototypes above 1100 °C to critically assess their potential for ultrahigh-temperature applications.

These two strategic directions are discussed in detail below:

Commercial single-crystal Ni-based superalloys (with a high amount of refractory elements) offer nearly unmatched creep resistance, combined with good toughness and oxidation resistance up to 1200 ºC [9, 146, 147, 205], albeit with the use of scarce alloying elements, particularly Ta, Re, and Ru. In applications where extreme high-temperature performance is not essential, there may be opportunities to develop advanced processing strategies for multi-phase FCC HEAs that offer adequate creep resistance while enabling shorter prototype-to-product transition times and greater potential for commercial implementation. Although the total cost reduction in the GRX-810 ODS-FCC HEA discussed above is not straightforward to assess due to the extra cost of powder production, it exemplifies a promising strategy that combines additive manufacturing with oxide dispersion strengthening to create high-strength structural materials potentially extending beyond the temperature capability of Ni-based single crystals. While the overall strength may be lower, similar to how MA754 compares with CMSX-4, the ODS approach retains its strength advantage up to ~1200 °C [206], where conventional γ′-strengthened superalloys lose effectiveness due to γ′ dissolution. In addition, the 3D-printability of ODS materials enables the rapid fabrication of near net-shape components, such as turbine blades, vanes, and rocket nozzles, for direct testing under application-relevant conditions [57, 58, 124, 143, 144, 156]. Furthermore, AM can provide beneficial microstructures for creep properties, such as elongated grains along the build direction that resembles directionally solidified alloys [143], in-situ oxidation during laser-powder interaction that can incorporate nanoscale oxides that pin the grain boundaries and strengthen the material [124, 207], as well as the high dislocation density or cell structure that can also improve strength. A fundamental understanding of the processing–structure–property relationships arising from these prospects remains incomplete, with several critical aspects yet to be fully explored. For instance, it is critical to understand the thermal stability of dislocation cell structures as influenced by local compositional fluctuations and stacking fault energies, both of which are affected by the nominal alloy composition and AM processing parameters, and how these factors collectively impact creep resistance at elevated temperatures. Recent studies have shown that thermal exposure can eliminate the dislocation structures and elemental segregation introduced by AM, resulting in a reduction in yield strength and creep resistance [208]. This degradation can be mitigated through ODS that effectively pins defects and stabilizes the microstructure [57, 143]. The same strategy also helps preserve the elongated grain structure by inhibiting recrystallization into an equiaxed morphology during annealing, HIP processing, or creep testing. Although ODS does not provide the same level of strengthening as the incorporation of higher than 60 vol% $\gamma'$ intermetallics, its capacity to maintain potentially favorable microstructures from AM should be fully leveraged.

BCC RHEAs may occupy a niche at temperatures beyond the operational limits of Ni-based superalloys, where the primary research objective shifts toward developing alloys capable of outperforming commercial refractory alloys. Current application demands for ultrahigh-temperature structural materials, such as those used in turbine engines, rocket propulsion systems, and hypersonic vehicles, have driven the development of Nb-based and lightweight Mo-based alloys, each facing distinct advantages and challenges.

Niobium has a density of 8.57 g/cm$^3$, comparable to that of Ni-based superalloys, making it an attractive base element for ultrahigh-temperature structural applications. Two prominent Nb-based alloys are C103 ($Nb_{92.5}Hf_{5.41}Ti_{2.0}$, in at. %) and WC3009 ($Nb_{75.1}Hf_{19.2}W_{5.6}$, in at. %). C103 exhibits a moderate yield strength of approximately 100 MPa at 1300 °C [209], while WC3009 offers a significantly higher yield strength of around 200 MPa at the same temperature [210]. However, the widespread use of WC3009 is limited by its high cost due to the substantial Hf content. Recent advances in NbTaTi-based RHEAs have shown excellent room-temperature tensile strength and fracture toughness [187, 211], and these alloys have also demonstrated compatibility with additive manufacturing processes [212, 213]. Despite these strengths, alloys such as HfNbTaTiZr and $Nb_{45}Ta_{25}Ti_{15}Hf_{15}$ currently suffer from insufficient creep resistance at elevated temperatures. In addition, the oxidation resistance of these alloys is a significant concern [214, 215]. Their creep performance could be improved through enhanced solid solution strengthening and more sluggish diffusion by incorporating minor amounts of Group VI elements (e.g., Mo, W), precipitation strengthening via niobium silicides [216] (as most other precipitates have solvus temperatures that are too low), and ODS enabled by powder metallurgy or additive manufacturing. Building on prior success with AM ODS-CrCoNi alloys and leveraging the proven printability of NbTaTi-based compositions, AM ODS-NbTaTi-based alloys represent a promising opportunity, particularly for applications where creep at or above 1100 °C is dominated by grain boundary diffusion. It is therefore prudent to envision the development of a multi-phase Nb-based RHEA with reduced Ta and Hf contents for lower cost and density, while targeting mechanical properties that match or exceed those of WC3009 at approximately 1300 °C. Environmental-barrier coating is mandatory for Nb-based alloys due to their poor oxidation resistance [217, 218], which could require extensive efforts that are beyond the scope of the current review.

Molybdenum-based alloys exhibit excellent high-temperature strength and creep resistance. Among them, the widely used TZM alloy demonstrates a yield strength exceeding 300 MPa at 1300 °C [219]. However, its application is constrained by poor oxidation resistance and high density. The ongoing development of Mo–Si–B alloys has led to significant improvements in oxidation resistance, particularly through the application of borosilicate coatings via pack cementation [220, 221]. These coatings enable self-healing, as Si and B in the alloy diffuse to damaged regions, thereby mitigating catastrophic oxidation and pesting [222]. Mo–Si–B alloys can also outperform TZM in creep resistance, with creep strengths around 200 MPa at 1300 °C for minimum creep rates in the range of $10^{-7}$ to $10^{-6}$ $s^{-1}$ [223]. Nevertheless, these alloys continue to suffer from limited tensile ductility at lower temperatures and relatively high density [224]. Recent efforts to incorporate substantial amounts of Ti have reduced the density of Mo–Si–B alloys to values below those of Ni-based superalloys while simultaneously enhancing creep resistance through solid solution strengthening within the Mo solid solution ($Mo_{ss}$) phase and eliminating the pesting behavior [225-227]. Parallel studies on MoNbTi-based RHEAs have demonstrated plastic deformability through multiple dislocation glide pathways enabled by the rugged potential energy landscape in HEAs [82]. Despite these advances, an intrinsic trade-off remains unresolved: high oxidation and creep resistance require increased Si content and a reduced $Mo_{ss}$ volume fraction, while good tensile ductility favors lower Si content and a higher $Mo_{ss}$ fraction [228]. With the recent demonstration of their additive manufacturability [229-231], these alloys present new opportunities in terms of the design of compositionally graded structures with spatially varying $Mo_{ss}$ and silicide volume fractions toward balancing oxidation resistance, creep strength, and tensile ductility.

The development of next-generation BCC–B2 refractory high entropy superalloys depends heavily on identifying B2 phases with solvus temperatures exceeding 1300 °C. Current BCC–B2 with Al and Zr additions are limited to operating below ~1050 °C and fail to surpass the performance of Ni-based superalloys at these temperatures. Promising new systems, such as those that contain HfRu [232, 233], ZrRu [234], and TaRe [235], have recently been reported to exhibit stable BCC–B2 structures beyond 1300 °C, suggesting potential for high-temperature creep resistance via B2 strengthening. However, the high cost and limited availability of elements like Ru and Re may restrict their widespread application, reserving these materials for only the most demanding applications where extreme performance can justify the expense.

## 5 Summary

We present a systematic review and a critical assessment of the current research progress of creep properties in high-entropy alloys (HEAs), with emphasis on both face-centered cubic (FCC) and body-centered cubic (BCC) systems, including both single-phase and multi-phase variants. While it was initially hypothesized that the compositional complexity of HEAs would confer exceptional high-temperature performance through mechanisms such as concentrated solid solution strengthening, severe lattice distortion, and sluggish diffusion, extensive creep studies have largely negated these assertions at least in single-phase alloys. Single-phase FCC HEAs, such as CrMnFeCoNi and its derivatives, exhibit creep resistance comparable to conventional austenitic stainless steels, significantly short of the performance offered by Ni-based superalloys. Their elevated-temperature deformation is primarily governed by thermally activated dislocation glide and grain boundary recovery mechanisms, with clear evidence of strength degradation at higher temperatures.

Enhanced creep performance (mostly in compression) has been observed in multi-phase FCC HEAs through strategies such as precipitation strengthening (e.g., FCC + $L1_2$) and oxide dispersion-strengthening (ODS). Notably, AM ODS-CrCoNi and NASA's GRX-810 represent major advances in creep-resistant HEAs, showing promising comparisons to commercial superalloys at temperatures approaching 1100 °C. However, despite these improvements, $\gamma + \gamma'$ high-entropy superalloys (HESAs) still owe much of their creep resistance to well-established superalloy design principles rather than novel HEA-specific mechanisms.

On the BCC front, refractory HEAs (RHEAs) such as MoNbTaWV demonstrate good strength in compression, but their practical applicability is severely hindered by poor tensile ductility due to grain boundary embrittlement. Ductile RHEAs like HfNbTaTiZr offer tensile ductility but suffer from significantly reduced creep resistance. Multi-phase BCC alloys, especially BCC-B2 dual-phase RHEAs, have emerged as promising candidates for high-temperature applications. Alloys such as $Ta_{27.3}Mo_{27.3}Ti_{27.3}Cr_8Al_{10}$ exhibit creep performance on par with CMSX-4 under lower stresses, yet limitations such as power-law breakdown, grain boundary embrittlement, and the lower solvus temperatures of B2 phases remain major challenges. High-solvus-temperature BCC-B2 RHEAs are currently under development, but the cost of their constituent elements may be prohibitively high for all but the most demanding applications.

Overall, recent advances in HEA research underscore the critical role of secondary phase strengthening, grain boundary engineering, and maintaining stable microstructures in improving creep resistance at elevated temperatures. Despite their unique compositional complexity, HEAs have not yet revealed fundamentally new creep mechanisms and better creep properties compared to conventional alloys. Moving forward, progress will depend on thoughtful alloy design, guided by high-throughput computational screening and enabled by advanced processing methods, to optimize phase stability, dislocation behavior, and resistance to environmental degradation. These efforts are essential for unlocking the full potential of HEAs for structural applications under extreme environments.

**Acknowledgements**

The study at the University of California, Davis led by MZ was partially supported by the U.S. Department of Energy Office of Science, Office of Basic Energy Sciences, under Contract No. DE-SC0025388. Work at the Molecular Foundry was supported by the Office of Science, Office of Basic Energy Sciences, of the U.S. Department of Energy under Contract No. DE-AC02-05CH11231.

UG acknowledges many years of support by the Deutsche Forschungsgemeinschaft (DFG) through the priority programme "Compositionally Complex Alloys - High Entropy Alloys (CCA - HEA)" with project number 388982456, GL181/57-1 and 57-2.

MH gratefully acknowledges financial support by the Deutsche Forschungsgemeinschaft (DFG), grant no. HE 1872/34-2.

## Appendix

**Supplementary Table 1 Secondary or minimum creep rates for single-phase FCC HEAs.**

| CrMnFeCoNi, FCC RX, Tensile creep, Ref. [115]. | | | | | |
|---|---|---|---|---|---|
| **T (°C)** | **$\sigma$ (MPa)** | **$\dot{\varepsilon}_{ss}$ (s$^{-1}$)** | **T (°C)** | **$\sigma$ (MPa)** | **$\dot{\varepsilon}_{ss}$ (s$^{-1}$)** |
| 600 | 206.8 | 2.03E-07 | 600 | 137.9 | 1.42E-08 |
| | 189.6 | 1.23E-07 | | 137.9 | 1.46E-08 |
| | 172.4 | 6.58E-08 | | 120.7 | 7.28E-09 |
| | 155.1 | 3.72E-08 | | 103.4 | 2.76E-09 |
| | | | | | |
| CrMnFeCoNi, FCC RX, Tensile creep, Ref. [88]. | | | | | |
| **T (°C)** | **$\sigma$ (MPa)** | **$\dot{\varepsilon}_{ss}$ (s$^{-1}$)** | **T (°C)** | **$\sigma$ (MPa)** | **$\dot{\varepsilon}_{ss}$ (s$^{-1}$)** |
| 500 | 200 | 1.59E-09 | 550 | 160 | 3.66E-09 |
| | 240 | 4.61E-09 | | 240 | 5.28E-08 |
| | 280 | 1.00E-08 | | 260 | 8.90E-08 |
| | 320 | 1.65E-08 | | 280 | 1.79E-07 |
| | 360 | 4.27E-08 | | 360 | 1.53E-08 |
| | 380 | 1.06E-07 | | 320 | 5.18E-07 |
| | 400 | 1.89E-07 | | 360 | 2.30E-06 |
| 600 | 140 | 2.63E-08 | | | |
| | 160 | 5.71E-08 | | | |
| | 180 | 1.08E-07 | | | |
| | 200 | 2.01E-07 | | | |
| | 220 | 4.52E-07 | | | |
| | 240 | 8.90E-07 | | | |
| | 280 | 4.27E-06 | | | |
| | 320 | 1.17E-05 | | | |
| | | | | | |
| CrMnFeCoNi, FCC RX, Tensile creep, Ref. [87]. | | | | | |
| **T (°C)** | **$\sigma$ (MPa)** | **$\dot{\varepsilon}_{ss}$ (s$^{-1}$)** | **T (°C)** | **$\sigma$ (MPa)** | **$\dot{\varepsilon}_{ss}$ (s$^{-1}$)** |
| 535 | 48 | 3.88E-08 | 650 | 22 | 6.96E-08 |
| | 64 | 7.98E-08 | | 26 | 1.55E-07 |
| | 96 | 2.31E-07 | | 32 | 5.37E-07 |
| 600 | 26 | 1.67E-08 | | 35 | 1.12E-06 |
| | 32 | 4.62E-08 | | 39 | 2.04E-06 |
| | 35 | 8.56E-08 | | 48 | 3.75E-06 |
| | 48 | 4.48E-07 | | 56 | 4.08E-06 |
| | 64 | 1.02E-06 | | 64 | 6.55E-06 |
| | 78 | 1.81E-06 | | 78 | 1.12E-05 |
| | 96 | 2.60E-06 | | 86 | 1.23E-05 |
| | | | | 92 | 1.36E-05 |
| | | | | 97 | 6.01E-05 |
| | | | | | |
| CrMnFeCoNi, FCC RX, Tensile creep, Ref. [116]. | | | | | |
| **T (°C)** | **$\sigma$ (MPa)** | **$\dot{\varepsilon}_{ss}$ (s$^{-1}$)** | **T (°C)** | **$\sigma$ (MPa)** | **$\dot{\varepsilon}_{ss}$ (s$^{-1}$)** |
| 500 | 140 | 6.32E-09 | 550 | 110 | 7.21E-09 |
| | 180 | 1.80E-08 | | 140 | 1.63E-08 |
| | 220 | 5.28E-08 | | 170 | 4.06E-08 |
| | 260 | 1.13E-07 | | 200 | 8.04E-08 |
| | 300 | 1.56E-07 | | 230 | 1.90E-07 |
| | 345 | 2.84E-07 | | 260 | 3.58E-07 |
| 600 | 80 | 9.71E-09 | 650 | 65 | 1.06E-08 |
| | 105 | 2.42E-08 | | 85 | 2.34E-08 |
| | 130 | 7.79E-08 | | 100 | 8.04E-08 |
| | 160 | 2.32E-07 | | 125 | 2.55E-07 |
| | 180 | 6.34E-07 | | 145 | 6.35E-07 |
| | 205 | 1.01E-06 | | 165 | 1.49E-06 |
| | 220 | 1.98E-06 | | 185 | 3.51E-06 |
| 700 | 50 | 2.26E-08 | | | |
| | 65 | 5.09E-08 | | | |
| | 80 | 9.18E-08 | | | |
| | 95 | 2.88E-07 | | | |
| | 110 | 6.21E-07 | | | |
| | 125 | 1.34E-06 | | | |
| | 130 | 1.85E-06 | | | |
| | | | | | |
| CrMnFeCoNi, FCC RX, Tensile creep, Ref. [117]. | | | | | |
| **T (°C)** | **$\sigma$ (MPa)** | **$\dot{\varepsilon}_{ss}$ ($s^{-1}$)** | **T (°C)** | **$\sigma$ (MPa)** | **$\dot{\varepsilon}_{ss}$ ($s^{-1}$)** |
| 650 | 50 | 1.77E-09 | 675 | 50 | 1.12E-08 |
| | 75 | 1.94E-08 | | 75 | 8.12E-08 |
| | 100 | 1.18E-07 | | 100 | 3.88E-07 |
| | 125 | 4.05E-07 | | 125 | 1.26E-06 |
| | 150 | 1.44E-06 | | 150 | 2.86E-06 |
| | 200 | 8.12E-06 | | | |
| 700 | 50 | 5.43E-08 | | | |
| | 75 | 2.67E-07 | | | |
| | 100 | 9.59E-07 | | | |
| | 125 | 2.66E-06 | | | |
| | 150 | 5.56E-06 | | | |
| | | | | | |
| CrMnFeCoNi, FCC RX, Tensile creep, Ref. [118]. | | | | | |
| **T (°C)** | **$\sigma$ (MPa)** | **$\dot{\varepsilon}_{ss}$ ($s^{-1}$)** | **T (°C)** | **$\sigma$ (MPa)** | **$\dot{\varepsilon}_{ss}$ ($s^{-1}$)** |
| 700 | 50 | 6.07E-09 | | | |
| | 70 | 3.73E-08 | | | |
| | 890 | 2.07E-07 | | | |
| | 110 | 6.48E-07 | | | |
| | | | | | |
| CrMnFeCoNi, FCC RX, Tensile creep, Ref. [92]. | | | | | |
| **T (°C)** | **$\sigma$ (MPa)** | **$\dot{\varepsilon}_{ss}$ ($s^{-1}$)** | **T (°C)** | **$\sigma$ (MPa)** | **$\dot{\varepsilon}_{ss}$ ($s^{-1}$)** |
| 750 | 40 | 3.37E-07 | 800 | 20 | 8.04E-08 |
| | 60 | 1.45E-06 | | 40 | 1.09E-06 |
| | 80 | 4.00E-06 | | 60 | 5.22E-06 |
| | 130 | 2.02E-05 | | 80 | 1.45E-05 |
| | 200 | 1.45E-4 | | 100 | 3.26E-05 |
| 850 | 20 | 2.82E-07 | | 130 | 1.14E-4 |
| | 30 | 1.37E-06 | 900 | 20 | 7.21E-07 |
| | 40 | 4.55E-06 | | 40 | 1.13E-05 |
| | 60 | 1.87E-05 | | 60 | 4.36E-05 |
| | 80 | 5.56E-05 | | 80 | 1.06E-4 |
| | | | | | |
| CrMnFeCoNi, FCC RX, Compressive creep, Ref. [119]. | | | | | |
| **T (°C)** | **$\sigma$ (MPa)** | **$\dot{\varepsilon}_{ss}$ ($s^{-1}$)** | **T (°C)** | **$\sigma$ (MPa)** | **$\dot{\varepsilon}_{ss}$ ($s^{-1}$)** |
| 800 | 10 | 7.03E-09 | 800 | 30 | 2.43E-07 |
| | 15 | 3.25E-08 | | 50 | 1.66E-06 |
| | 20 | 5.16E-08 | | 70 | 3.34E-06 |
| | | | | | |
| CrMnFeCoNi, FCC SX, Compressive creep, Ref. [119]. | | | | | |
| **T (°C)** | **$\sigma$ (MPa)** | **$\dot{\varepsilon}_{ss}$ ($s^{-1}$)** | **T (°C)** | **$\sigma$ (MPa)** | **$\dot{\varepsilon}_{ss}$ ($s^{-1}$)** |
| 800 | 14 | 6.18E-10 | 800 | 40 | 2.12E-07 |
| | 20 | 2.67E-09 | | 50 | 3.98E-07 |
| | 30 | 2.75E-08 | | 70 | 2.49E-06 |
| | | | | | |
| CrMnFeCoNi, FCC RX, Tensile creep, Ref. [119]. | | | | | |
| **T (°C)** | **$\sigma$ (MPa)** | **$\dot{\varepsilon}_{ss}$ ($s^{-1}$)** | **T (°C)** | **$\sigma$ (MPa)** | **$\dot{\varepsilon}_{ss}$ ($s^{-1}$)** |
| 980 | 5 | 3.95E-08 | 980 | 13 | 1.67E-06 |
| | 8 | 2.95E-07 | | 20 | 8.93E-06 |
| | 10 | 5.55E-07 | | | |
| | | | | | |
| CrMnFeCoNi, FCC SX, Tensile creep, Ref. [119]. | | | | | |
| **T (°C)** | **$\sigma$ (MPa)** | **$\dot{\varepsilon}_{ss}$ ($s^{-1}$)** | **T (°C)** | **$\sigma$ (MPa)** | **$\dot{\varepsilon}_{ss}$ ($s^{-1}$)** |
| 980 | 5 | 2.34E-09 | 980 | 20 | 5.74E-06 |
| | 8 | 6.44E-08 | | 25 | 2.49E-05 |
| | 13 | 8.97E-07 | | | |
| | | | | | |
| CrMnFeCoNi, FCC SX, Tensile creep, Ref. [120]. | | | | | |
| **T (°C)** | **$\sigma$ (MPa)** | **$\dot{\varepsilon}_{ss}$ ($s^{-1}$)** | **T (°C)** | **$\sigma$ (MPa)** | **$\dot{\varepsilon}_{ss}$ ($s^{-1}$)** |
| 700 | 50 | 4.62E-09 | 1100 | 2 | 8.50E-09 |
| | 64 | 2.30E-08 | | 3 | 1.27E-08 |
| | 78 | 1.49E-07 | | 4 | 6.97E-08 |
| | 100 | 5.23E-07 | | 5 | 3.37E-07 |
| | 120 | 2.13E-06 | | 6 | 1.59E-06 |
| 980 | 8 | 6.40E-08 | | 8 | 7.65E-06 |
| | 13 | 8.66E-07 | | | |
| | 20 | 5.38E-06 | | | |
| | 25 | 2.42E-05 | | | |
| | | | | | |

| CrFeCoNi, FCC RX, Tensile creep, Ref. [121]. | | | | | |
|---|---|---|---|---|---|
| **T (°C)** | **$\sigma$ (MPa)** | **$\dot{\varepsilon}_{ss}$ ($s^{-1}$)** | **T (°C)** | **$\sigma$ (MPa)** | **$\dot{\varepsilon}_{ss}$ ($s^{-1}$)** |
| 600 | 137.9 | 2.71E-09 | 612.5 | 137.9 | 9.31E-09 |
| | 189.6 | 2.59E-08 | | 172.4 | 3.28E-08 |
| 625 | 137.9 | 1.14E-08 | 650 | 137.9 | 4.47E-08 |
| | 155.1 | 3.28E-08 | | 172.4 | 2.23E-07 |
| 637.5 | 137.9 | 2.64E-08 | 662.5 | 120.7 | 2.86E-08 |

| CrFeCoNi, FCC RX, Tensile creep, Ref. [118]. | | | | | |
|---|---|---|---|---|---|
| **T (°C)** | **$\sigma$ (MPa)** | **$\dot{\varepsilon}_{ss}$ ($s^{-1}$)** | **T (°C)** | **$\sigma$ (MPa)** | **$\dot{\varepsilon}_{ss}$ ($s^{-1}$)** |
| 700 | 50 | 1.20E-08 | 700 | 90 | 3.60E-07 |
| | 70 | 8.09E-08 | | 140 | 3.28E-06 |

| CrFeCoNi, FCC RX, Tensile creep, Ref. [117]. | | | | | |
|---|---|---|---|---|---|
| **T (°C)** | **$\sigma$ (MPa)** | **$\dot{\varepsilon}_{ss}$ ($s^{-1}$)** | **T (°C)** | **$\sigma$ (MPa)** | **$\dot{\varepsilon}_{ss}$ ($s^{-1}$)** |
| 650 | 100 | 1.48E-08 | 700 | 50.003876 | 5.56E-09 |
| | 125 | 6.95E-08 | | 75.054508 | 5.69E-08 |
| | 150 | 2.79E-07 | | 100.06107 | 3.67E-07 |
| | 200 | 1.76E-06 | | 125.06031 | 1.20E-06 |
| 725 | 50 | 2.58E-08 | | 149.87646 | 3.73E-06 |
| | 75 | 2.50E-07 | | | |
| | 100 | 1.14E-06 | | | |
| | 125 | 4.26E-06 | | | |
| | 150 | 1.10E-05 | | | |

| CrCoNi, FCC RX, Tensile creep, Ref. [122]. | | | | | |
|---|---|---|---|---|---|
| **T (°C)** | **$\sigma$ (MPa)** | **$\dot{\varepsilon}_{ss}$ ($s^{-1}$)** | **T (°C)** | **$\sigma$ (MPa)** | **$\dot{\varepsilon}_{ss}$ ($s^{-1}$)** |
| 550 | 200 | 1.62E-07 | 55 | 260 | 2.69E-07 |
| | 220 | 1.96E-07 | | 280 | 5.64E-07 |

| CrCoNi, FCC RX, Tensile creep, Ref. [123]. | | | | | |
|---|---|---|---|---|---|
| **T (°C)** | **$\sigma$ (MPa)** | **$\dot{\varepsilon}_{ss}$ ($s^{-1}$)** | **T (°C)** | **$\sigma$ (MPa)** | **$\dot{\varepsilon}_{ss}$ ($s^{-1}$)** |
| 700 | 50 | 2.55E-09 | 750 | 40 | 1.44E-08 |
| | 70 | 2.45E-08 | | 55 | 7.41E-08 |
| | 90 | 7.71E-08 | | 70 | 1.84E-07 |
| | 110 | 1.97E-07 | | 110 | 2.723E-06 |
| | 130 | 5.87E-07 | | | |
| 800 | 30 | 1.99E-08 | | | |
| | 50 | 2.13E-07 | | | |
| | 70 | 1.248E-06 | | | |

| CrCoNi, FCC RX, Tensile creep, Ref. [124]. | | | | | |
|---|---|---|---|---|---|
| **T (°C)** | **$\sigma$ (MPa)** | **$\dot{\varepsilon}_{ss}$ ($s^{-1}$)** | **T (°C)** | **$\sigma$ (MPa)** | **$\dot{\varepsilon}_{ss}$ ($s^{-1}$)** |
| 750 | 60 | 4.20E-07 | 800 | 60 | 1.86E-06 |
| | 80 | 1.65E-06 | | 80 | 5.07E-06 |
| | 130 | 1.32E-05 | | 130 | 5.94E-05 |
| | 200 | 9.57E-05 | | 200 | 2.80E-04 |
| 850 | 40 | 1.18E-06 | 900 | 40 | 2.53E-06 |
| | 60 | 5.39E-06 | | 60 | 1.52E-05 |
| | 80 | 1.99E-05 | | 80 | 6.81E-05 |
| | 130 | 1.99E-04 | | 1308 | 6.34E-04 |
| | 200 | 1.14E-03 | | | |
| | | | | | |
| CrCoNi, FCC AM, Tensile creep, Ref. [124]. | | | | | |
| **T (°C)** | **$\sigma$ (MPa)** | **$\dot{\varepsilon}_{ss}$ ($s^{-1}$)** | **T (°C)** | **$\sigma$ (MPa)** | **$\dot{\varepsilon}_{ss}$ ($s^{-1}$)** |
| 750 | 60 | 4.29E-08 | 800 | 60 | 2.72E-07 |
| | 80 | 1.85E-07 | | 80 | 1.28E-06 |
| | 130 | 3.74E-06 | | 130 | 2.18E-05 |
| | 200 | 6.42E-05 | | 200 | 4.11E-04 |
| 850 | 40 | 1.32E-07 | 900 | 40 | 5.30E-07 |
| | 60 | 1.42E-06 | | 60 | 6.40E-06 |
| | 80. | 6.65E-06 | | 80 | 2.85E-05 |
| | 130 | 1.21E-04 | | 130 | 5.13E-04 |
| | 200 | 1.91E-03 | | | |
| | | | | | |
| $Ni_{42.5}Co_{21.3}Fe_{21.3}V_{10.6}Mo_{4.3}$, FCC RX, Tensile creep, Ref. [125]. | | | | | |
| **T (°C)** | **$\sigma$ (MPa)** | **$\dot{\varepsilon}_{ss}$ ($s^{-1}$)** | **T (°C)** | **$\sigma$ (MPa)** | **$\dot{\varepsilon}_{ss}$ ($s^{-1}$)** |
| 700 | 150 | 1.89E-08 | | | |
| | 175 | 2.71E-08 | | | |
| | 200 | 4.66E-08 | | | |

**Supplementary Table 2 Secondary or minimum creep rates for multi-phase FCC HEAs.**

| $Al_{7.5}Cr_{18.5}Mn_{18.5}Fe_{18.5}Co_{18.5}Ni_{18.5}$, FCC + BCC, Tensile creep, Ref. [136]. | | | | | |
|---|---|---|---|---|---|
| **T (°C)** | **$\sigma$ (MPa)** | **$\dot{\varepsilon}_{ss}$ ($s^{-1}$)** | **T (°C)** | **$\sigma$ (MPa)** | **$\dot{\varepsilon}_{ss}$ ($s^{-1}$)** |
| 600 | 120 | 2.31E-09 | 650 | 80 | 1.64E-08 |
| | 160 | 8.18E-09 | | 120 | 3.30E-08 |
| | 200 | 1.53E-08 | | 160 | 5.10E-08 |
| | 240 | 4.15E-08 | | 200 | 1.47E-07 |
| | 280 | 8.77E-08 | | 240 | 3.72E-07 |
| | 320 | 1.31E-07 | | | |
| 700 | 40 | 1.70E-08 | | | |
| | 80 | 9.93E-08 | | | |
| | 120 | 1.85E-07 | | | |
| | 160 | 1.14E-06 | | | |
| | 200 | 2.17E-06 | | | |
| $Al_{10.2}Cr_{17.9}Mn_{17.9}Fe_{17.9}Co_{17.9}Ni_{17.9}$ , FCC + BCC, Tensile creep, Ref. [136]. | | | | | |
| **T (°C)** | **$\sigma$ (MPa)** | **$\dot{\varepsilon}_{ss}$ ($s^{-1}$)** | **T (°C)** | **$\sigma$ (MPa)** | **$\dot{\varepsilon}_{ss}$ ($s^{-1}$)** |
| 600 | 160 | 8.24E-08 | 650 | 80 | 1.70E-07 |
| | 200 | 1.71E-07 | | 120 | 4.47E-07 |
| | 240 | 5.13E-07 | | 160 | 1.24E-06 |
| | 280 | 1.69E-06 | | 200 | 3.17E-06 |
| | 320 | 4.45E-06 | | 240 | 8.76E-06 |
| 700 | 40 | 9.92E-08 | | | |
| | 80 | 5.70E-07 | | | |
| | 120 | 1.39E-06 | | | |
| | 160 | 3.60E-06 | | | |
| | 200 | 1.14E-05 | | | |
| | | | | | |
| $Al_9Cr_{9.1}Mn_{27.3}Fe_{45.5}Co_{9.1}$, FCC + BCC, Tensile creep, Ref. [137]. | | | | | |
| **T (°C)** | **$\sigma$ (MPa)** | **$\dot{\varepsilon}_{ss}$ ($s^{-1}$)** | **T (°C)** | **$\sigma$ (MPa)** | **$\dot{\varepsilon}_{ss}$ ($s^{-1}$)** |
| 650 | 50 | 5.21E-08 | | | |
| | 70 | 2.76E-07 | | | |
| | 90 | 9.84E-07 | | | |
| | | | | | |
| $Al_{6.8}Cr_{23.3}Fe_{23.3}Co_{23.3}Ni_{23.3}$, FCC + B2 + $L1_2$, Tensile creep, Ref. [138]. | | | | | |
| **T (°C)** | **$\sigma$ (MPa)** | **$\dot{\varepsilon}_{ss}$ ($s^{-1}$)** | **T (°C)** | **$\sigma$ (MPa)** | **$\dot{\varepsilon}_{ss}$ ($s^{-1}$)** |
| 700 | 65 | 2.07E-09 | 730 | 50 | 1.69E-09 |
| | 70 | 1.91E-09 | | 65 | 6.53E-09 |
| | 75 | 4.56E-09 | | 70 | 7.15E-09 |
| | 80 | 2.96E-09 | | 75 | 1.28E-08 |
| | 85 | 4.91E-09 | | 110 | 6.49E-08 |
| | 100 | 5.87E-09 | | | |
| | 110 | 1.05E-08 | | | |
| 760 | 40 | 2.50E-09 | | | |
| | 50 | 3.12E-09 | | | |
| | 60 | 1.63E-08 | | | |
| | 65 | 3.30E-08 | | | |
| | 75 | 2.41E-07 | | | |
| | 110 | 8.39E-07 | | | |
| | | | | | |
| $Al_{16.4}Cr_{16.4}Fe_{16.4}Co_{16.4}Ni_{34.4}$, FCC + B2 + $L1_2$, Tensile creep, Ref. [139]. | | | | | |
| **T (°C)** | **$\sigma$ (MPa)** | **$\dot{\varepsilon}_{ss}$ ($s^{-1}$)** | **T (°C)** | **$\sigma$ (MPa)** | **$\dot{\varepsilon}_{ss}$ ($s^{-1}$)** |
| 700 | 80 | 3.87E-09 | 800 | 50 | 6.58E-08 |
| | 160 | 3.33E-08 | | 100 | 2.53E-06 |
| | 240 | 1.78E-07 | | 150 | 5.24E-06 |
| 900 | 20 | 3.47E-08 | | | |
| | 40 | 2.27E-06 | | | |
| | 60 | 3.40E-06 | | | |
| | | | | | |

$Ni_{36}Al_{10}Co_{25}Cr_{8}Fe_{15}Ti_{6}$, FCC + $L1_2$, Tensile creep, Ref. [140].

| **T (°C)** | ***σ* (MPa)** | **$\dot{\varepsilon}_{ss}$ (s$^{-1}$)** | **T (°C)** | ***σ* (MPa)** | **$\dot{\varepsilon}_{ss}$ (s$^{-1}$)** |
|---|---|---|---|---|---|
| 750 | 325 | 2.30E-08 | | | |
| | 375 | 1.06E-07 | | | |
| | 425 | 4.18E-07 | | | |

$Ni_{47.9}Al_{10.2}Co_{16.9}Cr_{7.4}Fe_{8.9}Ti_{5.8}Mo_{0.9}Nb_{1.2}W_{0.4}C_{0.4}$, FCC + $L1_2$, Tensile creep, Ref. [40].

| **T (°C)** | ***σ* (MPa)** | **$\dot{\varepsilon}_{ss}$ (s$^{-1}$)** | **T (°C)** | ***σ* (MPa)** | **$\dot{\varepsilon}_{ss}$ (s$^{-1}$)** |
|---|---|---|---|---|---|
| 750 | 159 | 2.06E-09 | 850 | 159 | 3.70E-08 |
| 982 | 159 | 1.23E-06 | | | |

$Cr_{19}Fe_{19}Co_{19}Ni_{38}Mo_{5}$, FCC + μ phase, Tensile creep, Ref. [141].

| **T (°C)** | ***σ* (MPa)** | **$\dot{\varepsilon}_{ss}$ (s$^{-1}$)** | **T (°C)** | ***σ* (MPa)** | **$\dot{\varepsilon}_{ss}$ (s$^{-1}$)** |
|---|---|---|---|---|---|
| 650 | 172 | 1.06E-08 | | | |

$Cr_{18.5}Fe_{14.8}Co_{14.8}Ni_{44.4}Mo_{7.5}$, FCC + μ phase, Tensile creep, Ref. [141].

| **T (°C)** | ***σ* (MPa)** | **$\dot{\varepsilon}_{ss}$ (s$^{-1}$)** | **T (°C)** | ***σ* (MPa)** | **$\dot{\varepsilon}_{ss}$ (s$^{-1}$)** |
|---|---|---|---|---|---|
| 650 | 172 | 1.65E-09 | | | |

$Cr_{18.8}Fe_{12.5}Co_{12.5}Ni_{50.0}Mo_{6.3}$, FCC + μ phase, Tensile creep, Ref. [141].

| **T (°C)** | ***σ* (MPa)** | **$\dot{\varepsilon}_{ss}$ (s$^{-1}$)** | **T (°C)** | ***σ* (MPa)** | **$\dot{\varepsilon}_{ss}$ (s$^{-1}$)** |
|---|---|---|---|---|---|
| 650 | 172 | 4.22E-09 | | | |

$Y_2O_3$-$TiO_2$-CrMnFeCoNi, ODS-FCC, Compressive creep, Ref. [142].

| **T (°C)** | ***σ* (MPa)** | **$\dot{\varepsilon}_{ss}$ (s$^{-1}$)** | **T (°C)** | ***σ* (MPa)** | **$\dot{\varepsilon}_{ss}$ (s$^{-1}$)** |
|---|---|---|---|---|---|
| 700 | 35 | 1.06E-09 | 750 | 29 | 1.90E-09 |
| | 50 | 2.03E-09 | | 35 | 2.41E-09 |
| | 58 | 7.80E-09 | | 46 | 7.49E-09 |
| | 64 | 1.80E-08 | | 52 | 2.05E-08 |
| | 68 | 4.20E-08 | | 60 | 1.12E-07 |
| | 75 | 4.14E-07 | | 73 | 4.03E-06 |
| | 85 | 1.56E-06 | | 84 | 2.60E-05 |
| | 100 | 1.14E-05 | | | |
| 800 | 10 | 8.78E-10 | | | |
| | 24 | 5.37E-09 | | | |
| | 30 | 1.23E-08 | | | |
| | 32 | 7.20E-09 | | | |
| | 38 | 6.97E-08 | | | |
| | 48 | 4.11E-07 | | | |
| | 57 | 4.639E-06 | | | |
| | 67 | 4.191E-05 | | | |

$Y_2O_3$-CrCoNi, ODS-FCC, Tensile creep, Ref. [143].

| **T (°C)** | ***σ* (MPa)** | **$\dot{\varepsilon}_{ss}$ (s$^{-1}$)** | **T (°C)** | ***σ* (MPa)** | **$\dot{\varepsilon}_{ss}$ (s$^{-1}$)** |
|---|---|---|---|---|---|
| 750 | 80 | 1.99E-08 | 800 | 60 | 3.65E-08 |
| | 100 | 6.95E-08 | | 80 | 1.45E-07 |
| | 130 | 4.52E-07 | | 130 | 4.68E-06 |
| | 200 | 8.94E-06 | | 200 | 7.30E-05 |
| 850 | 60 | 2.00E-07 | 900 | 60 | 8.33E-07 |
| | 80 | 1.03E-06 | | 80 | 4.99E-06 |
| | 130 | 2.08E-05 | | 130 | 9.10E-05 |
| | 200 | 5.08E-04 | | 200 | 2.02E-03 |
| | | | | | |
| C-CrMnFeCoNi, Carbide-FCC, Tensile creep, Ref. [144]. | | | | | |
| **T (°C)** | **$\sigma$ (MPa)** | **$\dot{\varepsilon}_{ss}$ ($s^{-1}$)** | **T (°C)** | **$\sigma$ (MPa)** | **$\dot{\varepsilon}_{ss}$ ($s^{-1}$)** |
| 600 | 175 | 3.85E-09 | 600 | 275 | 2.13E-08 |
| | 200 | 5.99E-09 | | 300 | 4.18E-08 |
| | 225 | 8.41E-09 | | 325 | 7.22E-08 |
| | 250 | 1.15E-08 | | | |

**Supplementary Table 3 Secondary or minimum creep rates for single-phase BCC HEAs.**

| HfNbTaTiZr, BCC RX, Tensile creep, Ref. [90]. | | | | | |
|---|---|---|---|---|---|
| **T (°C)** | **$\sigma$ (MPa)** | **$\dot{\varepsilon}_{ss}$ ($s^{-1}$)** | **T (°C)** | **$\sigma$ (MPa)** | **$\dot{\varepsilon}_{ss}$ ($s^{-1}$)** |
| 980 | 10 | 1.33E-07 | 1100 | 5 | 5.65E-07 |
| | 20 | 7.38E-07 | | 10 | 3.29E-06 |
| | 40 | 1.94E-06 | | 20 | 1.05E-05 |
| | 60 | 7.88E-06 | | 30 | 1.95E-05 |
| | 80 | 2.19E-05 | | | |
| | 100 | 5.63E-05 | | | |
| | 120 | 2.07E-04 | | | |
| | | | | | |
| HfNbTaTiZr, BCC RX, Tensile creep, Ref. [91]. | | | | | |
| **T (°C)** | **$\sigma$ (MPa)** | **$\dot{\varepsilon}_{ss}$ ($s^{-1}$)** | **T (°C)** | **$\sigma$ (MPa)** | **$\dot{\varepsilon}_{ss}$ ($s^{-1}$)** |
| 1100 | 10 | 3.79E-07 | 1150 | 5 | 1.80E-07 |
| | 10 | 5.99E-07 | | 5 | 2.08E-07 |
| | 20 | 2.76E-06 | | 10 | 9.76E-07 |
| | 20 | 2.09E-06 | | 10 | 7.55E-07 |
| | 30 | 7.74E-06 | | 20 | 5.04E-06 |
| | 30 | 8.10E-06 | | 20 | 7.52E-06 |
| 1200 | 5 | 2.76E-07 | 1250 | 5 | 7.87E-07 |
| | 5 | 3.28E-07 | | 5 | 8.73E-07 |
| | 10 | 1.19E-06 | | 10 | 4.24E-06 |
| | 10 | 2.52E-06 | | 10 | 4.96E-06 |
| | 20 | 1.05E-05 | | 20 | 2.67E-05 |
| | 20 | 1.91E-05 | | | |
| | | | | | |
| $Nb_{45}Ta_{25}Ti_{15}Hf_{15}$, BCC RX, Tensile creep, Ref. [173]. | | | | | |
| **T (°C)** | **$\sigma$ (MPa)** | **$\dot{\varepsilon}_{ss}$ ($s^{-1}$)** | **T (°C)** | **$\sigma$ (MPa)** | **$\dot{\varepsilon}_{ss}$ ($s^{-1}$)** |
| 900 | 50 | 1.64E-08 | 900 | 250 | 1.09E-05 |
| | 100 | 5.30E-07 | | 300 | 3.06E-05 |

**Supplementary Table 4 Secondary or minimum creep rates for multi-phase BCC HEAs.**

| $Ta_{27.3}Mo_{27.3}Ti_{27.3}Cr_8Al_{10}$, BCC + B2, Compressive creep, Ref. [93]. | | | | | |
|---|---|---|---|---|---|
| **T (°C)** | **$\sigma$ (MPa)** | **$\dot{\varepsilon}_{ss}$ ($s^{-1}$)** | **T (°C)** | **$\sigma$ (MPa)** | **$\dot{\varepsilon}_{ss}$ ($s^{-1}$)** |
| 1030 | 75 | 1.31E-09 | 1030 | 125 | 3.57E-08 |
| | 75 | 2.51E-09 | | 150 | 4.21E-07 |
| | 100 | 3.36E-09 | | 150 | 5.64E-07 |
| | 125 | 9.97E-09 | | 150 | 9.62E-07 |
| | 125 | 1.28E-08 | | 175 | 7.41E-06 |
| | 125 | 1.91E-08 | | | |
| | | | | | |
| $Al_{23.5}Ti_{23.5}V_{23.5}Nb_{23.5}Zr_6$, B2 + $Zr_5Al_3$ + $Nb_2Al$, Compressive creep, Ref. [178]. | | | | | |
| **T (°C)** | **$\sigma$ (MPa)** | **$\dot{\varepsilon}_{ss}$ ($s^{-1}$)** | **T (°C)** | **$\sigma$ (MPa)** | **$\dot{\varepsilon}_{ss}$ ($s^{-1}$)** |
| 800 | 100 | 7.75E-07 | 800 | 315 | 2.51E-05 |
| | 150 | 2.32E-06 | | 325 | 3.35E-05 |
| | 200 | 2.75E-06 | | 340 | 3.77E-05 |
| | 300 | 2.49E-05 | | | |
| | | | | | |
| $Ti_{33.3}Ta_{11.1}Nb_{22.2}Zr_{22.2}Mo_{11.2}$, BCC + Zr-rich precipitates, Tensile creep, Ref. [179]. | | | | | |
| **T (°C)** | **$\sigma$ (MPa)** | **$\dot{\varepsilon}_{ss}$ ($s^{-1}$)** | **T (°C)** | **$\sigma$ (MPa)** | **$\dot{\varepsilon}_{ss}$ ($s^{-1}$)** |
| 650 | 30 | 9.947E-09 | 700 | 30 | 3.62E-08 |
| | 60 | 1.35E-07 | | 60 | 5.93E-07 |
| | 90 | 6.14E-07 | | 90 | 1.87E-06 |
| | 120 | 8.82E-07 | | 120 | 3.93E-06 |
| | 150 | 1.74E-06 | | 150 | 9.15E-06 |
| 750 | 30 | 1.15E-07 | | | |
| | 60 | 1.62E-06 | | | |
| | 90 | 6.99E-06 | | | |
| | 120 | 1.38E-05 | | | |
| | 150 | 2.74E-05 | | | |

## References

[1] B.A. Pint, J.R. DiStefano, I.G. Wright, Oxidation resistance: One barrier to moving beyond Ni-base superalloys, Materials Science and Engineering: A 415(1) (2006) 255-263.
[2] A. Sato, Y.L. Chiu, R.C. Reed, Oxidation of nickel-based single-crystal superalloys for industrial gas turbine applications, Acta Materialia 59(1) (2011) 225-240.
[3] B. Pieraggi, F. Dabosi, High-temperature oxidation of a single crystal Ni-base superalloy, Materials and Corrosion 38(10) (1987) 584-590.
[4] L. Garimella, P. Liaw, D. Klarstrom, Fatigue behavior in nickel-based superalloys: A literature review, Jom 49(7) (1997) 67-71.
[5] M. Gell, G. Leverant, C. Wells, The fatigue strength of nickel-base superalloys, ASTM STP 467 (1970) 113-153.
[6] R.V. Miner, J. Gayda, R.D. Maier, Fatigue and creep-fatigue deformation of several nickel-base superalloys at 650 °c, Metallurgical Transactions A 13(10) (1982) 1755-1765.
[7] J. Cormier, Thermal cycling creep resistance of Ni-based single crystal superalloys, Superalloys 2016 (2016) 385-394.
[8] E. Fleischmann, C. Konrad, J. Preußner, R. Völkl, E. Affeldt, U. Glatzel, Influence of Solid Solution Hardening on Creep Properties of Single-Crystal Nickel-Based Superalloys, Metallurgical and Materials Transactions A 46(3) (2015) 1125-1130.
[9] T.M. Pollock, A.S. Argon, Creep resistance of CMSX-3 nickel base superalloy single crystals, Acta Metallurgica et Materialia 40(1) (1992) 1-30.
[10] A.-C. Yeh, A. Sato, T. Kobayashi, H. Harada, On the creep and phase stability of advanced Ni-base single crystal superalloys, Materials Science and Engineering: A 490(1) (2008) 445-451.
[11] R.C. Reed, T. Tao, N. Warnken, Alloys-By-Design: Application to nickel-based single crystal superalloys, Acta Materialia 57(19) (2009) 5898-5913.
[12] V. Sass, U. Glatzel, M. Feller-Kniepmeier, Anisotropic creep properties of the nickel-base superalloy CMSX-4, Acta Materialia 44(5) (1996) 1967-1977.
[13] D.M. Bahr[1], W.S. Johnson, Temperature dependent fracture toughness of a single crystal nickel superalloy, Fatigue and Fracture Mechanics 34(4) (2005) 340.
[14] Z.X. Wen, Z.F. Yue, Fracture behaviour of the compact tension specimens of nickel-based single crystal superalloys at high temperatures, Materials Science and Engineering: A 456(1) (2007) 189-201.
[15] M. Heilmaier, M. Krüger, H. Saage, J. Rösler, D. Mukherji, U. Glatzel, R. Völkl, R. Hüttner, G. Eggeler, C. Somsen, T. Depka, H.J. Christ, B. Gorr, S. Burk, Metallic materials for structural applications beyond nickel-based superalloys, JOM 61(7) (2009) 61-67.
[16] D. Mukherji, J. Rösler, P. Strunz, R. Gilles, G. Schumacher, S. Piegert, Beyond Ni-based superalloys: Development of CoRe-based alloys for gas turbine applications at very high temperatures, International journal of materials research 102(9) (2011) 1125.
[17] J.W. Yeh, S.K. Chen, S.J. Lin, J.Y. Gan, T.S. Chin, T.T. Shun, C.H. Tsau, S.Y. Chang, Nanostructured High-Entropy Alloys with Multiple Principal Elements: Novel Alloy Design Concepts and Outcomes, Advanced Engineering Materials 6(5) (2004) 299-303.
[18] B. Cantor, Multicomponent and High Entropy Alloys, Entropy 16(9) (2014) 4749-4768.
[19] E.P. George, D. Raabe, R.O. Ritchie, High-entropy alloys, Nature Reviews Materials 4(8) (2019) 515-534.
[20] D.B. Miracle, O.N. Senkov, A critical review of high entropy alloys and related concepts, Acta Materialia 122 (2017) 448-511.

[21] M.-H. Tsai, J.-W. Yeh, High-Entropy Alloys: A Critical Review, Materials Research Letters 2(3) (2014) 107-123.
[22] S. Gorsse, J.-P. Couzinié, D.B. Miracle, From high-entropy alloys to complex concentrated alloys, Comptes Rendus Physique 19(8) (2018) 721-736.
[23] W. Li, D. Xie, D. Li, Y. Zhang, Y. Gao, P.K. Liaw, Mechanical behavior of high-entropy alloys, Progress in Materials Science 118 (2021) 100777.
[24] B. Cantor, I.T.H. Chang, P. Knight, A.J.B. Vincent, Microstructural development in equiatomic multicomponent alloys, Materials Science and Engineering: A 375-377 (2004) 213-218.
[25] Y.H. Jo, S. Jung, W.M. Choi, S.S. Sohn, H.S. Kim, B.J. Lee, N.J. Kim, S. Lee, Cryogenic strength improvement by utilizing room-temperature deformation twinning in a partially recrystallized VCrMnFeCoNi high-entropy alloy, Nature Communications 8(1) (2017) 15719.
[26] A. Shabani, M.R. Toroghinejad, A. Shafyei, R.E. Logé, Microstructure and Mechanical Properties of a Multiphase FeCrCuMnNi High-Entropy Alloy, Journal of Materials Engineering and Performance 28(4) (2019) 2388-2398.
[27] B. Gludovatz, E.P. George, R.O. Ritchie, Processing, Microstructure and Mechanical Properties of the CrMnFeCoNi High-Entropy Alloy, JOM 67(10) (2015) 2262-2270.
[28] G. Laplanche, A. Kostka, C. Reinhart, J. Hunfeld, G. Eggeler, E.P. George, Reasons for the superior mechanical properties of medium-entropy CrCoNi compared to high-entropy CrMnFeCoNi, Acta Materialia 128 (2017) 292-303.
[29] B. Gludovatz, A. Hohenwarter, D. Catoor, E.H. Chang, E.P. George, R.O. Ritchie, A fracture-resistant high-entropy alloy for cryogenic applications, Science 345(6201) (2014) 1153-1158.
[30] B. Gludovatz, A. Hohenwarter, K.V.S. Thurston, H. Bei, Z. Wu, E.P. George, R.O. Ritchie, Exceptional damage-tolerance of a medium-entropy alloy CrCoNi at cryogenic temperatures, Nature Communications 7(1) (2016) 10602.
[31] D. Liu, Q. Yu, S. Kabra, M. Jiang, P. Forna-Kreutzer, R. Zhang, M. Payne, F. Walsh, B. Gludovatz, M. Asta, A.M. Minor, E.P. George, R.O. Ritchie, Exceptional fracture toughness of CrCoNi-based medium- and high-entropy alloys at 20 kelvin, Science 378(6623) (2022) 978-983.
[32] F. Otto, A. Dlouhý, C. Somsen, H. Bei, G. Eggeler, E.P. George, The influences of temperature and microstructure on the tensile properties of a CoCrFeMnNi high-entropy alloy, Acta Materialia 61(15) (2013) 5743-5755.
[33] M. Tokarewicz, M. Grądzka-Dahlke, Review of Recent Research on AlCoCrFeNi High-Entropy Alloy, Metals 11(8) (2021) 1302.
[34] X. Sun, H. Zhang, S. Lu, X. Ding, Y. Wang, L. Vitos, Phase selection rule for Al-doped CrMnFeCoNi high-entropy alloys from first-principles, Acta Materialia 140 (2017) 366-374.
[35] X. Li, Z. Li, Z. Wu, S. Zhao, W. Zhang, H. Bei, Y. Gao, Strengthening in Al-, Mo- or Ti-doped CoCrFeNi high entropy alloys: A parallel comparison, Journal of Materials Science & Technology 94 (2021) 264-274.
[36] G. Talluri, R.S. Maurya, B.S. Murty, Composition design of eutectic high-entropy alloys: a review, Journal of Materials Science 60(3) (2025) 1156-1186.
[37] Y.-T. Chen, Y.-J. Chang, H. Murakami, S. Gorsse, A.-C. Yeh, Designing high entropy superalloys for elevated temperature application, Scripta Materialia 187 (2020) 177-182.
[38] Y.-T. Chen, Y.-J. Chang, H. Murakami, T. Sasaki, K. Hono, C.-W. Li, K. Kakehi, J.-W. Yeh, A.-C. Yeh, Hierarchical microstructure strengthening in a single crystal high entropy superalloy, Scientific Reports 10(1) (2020) 12163.

[39] M. Joele, W.R. Matizamhuka, A Review on the High Temperature Strengthening Mechanisms of High Entropy Superalloys (HESA), Materials 14(19) (2021) 5835.
[40] T.-K. Tsao, A.-C. Yeh, C.-M. Kuo, K. Kakehi, H. Murakami, J.-W. Yeh, S.-R. Jian, The High Temperature Tensile and Creep Behaviors of High Entropy Superalloy, Scientific Reports 7(1) (2017) 12658.
[41] A.-C. Yeh, T.-K. Tsao, Y.-J. Chang, K. Chang, J.-W. Yeh, M. Chiou, S.-R. Jian, C.-M. Kuo, W. Wang, H. Murakami, Developing new type of high temperature alloys–high entropy superalloys, Int. J. Metall. Mater. Eng 1(107) (2015) 1-4.
[42] M.C. Gao, C.S. Carney, Ö.N. Doğan, P.D. Jablonksi, J.A. Hawk, D.E. Alman, Design of Refractory High-Entropy Alloys, JOM 67(11) (2015) 2653-2669.
[43] O.N. Senkov, S. Gorsse, D.B. Miracle, High temperature strength of refractory complex concentrated alloys, Acta Materialia 175 (2019) 394-405.
[44] O.N. Senkov, D.B. Miracle, K.J. Chaput, J.-P. Couzinie, Development and exploration of refractory high entropy alloys—A review, Journal of Materials Research 33(19) (2018) 3092-3128.
[45] O.N. Senkov, G.B. Wilks, D.B. Miracle, C.P. Chuang, P.K. Liaw, Refractory high-entropy alloys, Intermetallics 18(9) (2010) 1758-1765.
[46] O.N. Senkov, J.M. Scott, S.V. Senkova, F. Meisenkothen, D.B. Miracle, C.F. Woodward, Microstructure and elevated temperature properties of a refractory TaNbHfZrTi alloy, Journal of Materials Science 47(9) (2012) 4062-4074.
[47] O.N. Senkov, G.B. Wilks, J.M. Scott, D.B. Miracle, Mechanical properties of Nb25Mo25Ta25W25 and V20Nb20Mo20Ta20W20 refractory high entropy alloys, Intermetallics 19(5) (2011) 698-706.
[48] P. Kumar, X. Gou, D.H. Cook, M.I. Payne, N.J. Morrison, W. Wang, M. Zhang, M. Asta, A.M. Minor, R. Cao, Y. Li, R.O. Ritchie, Degradation of the mechanical properties of NbMoTaW refractory high-entropy alloy in tension, Acta Materialia 279 (2024) 120297.
[49] O.N. Senkov, D.B. Miracle, S.I. Rao, Correlations to improve room temperature ductility of refractory complex concentrated alloys, Materials Science and Engineering: A 820 (2021) 141512.
[50] L.H. Mills, M.G. Emigh, C.H. Frey, N.R. Philips, S.P. Murray, J. Shin, D.S. Gianola, T.M. Pollock, Temperature-dependent tensile behavior of the HfNbTaTiZr multi-principal element alloy, Acta Materialia 245 (2023) 118618.
[51] D.B. Miracle, M.-H. Tsai, O.N. Senkov, V. Soni, R. Banerjee, Refractory high entropy superalloys (RSAs), Scripta Materialia 187 (2020) 445-452.
[52] O.N. Senkov, D. Isheim, D.N. Seidman, A.L. Pilchak, Development of a Refractory High Entropy Superalloy, Entropy 18(3) (2016) 102.
[53] A.J. Knowles, C.H. Zenk, BCC-superalloys: Perspectives and challenges, Scripta Materialia 267 (2025) 116761.
[54] B. Gwalani, R.M. Pohan, O.A. Waseem, T. Alam, S.H. Hong, H.J. Ryu, R. Banerjee, Strengthening of Al0.3CoCrFeMnNi-based ODS high entropy alloys with incremental changes in the concentration of Y2O3, Scripta Materialia 162 (2019) 477-481.
[55] H. Hadraba, Z. Chlup, A. Dlouhy, F. Dobes, P. Roupcova, M. Vilemova, J. Matejicek, Oxide dispersion strengthened CoCrFeNiMn high-entropy alloy, Materials Science and Engineering: A 689 (2017) 252-256.
[56] M. Li, Y. Guo, H. Wang, J. Shan, Y. Chang, Microstructures and mechanical properties of oxide dispersion strengthened CoCrFeNi high-entropy alloy produced by mechanical alloying and spark plasma sintering, Intermetallics 123 (2020) 106819.

[57] T.M. Smith, C.A. Kantzos, N.A. Zarkevich, B.J. Harder, M. Heczko, P.R. Gradl, A.C. Thompson, M.J. Mills, T.P. Gabb, J.W. Lawson, A 3D printable alloy designed for extreme environments, Nature 617(7961) (2023) 513-518.
[58] T.M. Smith, A.C. Thompson, T.P. Gabb, C.L. Bowman, C.A. Kantzos, Efficient production of a high-performance dispersion strengthened, multi-principal element alloy, Scientific Reports 10(1) (2020) 9663.
[59] T. Liao, Y. Cao, Q. Huang, A. Fu, J. Li, Q. Fang, J. Qiu, B. Liu, Y. Liu, Multiscale oxide dispersion strengthened refractory high entropy alloys with superior mechanical properties, International Journal of Refractory Metals and Hard Materials 123 (2024) 106796.
[60] T. Liao, Y.-K. Cao, W.-M. Guo, Q.-H. Fang, J. Li, B. Liu, Microstructure and mechanical property of NbTaTiV refractory high-entropy alloy with different Y2O3 contents, Rare Metals 41(10) (2022) 3504-3514.
[61] M. Feuerbacher, M. Heidelmann, C. Thomas, Hexagonal High-entropy Alloys, Materials Research Letters 3(1) (2015) 1-6.
[62] M.C. Gao, B. Zhang, S.M. Guo, J.W. Qiao, J.A. Hawk, High-Entropy Alloys in Hexagonal Close-Packed Structure, Metallurgical and Materials Transactions A 47(7) (2016) 3322-3332.
[63] R. Feng, C. Zhang, M.C. Gao, Z. Pei, F. Zhang, Y. Chen, D. Ma, K. An, J.D. Poplawsky, L. Ouyang, Y. Ren, J.A. Hawk, M. Widom, P.K. Liaw, High-throughput design of high-performance lightweight high-entropy alloys, Nature Communications 12(1) (2021) 4329.
[64] A. Kumar, M. Gupta, An Insight into Evolution of Light Weight High Entropy Alloys: A Review, Metals, 2016.
[65] O. Maulik, D. Kumar, S. Kumar, S.K. Dewangan, V. Kumar, Structure and properties of lightweight high entropy alloys: a brief review, Materials Research Express 5(5) (2018) 052001.
[66] W. Blum, P. Eisenlohr, F. Breutinger, Understanding creep—a review, Metallurgical and Materials Transactions A 33(2) (2002) 291-303.
[67] O.D. Sherby, R.L. Orr, J.E. Dorn, Creep correlations of metals at elevated temperatures, JOM 6(1) (1954) 71-80.
[68] S. Takeuchi, A.S. Argon, Steady-state creep of single-phase crystalline matter at high temperature, Journal of Materials Science 11(8) (1976) 1542-1566.
[69] J. Weertman, Theory of Steady-State Creep Based on Dislocation Climb, Journal of Applied Physics 26(10) (1955) 1213-1217.
[70] J. Weertman, Steady-State Creep of Crystals, Journal of Applied Physics 28(10) (1957) 1185-1189.
[71] M.E. Kassner, M.T. Pérez-Prado, Five-power-law creep in single phase metals and alloys, Progress in Materials Science 45(1) (2000) 1-102.
[72] O.D. Sherby, P.M. Burke, Mechanical behavior of crystalline solids at elevated temperature, Progress in Materials Science 13 (1968) 323-390.
[73] R. Feng, M.C. Gao, C. Zhang, W. Guo, J.D. Poplawsky, F. Zhang, J.A. Hawk, J.C. Neuefeind, Y. Ren, P.K. Liaw, Phase stability and transformation in a light-weight high-entropy alloy, Acta Materialia 146 (2018) 280-293.
[74] H.-W. Luan, Y. Shao, J.-F. Li, W.-L. Mao, Z.-D. Han, C. Shao, K.-F. Yao, Phase stabilities of high entropy alloys, Scripta Materialia 179 (2020) 40-44.
[75] F. Otto, Y. Yang, H. Bei, E.P. George, Relative effects of enthalpy and entropy on the phase stability of equiatomic high-entropy alloys, Acta Materialia 61(7) (2013) 2628-2638.

[76] J.E. Saal, I.S. Berglund, J.T. Sebastian, P.K. Liaw, G.B. Olson, Equilibrium high entropy alloy phase stability from experiments and thermodynamic modeling, Scripta Materialia 146 (2018) 5-8.
[77] Y.F. Ye, C.T. Liu, Y. Yang, A geometric model for intrinsic residual strain and phase stability in high entropy alloys, Acta Materialia 94 (2015) 152-161.
[78] D.L. Beke, G. Erdélyi, On the diffusion in high-entropy alloys, Materials Letters 164 (2016) 111-113.
[79] J. Dąbrowa, M. Danielewski, State-of-the-Art Diffusion Studies in the High Entropy Alloys, Metals, 2020.
[80] V. Verma, C.H. Belcher, D. Apelian, E.J. Lavernia, Diffusion in High Entropy Alloy Systems – A Review, Progress in Materials Science 142 (2024) 101245.
[81] B. Xu, J. Zhang, S. Ma, Y. Xiong, S. Huang, J.J. Kai, S. Zhao, Revealing the crucial role of rough energy landscape on self-diffusion in high-entropy alloys based on machine learning and kinetic Monte Carlo, Acta Materialia 234 (2022) 118051.
[82] F. Wang, G.H. Balbus, S. Xu, Y. Su, J. Shin, P.F. Rottmann, K.E. Knipling, J.-C. Stinville, L.H. Mills, O.N. Senkov, I.J. Beyerlein, T.M. Pollock, D.S. Gianola, Multiplicity of dislocation pathways in a refractory multiprincipal element alloy, Science 370(6512) (2020) 95-101.
[83] S. Zhao, Y. Zhang, W.J. Weber, Engineering defect energy landscape of CoCrFeNi high-entropy alloys by the introduction of additional dopants, Journal of Nuclear Materials 561 (2022) 153573.
[84] S. Kang, A. Tamm, Density functional study of atomic arrangements in CrMnFeCoNi high-entropy alloy and their impact on vacancy formation energy and segregation, Computational Materials Science 230 (2023) 112456.
[85] S. Yin, J. Ding, M. Asta, R.O. Ritchie, Ab initio modeling of the energy landscape for screw dislocations in body-centered cubic high-entropy alloys, npj Computational Materials 6(1) (2020) 110.
[86] X. Wang, F. Maresca, P. Cao, The hierarchical energy landscape of screw dislocation motion in refractory high-entropy alloys, Acta Materialia 234 (2022) 118022.
[87] Y.B. Kang, S.H. Shim, K.H. Lee, S.I. Hong, Dislocation creep behavior of CoCrFeMnNi high entropy alloy at intermediate temperatures, Materials Research Letters 6(12) (2018) 689-695.
[88] C. Cao, J. Fu, T. Tong, Y. Hao, P. Gu, H. Hao, L. Peng, Intermediate-Temperature Creep Deformation and Microstructural Evolution of an Equiatomic FCC-Structured CoCrFeNiMn High-Entropy Alloy, Entropy, 2018.
[89] K.Y. Tsai, M.H. Tsai, J.W. Yeh, Sluggish diffusion in Co–Cr–Fe–Mn–Ni high-entropy alloys, Acta Materialia 61(13) (2013) 4887-4897.
[90] C. Gadelmeier, Y. Yang, U. Glatzel, E.P. George, Creep strength of refractory high-entropy alloy TiZrHfNbTa and comparison with Ni-base superalloy CMSX-4, Cell Reports Physical Science 3(8) (2022).
[91] C.-J. Liu, C. Gadelmeier, S.-L. Lu, J.-W. Yeh, H.-W. Yen, S. Gorsse, U. Glatzel, A.-C. Yeh, Tensile creep behavior of HfNbTaTiZr refractory high entropy alloy at elevated temperatures, Acta Materialia 237 (2022) 118188.
[92] M. Zhang, E.P. George, J.C. Gibeling, Tensile creep properties of a CrMnFeCoNi high-entropy alloy, Scripta Materialia 194 (2021) 113633.
[93] L. Yang, S. Sen, D. Schliephake, R.J. Vikram, S. Laube, A. Pramanik, A. Chauhan, S. Neumeier, M. Heilmaier, A. Kauffmann, Creep behavior of a precipitation-strengthened A2-B2 refractory high entropy alloy, Acta Materialia 288 (2025) 120827.

[94] S. Chen, W. Li, X. Xie, J. Brechtl, B. Chen, P. Li, G. Zhao, F. Yang, J. Qiao, P.K. Liaw, Nanoscale serration and creep characteristics of Al0.5CoCrCuFeNi high-entropy alloys, Journal of Alloys and Compounds 752 (2018) 464-475.
[95] Q. Fan, K. Gan, D. Yan, Z. Li, Nanoindentation creep behavior of diverse microstructures in a pre-strained interstitial high-entropy alloy by high-throughput mapping, Materials Science and Engineering: A 856 (2022) 143988.
[96] K. Han, Y. Zhu, C. Wang, P. Gao, S. Liang, L. Zhao, M. Huang, Z. Li, Nanoindentation creep behavior of high entropy superalloy with $\gamma/\gamma'$ microstructure under various loads and temperatures, Journal of Alloys and Compounds 1031 (2025) 180985.
[97] Z.M. Jiao, Z.H. Wang, R.F. Wu, J.W. Qiao, Strain rate sensitivity of nanoindentation creep in an AlCoCrFeNi high-entropy alloy, Applied Physics A 122(9) (2016) 794.
[98] D.-H. Lee, I.-C. Choi, M. Kawasaki, T.G. Langdon, J.-i. Jang, A Review of Recent Research on Nanoindentation of High-Entropy Alloys Processed by High-Pressure Torsion, MATERIALS TRANSACTIONS 64(7) (2023) 1551-1565.
[99] D.-H. Lee, M.-Y. Seok, Y. Zhao, I.-C. Choi, J. He, Z. Lu, J.-Y. Suh, U. Ramamurty, M. Kawasaki, T.G. Langdon, J.-i. Jang, Spherical nanoindentation creep behavior of nanocrystalline and coarse-grained CoCrFeMnNi high-entropy alloys, Acta Materialia 109 (2016) 314-322.
[100] P.H. Lin, H.S. Chou, J.C. Huang, W.S. Chuang, J.S.C. Jang, T.G. Nieh, Elevated-temperature creep of high-entropy alloys via nanoindentation, MRS Bulletin 44(11) (2019) 860-866.
[101] S.G. Ma, Creep resistance and strain-rate sensitivity of a CoCrFeNiAl0.3 high-entropy alloy by nanoindentation, Materials Research Express 6(12) (2019) 126508.
[102] Y. Ma, Y.H. Feng, T.T. Debela, G.J. Peng, T.H. Zhang, Nanoindentation study on the creep characteristics of high-entropy alloy films: fcc versus bcc structures, International Journal of Refractory Metals and Hard Materials 54 (2016) 395-400.
[103] Y. Ma, G.J. Peng, D.H. Wen, T.H. Zhang, Nanoindentation creep behavior in a CoCrFeCuNi high-entropy alloy film with two different structure states, Materials Science and Engineering: A 621 (2015) 111-117.
[104] Y. Sun, Y. Huo, W. Yu, Z. Yan, Z. Wang, Z. Li, Z. Wang, H. Chen, A. Jiang, X. Wang, Microstructure and nanoindentation creep behavior of NiAlCrFeMo high-entropy alloy, Journal of Alloys and Compounds 1020 (2025) 179330.
[105] S. Tan, X. Liu, Z. Wang, Nanoindentation mechanical properties of CoCrFeNi high entropy alloy films, Materials Technology 37(9) (2022) 1097-1108.
[106] Z. Wang, S. Guo, Q. Wang, Z. Liu, J. Wang, Y. Yang, C.T. Liu, Nanoindentation characterized initial creep behavior of a high-entropy-based alloy CoFeNi, Intermetallics 53 (2014) 183-186.
[107] Z. Xu, H. Zhang, W. Li, A. Mao, L. Wang, G. Song, Y. He, Microstructure and nanoindentation creep behavior of CoCrFeMnNi high-entropy alloy fabricated by selective laser melting, Additive Manufacturing 28 (2019) 766-771.
[108] L. Zhang, P. Yu, H. Cheng, H. Zhang, H. Diao, Y. Shi, B. Chen, P. Chen, R. Feng, J. Bai, Q. Jing, M. Ma, P.K. Liaw, G. Li, R. Liu, Nanoindentation Creep Behavior of an Al0.3CoCrFeNi High-Entropy Alloy, Metallurgical and Materials Transactions A 47(12) (2016) 5871-5875.
[109] P.F. Zhou, D.H. Xiao, G. Li, M. Song, Nanoindentation Creep Behavior of CoCrFeNiMn High-Entropy Alloy under Different High-Pressure Torsion Deformations, Journal of Materials Engineering and Performance 28(5) (2019) 2620-2629.
[110] B.R. Anne, S. Shaik, M. Tanaka, A. Basu, A crucial review on recent updates of oxidation behavior in high entropy alloys, SN Applied Sciences 3(3) (2021) 366.

[111] B. Gorr, F. Müller, M. Azim, H.-J. Christ, T. Müller, H. Chen, A. Kauffmann, M. Heilmaier, High-Temperature Oxidation Behavior of Refractory High-Entropy Alloys: Effect of Alloy Composition, Oxidation of Metals 88(3) (2017) 339-349.
[112] B. Gorr, S. Schellert, F. Müller, H.-J. Christ, A. Kauffmann, M. Heilmaier, Current Status of Research on the Oxidation Behavior of Refractory High Entropy Alloys, Advanced Engineering Materials 23(5) (2021) 2001047.
[113] P. Kumar, T.-N. Lam, P.K. Tripathi, S.S. Singh, P.K. Liaw, E.W. Huang, Recent progress in oxidation behavior of high-entropy alloys: A review, APL Materials 10(12) (2022) 120701.
[114] F. Müller, B. Gorr, H.-J. Christ, J. Müller, B. Butz, H. Chen, A. Kauffmann, M. Heilmaier, On the oxidation mechanism of refractory high entropy alloys, Corrosion Science 159 (2019) 108161.
[115] K.A. Rozman, M. Detrois, T. Liu, M.C. Gao, P.D. Jablonski, J.A. Hawk, Long-Term Creep Behavior of a CoCrFeNiMn High-Entropy Alloy, Journal of Materials Engineering and Performance 29(9) (2020) 5822-5839.
[116] C. Song, G. Li, G. Li, G. Zhang, B. Cai, Tensile creep behavior and mechanism of CoCrFeMnNi high entropy alloy, Micron 150 (2021) 103144.
[117] M.-G. Jo, J.-Y. Suh, M.-Y. Kim, H.-J. Kim, W.-S. Jung, D.-I. Kim, H.N. Han, High temperature tensile and creep properties of CrMnFeCoNi and CrFeCoNi high-entropy alloys, Materials Science and Engineering: A 838 (2022) 142748.
[118] D. Xie, R. Feng, P.K. Liaw, H. Bei, Y. Gao, Long-term tensile creep behavior of a family of FCC-structured multi-component equiatomic solid solution alloys, Scripta Materialia 212 (2022) 114556.
[119] T. Záležák, C. Gadelmeier, Š. Gamanov, U. Glatzel, H. Inui, E. George, A. Dlouhý, Creep strength variations related to grain boundaries in the equiatomic CoCrFeMnNi high-entropy alloy, Scripta Materialia 249 (2024) 116165.
[120] C. Gadelmeier, S. Haas, T. Lienig, A. Manzoni, M. Feuerbacher, U. Glatzel, Temperature Dependent Solid Solution Strengthening in the High Entropy Alloy CrMnFeCoNi in Single Crystalline State, Metals, 2020.
[121] K.A. Rozman, M. Detrois, M.C. Gao, P.D. Jablonski, J.A. Hawk, Long-Term Creep Behavior of a CoCrFeNi Medium-Entropy Alloy, Journal of Materials Engineering and Performance 31(11) (2022) 9220-9235.
[122] K. Lu, J. Aktaa, Short-term tensile creep behavior of CoCrNi-based multi-principal element alloys, Intermetallics 175 (2024) 108500.
[123] D. Xie, R. Feng, P.K. Liaw, H. Bei, Y. Gao, Tensile creep behavior of an equiatomic CoCrNi medium entropy alloy, Intermetallics 121 (2020) 106775.
[124] G. Sahragard-Monfared, M. Zhang, T.M. Smith, A.M. Minor, E.P. George, J.C. Gibeling, The influence of processing methods on creep of wrought and additively manufactured CrCoNi multi-principal element alloys, Acta Materialia 261 (2023) 119403.
[125] W. Jiang, Y. Cao, S. Yuan, Y. Zhang, Y. Zhao, Creep properties and deformation mechanisms of a Ni2Co1Fe1V0.5Mo0.2 medium-entropy alloy, Acta Materialia 245 (2023) 118590.
[126] M. Vaidya, K.G. Pradeep, B.S. Murty, G. Wilde, S.V. Divinski, Bulk tracer diffusion in CoCrFeNi and CoCrFeMnNi high entropy alloys, Acta Materialia 146 (2018) 211-224.
[127] K.L. Murty, G. Dentel, J. Britt, Effect of temperature on transitions in creep mechanisms in class-A alloys, Materials Science and Engineering: A 410-411 (2005) 28-31.
[128] R. Raj, M.F. Ashby, On grain boundary sliding and diffusional creep, Metallurgical Transactions 2(4) (1971) 1113-1127.

[129] T.G. Langdon, Grain boundary sliding as a deformation mechanism during creep, The Philosophical Magazine: A Journal of Theoretical Experimental and Applied Physics 22(178) (1970) 689-700.
[130] F. Otto, A. Dlouhý, K.G. Pradeep, M. Kuběnová, D. Raabe, G. Eggeler, E.P. George, Decomposition of the single-phase high-entropy alloy CrMnFeCoNi after prolonged anneals at intermediate temperatures, Acta Materialia 112 (2016) 40-52.
[131] Z. Wu, H. Bei, G.M. Pharr, E.P. George, Temperature dependence of the mechanical properties of equiatomic solid solution alloys with face-centered cubic crystal structures, Acta Materialia 81 (2014) 428-441.
[132] B. Yin, F. Maresca, W.A. Curtin, Vanadium is an optimal element for strengthening in both fcc and bcc high-entropy alloys, Acta Materialia 188 (2020) 486-491.
[133] A.K. Mukherjee, J.E. Bird, J.E. Dorn, Experimental correlations for high-temperature creep, (1968).
[134] G. Laplanche, P. Gadaud, C. Bärsch, K. Demtröder, C. Reinhart, J. Schreuer, E.P. George, Elastic moduli and thermal expansion coefficients of medium-entropy subsystems of the CrMnFeCoNi high-entropy alloy, Journal of Alloys and Compounds 746 (2018) 244-255.
[135] G. Laplanche, P. Gadaud, O. Horst, F. Otto, G. Eggeler, E.P. George, Temperature dependencies of the elastic moduli and thermal expansion coefficient of an equiatomic, single-phase CoCrFeMnNi high-entropy alloy, Journal of Alloys and Compounds 623 (2015) 348-353.
[136] C.M. Cao, J. Xu, W. Tong, Y.X. Hao, P. Gu, L.M. Peng, Creep behaviour and microstructural evolution of AlxCrMnFeCoNi high-entropy alloys, Materials Science and Technology 35(10) (2019) 1283-1290.
[137] Z.Y. You, Z.Y. Tang, F.B. Chu, L. Zhao, H.W. Zhang, D.D. Cao, L. Jiang, H. Ding, Elevated-temperature creep properties and deformation mechanisms of a non-equiatomic FeMnCoCrAl high-entropy alloy, Journal of Materials Research and Technology 30 (2024) 3822-3830.
[138] S. Chen, J. Qiao, H. Diao, T. Yang, J. Poplawsky, W. Li, F. Meng, Y. Tong, L. Jiang, P.K. Liaw, Y. Gao, Extraordinary creep resistance in a non-equiatomic high-entropy alloy from the optimum solid-solution strengthening and stress-assisted precipitation process, Acta Materialia 244 (2023) 118600.
[139] Y. Li, W. Chen, C. Lu, H. Li, W. Zheng, Y. Ma, Y. Jin, W. Jin, Z. Gao, J. Yang, Y. He, Microstructural evolution mediated creep deformation mechanism for the AlCoCrFeNi2.1 eutectic high-entropy alloy under different testing conditions, Materials Science and Engineering: A 857 (2022) 144100.
[140] S. Haas, A.M. Manzoni, M. Holzinger, U. Glatzel, Influence of high melting elements on microstructure, tensile strength and creep resistance of the compositionally complex alloy Al10Co25Cr8Fe15Ni36Ti6, Materials Chemistry and Physics 274 (2021) 125163.
[141] K.A. Rozman, M. Detrois, P.D. Jablonski, J.A. Hawk, M.C. Gao, Improved creep performance of CoCrFeNi high entropy alloys by Mo addition, Materials Chemistry and Physics 344 (2025) 130999.
[142] F. Dobeš, H. Hadraba, Z. Chlup, A. Dlouhý, M. Vilémová, J. Matějíček, Compressive creep behavior of an oxide-dispersion-strengthened CoCrFeMnNi high-entropy alloy, Materials Science and Engineering: A 732 (2018) 99-104.
[143] G. Sahragard-Monfared, M. Zhang, T.M. Smith, A.M. Minor, J.C. Gibeling, Superior tensile creep behavior of a novel oxide dispersion strengthened CrCoNi multi-principal element alloy, Acta Materialia 255 (2023) 119032.

[144] Y.K. Kim, K. Ram Lim, K.A. Lee, Superior resistance to high–temperature creep in an additively manufactured precipitation–hardened CrMnFeCoNi high–entropy alloy nanocomposite, Materials & Design 227 (2023) 111761.
[145] A. Sengupta, S.K. Putatunda, L. Bartosiewicz, J. Hangas, P.J. Nailos, M. Peputapeck, F.E. Alberts, Tensile behavior of a new single-crystal nickel-based superalloy (CMSX-4) at room and elevated temperatures, Journal of Materials Engineering and Performance 3(1) (1994) 73-81.
[146] K. Harris, J.B. Wahl, Improved single crystal superalloys, CMSX-4 (SLS)[La+ Y] and CMSX-486, Superalloys 2004 (2004) 45-52.
[147] G.L. Erickson, The development and application of CMSX-10, Superalloys 1996 (1996) 35-44.
[148] M. Huang, L. Zhuo, Z. Liu, X. Lu, Z. Shi, J. Li, J. Zhu, Misorientation related microstructure at the grain boundary in a nickel-based single crystal superalloy, Materials Science and Engineering: A 640 (2015) 394-401.
[149] E.J. Lavernia, J.D. Ayers, T.S. Srivatsan, Rapid solidification processing with specific application to aluminium alloys, International Materials Reviews 37(1) (1992) 1-44.
[150] M.S. Nagorka, C.G. Levi, G.E. Lucas, S.D. Ridder, The potential of rapid solidification in oxide-dispersion-strengthened copper alloy development, Materials Science and Engineering: A 142(2) (1991) 277-289.
[151] S. Jha, R. Ray, Dispersion strengthened NiAl alloys produced by rapid solidification processing, Journal of materials science letters 7(3) (1988) 285-288.
[152] M.B. Wilms, S.-K. Rittinghaus, M. Goßling, B. Gökce, Additive manufacturing of oxide-dispersion strengthened alloys: Materials, synthesis and manufacturing, Progress in Materials Science 133 (2023) 101049.
[153] B. Reppich, M. Heilmaier, K. Liebig, G. Schumann, K.-D. Stein, T. Woller, Microstructural modelling of the creep behaviour of particle-strengthened superalloys, Steel Research 61(6) (1990) 251-257.
[154] M. McLean, On the threshold stress for dislocation creep in particle strengthened alloys, Acta Metallurgica 33(4) (1985) 545-556.
[155] J. Rösler, E. Arzt, A new model-based creep equation for dispersion strengthened materials, Acta Metallurgica et Materialia 38(4) (1990) 671-683.
[156] R. Zhou, Y. Liu, C. Zhou, S. Li, W. Wu, M. Song, B. Liu, X. Liang, P.K. Liaw, Microstructures and mechanical properties of C-containing FeCoCrNi high-entropy alloy fabricated by selective laser melting, Intermetallics 94 (2018) 165-171.
[157] T. Huang, L. Jiang, C. Zhang, H. Jiang, Y. Lu, T. Li, Effect of carbon addition on the microstructure and mechanical properties of CoCrFeNi high entropy alloy, Science China Technological Sciences 61(1) (2018) 117-123.
[158] M. Zhang, E.P. George, J.C. Gibeling, Elevated-temperature Deformation Mechanisms in a CrMnFeCoNi High-Entropy Alloy, Acta Materialia 218 (2021) 117181.
[159] U.F. Kocks, A.S. Argon, M.F. Ashby, Thermodynamics and kinetics of slip, 1st ed. ed., Pergamon Press, Oxford ;, 1975.
[160] G. Taylor, Thermally-activated deformation of BCC metals and alloys, Progress in Materials Science 36 (1992) 29-61.
[161] G. Laplanche, J. Bonneville, C. Varvenne, W.A. Curtin, E.P. George, Thermal activation parameters of plastic flow reveal deformation mechanisms in the CrMnFeCoNi high-entropy alloy, Acta Materialia 143 (2018) 257-264.

[162] Z. Wu, Y. Gao, H. Bei, Thermal activation mechanisms and Labusch-type strengthening analysis for a family of high-entropy and equiatomic solid-solution alloys, Acta Materialia 120 (2016) 108-119.
[163] S.I. Hong, J. Moon, S.K. Hong, H.S. Kim, Thermally activated deformation and the rate controlling mechanism in CoCrFeMnNi high entropy alloy, Materials Science and Engineering: A 682 (2017) 569-576.
[164] C. Varvenne, A. Luque, W.A. Curtin, Theory of strengthening in fcc high entropy alloys, Acta Materialia 118 (2016) 164-176.
[165] C. Varvenne, G.P.M. Leyson, M. Ghazisaeidi, W.A. Curtin, Solute strengthening in random alloys, Acta Materialia 124 (2017) 660-683.
[166] B. Cao, T. Yang, W.-h. Liu, C.T. Liu, Precipitation-hardened high-entropy alloys for high-temperature applications: A critical review, MRS Bulletin 44(11) (2019) 854-859.
[167] H. Mughrabi, U. Tetzlaff, Microstructure and High-Temperature Strength of Monocrystalline Nickel-Base Superalloys, Advanced Engineering Materials 2(6) (2000) 319-326.
[168] A. Suzuki, H. Inui, T.M. Pollock, L12-Strengthened Cobalt-Base Superalloys, Annual Review of Materials Research 45(Volume 45, 2015) (2015) 345-368.
[169] T.M. Smith, R.R. Unocic, H. Deutchman, M.J. Mills, Creep deformation mechanism mapping in nickel base disk superalloys, Materials at High Temperatures 33(4-5) (2016) 372-383.
[170] Y.M. Eggeler, K.V. Vamsi, T.M. Pollock, Precipitate Shearing, Fault Energies, and Solute Segregation to Planar Faults in Ni-, CoNi-, and Co-Base Superalloys, Annual Review of Materials Research 51(Volume 51, 2021) (2021) 209-240.
[171] R.F. Singer, E. Arzt, Structure, processing and properties of ODS superalloys, (1986).
[172] E. Arzt, Creep of dispersion strengthened materials: a critical assessment, (1991).
[173] G. Sahragard-Monfared, C.H. Belcher, S. Bajpai, M. Wirth, A. Devaraj, D. Apelian, E.J. Lavernia, R.O. Ritchie, A.M. Minor, J.C. Gibeling, C. Zhang, M. Zhang, Tensile creep behavior of the Nb45Ta25Ti15Hf15 refractory high entropy alloy, Acta Materialia 272 (2024) 119940.
[174] X. Shen, S. Xin, S. Ding, Y. He, W. Dong, B. Sun, X. Cai, T. Shen, Intermediate-temperature creep behaviors of an equiatomic VNbMoTaW refractory high-entropy alloy, Journal of Materials Research and Technology 24 (2023) 4796-4807.
[175] Z. Wang, H. Wu, Y. Wu, H. Huang, X. Zhu, Y. Zhang, H. Zhu, X. Yuan, Q. Chen, S. Wang, X. Liu, H. Wang, S. Jiang, M.J. Kim, Z. Lu, Solving oxygen embrittlement of refractory high-entropy alloy via grain boundary engineering, Materials Today 54 (2022) 83-89.
[176] P.P.P.O. Borges, R.O. Ritchie, M. Asta, Electronic descriptors for dislocation deformation behavior and intrinsic ductility in bcc high-entropy alloys, Science Advances 10(38) eadp7670.
[177] F. Hinrichs, G. Winkens, L.K. Kramer, G. Falcão, E.M. Hahn, D. Schliephake, M.K. Eusterholz, S. Sen, M.C. Galetz, H. Inui, A. Kauffmann, M. Heilmaier, A ductile chromium–molybdenum alloy resistant to high-temperature oxidation, Nature 646(8084) (2025) 331-337.
[178] P. Kral, W. Blum, J. Dvorak, N. Yurchenko, N. Stepanov, S. Zherebtsov, L. Kuncicka, M. Kvapilova, V. Sklenicka, Creep behavior of an AlTiVNbZr0.25 high entropy alloy at 1073 K, Materials Science and Engineering: A 783 (2020) 139291.
[179] J. Feng, B. Wang, Y. Zhang, P. Zhang, C. Liu, X. Ma, K. Wang, L. Xie, N. Li, L. Wang, High-temperature creep mechanism of Ti-Ta-Nb-Mo-Zr refractory high-entropy alloys prepared by laser powder bed fusion technology, International Journal of Plasticity 181 (2024) 104080.
[180] V. Soni, B. Gwalani, T. Alam, S. Dasari, Y. Zheng, O.N. Senkov, D. Miracle, R. Banerjee, Phase inversion in a two-phase, BCC+B2, refractory high entropy alloy, Acta Materialia 185 (2020) 89-97.

[181] V. Soni, O.N. Senkov, B. Gwalani, D.B. Miracle, R. Banerjee, Microstructural Design for Improving Ductility of An Initially Brittle Refractory High Entropy Alloy, Scientific Reports 8(1) (2018) 8816.
[182] P. Kumar, S.J. Kim, Q. Yu, J. Ell, M. Zhang, Y. Yang, J.Y. Kim, H.-K. Park, A.M. Minor, E.S. Park, R.O. Ritchie, Compressive vs. tensile yield and fracture toughness behavior of a body-centered cubic refractory high-entropy superalloy Al0.5Nb1.25Ta1.25TiZr at temperatures from ambient to 1200°C, Acta Materialia 245 (2023) 118620.
[183] K. Ma, T. Blackburn, J.P. Magnussen, M. Kerbstadt, P.A. Ferreirós, T. Pinomaa, C. Hofer, D.G. Hopkinson, S.J. Day, P.A.J. Bagot, M.P. Moody, M.C. Galetz, A.J. Knowles, Chromium-based bcc-superalloys strengthened by iron supplements, Acta Materialia 257 (2023) 119183.
[184] H. Suzuki, Solid solution hardening in body-centred cubic alloys, Dislocations in solids 4 (1980) 191-217.
[185] Y.W. Wang, Q.F. He, Z.H. Ye, Y. Liao, C.W. Li, Q. Wang, J.F. Gu, Ductilization of single-phase refractory high-entropy alloys via activation of edge dislocation, Acta Materialia 284 (2025) 120614.
[186] T. Tsuru, S. Han, S. Matsuura, Z. Chen, K. Kishida, I. Iobzenko, S.I. Rao, C. Woodward, E.P. George, H. Inui, Intrinsic factors responsible for brittle versus ductile nature of refractory high-entropy alloys, Nature Communications 15(1) (2024) 1706.
[187] C. Lee, F. Maresca, R. Feng, Y. Chou, T. Ungar, M. Widom, K. An, J.D. Poplawsky, Y.-C. Chou, P.K. Liaw, W.A. Curtin, Strength can be controlled by edge dislocations in refractory high-entropy alloys, Nature Communications 12(1) (2021) 5474.
[188] F. Maresca, W.A. Curtin, Theory of screw dislocation strengthening in random BCC alloys from dilute to "High-Entropy" alloys, Acta Materialia 182 (2020) 144-162.
[189] S.I. Rao, C. Woodward, B. Akdim, O.N. Senkov, D. Miracle, Theory of solid solution strengthening of BCC Chemically Complex Alloys, Acta Materialia 209 (2021) 116758.
[190] H. Hattendorf, A.R. Büchner, A review of Suzuki's solid solution hardening theory for substitutional bcc alloys, International Journal of Materials Research 83(9) (1992) 690-698.
[191] A. Büchner, W. Wunderlich, A new discussion of the interaction energy in the solid solution hardening of BCC iron alloys, physica status solidi (a) 135(2) (1993) 391-403.
[192] S.I. Rao, W. Wang, M. Asta, R.O. Ritchie, M. Zhang, Rationalization of the tensile creep behavior of the Nb45Ta25Ti15Hf15 bcc refractory complex concentrated alloy using Rao-Suzuki screw dislocation glide model, Scripta Materialia 265 (2025) 116752.
[193] F. Maresca, W.A. Curtin, Mechanistic origin of high strength in refractory BCC high entropy alloys up to 1900K, Acta Materialia 182 (2020) 235-249.
[194] J.-P. Couzinié, M. Heczko, V. Mazánová, O.N. Senkov, M. Ghazisaeidi, R. Banerjee, M.J. Mills, High-temperature deformation mechanisms in a BCC+B2 refractory complex concentrated alloy, Acta Materialia 233 (2022) 117995.
[195] S. Purushothaman, J.K. Tien, Role of back stress in the creep behavior of particle strengthened alloys, Acta Metallurgica 26(4) (1978) 519-528.
[196] R. Srinivasan, G.F. Eggeler, M.J. Mills, $\gamma'$-cutting as rate-controlling recovery process during high-temperature and low-stress creep of superalloy single crystals, Acta Materialia 48(20) (2000) 4867-4878.
[197] M. Probst-Hein, A. Dlouhy, G. Eggeler, Interface dislocations in superalloy single crystals, Acta Materialia 47(8) (1999) 2497-2510.
[198] O.K. Chopra, K. Natesan, Interpretation of high-temperature creep of type 304 stainless steel, Metallurgical Transactions A 8(4) (1977) 633-638.

[199] A. Purohit, W. Burke, Elevated temperature creep behavior of Inconel alloy 625, Argonne National Lab., IL (USA), 1984.
[200] L.S. Bowling, N.R. Philips, D.E. Matejczyk, J.M. Skelton, J.M. Fitz-Gerald, S.R. Agnew, A unified model of tensile and creep deformation for use in niobium alloy materials selection and design for high-temperature applications, Materialia 38 (2024) 102210.
[201] H.A. Calderon, G. Kostorz, G. Ullrich, Microstructure and plasticity of two molybdenum-base alloys (TZM), Materials Science and Engineering: A 160(2) (1993) 189-199.
[202] H.J. Frost, M.F. Ashby, Deformation-mechanism maps : the plasticity and creep of metals and ceramics, 1st ed. ed., Pergamon Press, Oxford [Oxfordshire] ;, 1982.
[203] G. Laplanche, P. Gadaud, L. Perrière, I. Guillot, J.P. Couzinié, Temperature dependence of elastic moduli in a refractory HfNbTaTiZr high-entropy alloy, Journal of Alloys and Compounds 799 (2019) 538-545.
[204] X. Shen, B. Sun, S. Xin, S. Ding, T. Shen, Creep in a nanocrystalline VNbMoTaW refractory high-entropy alloy, Journal of Materials Science & Technology 187 (2024) 221-229.
[205] K. Kawagishi, A.-C. Yeh, T. Yokokawa, T. Kobayashi, Y. Koizumi, H. Harada, Development of an oxidation-resistant high-strength sixth-generation single-crystal superalloy TMS-238, Superalloys 9 (2012) 189-195.
[206] J.J. Stephens, W.D. Nix, The effect of grain morphology on longitudinal creep properties of INCONEL MA 754 at elevated temperatures, Metallurgical Transactions A 16(7) (1985) 1307-1324.
[207] Y.-K. Kim, S. Yang, K.-A. Lee, Compressive creep behavior of selective laser melted CoCrFeMnNi high-entropy alloy strengthened by in-situ formation of nano-oxides, Additive Manufacturing 36 (2020) 101543.
[208] Y. Li, T. Krajňák, P. Podaný, J. Veselý, J. Džugan, Thermal stability of dislocation structure and its effect on creep property in austenitic 316L stainless steel manufactured by directed energy deposition, Materials Science and Engineering: A 873 (2023) 144981.
[209] E. Brizes, J. Milner, A Comparison of Niobium Alloys C103 and Nb521, TMS Annual Meeting, 2025.
[210] M.G. Hebsur, R.H. Titran, Tensile and creep rupture behavior of P/M processed Nb-base alloy, WC-3009, TMS-AIME Fall Meeting, 1988.
[211] D.H. Cook, P. Kumar, M.I. Payne, C.H. Belcher, P. Borges, W. Wang, F. Walsh, Z. Li, A. Devaraj, M. Zhang, M. Asta, A.M. Minor, E.J. Lavernia, D. Apelian, R.O. Ritchie, Kink bands promote exceptional fracture resistance in a NbTaTiHf refractory medium-entropy alloy, Science 384(6692) (2024) 178-184.
[212] Y. An, Y. Liu, S. Liu, B. Zhang, G. Yang, C. Zhang, X. Tan, J. Ding, E. Ma, Additive manufacturing of a strong and ductile oxygen-doped NbTiZr medium-entropy alloy, Materials Futures 4(1) (2025) 015001.
[213] F. Wang, T. Yuan, R. Li, S. Lin, Z. Xie, L. Li, V. Cristino, R. Xu, B. Liu, Comparative study on microstructures and mechanical properties of ultra ductility single-phase Nb40Ti40Ta20 refractory medium entropy alloy by selective laser melting and vacuum arc melting, Journal of Alloys and Compounds 942 (2023) 169065.
[214] O.N. Senkov, J. Gild, T.M. Butler, Microstructure, mechanical properties and oxidation behavior of NbTaTi and NbTaZr refractory alloys, Journal of Alloys and Compounds 862 (2021) 158003.

[215] A. Günen, K.M. Döleker, E. Kanca, M.A. Akhtar, K. Patel, S. Mukherjee, Oxidation resistance of aluminized refractory HfNbTaTiZr high entropy alloy, Journal of Alloys and Compounds 999 (2024) 175100.
[216] P. Tsakiropoulos, On the Nb5Si3 Silicide in Metallic Ultra-High Temperature Materials, Metals, 2023.
[217] R. Smith, The development of oxidation-resistant niobium alloys, Journal of the Less Common Metals 2(2) (1960) 191-206.
[218] R.A. Perkins, G.H. Meier, The oxidation behavior and protection of niobium, JOM 42(8) (1990) 17-21.
[219] Rembar, When To Use TZM Alloy Instead of Pure Molybdenum. https://www.rembar.com/when-to-use-tzm-alloy-instead-of-pure-molybdenum/, 2020).
[220] A. Lange, M. Heilmaier, T.A. Sossamann, J.H. Perepezko, Oxidation behavior of pack-cemented Si–B oxidation protection coatings for Mo–Si–B alloys at 1300°C, Surface and Coatings Technology 266 (2015) 57-63.
[221] G. Ouyang, P.K. Ray, S. Thimmaiah, M.J. Kramer, M. Akinc, P. Ritt, J.H. Perepezko, Oxidation resistance of a Mo-W-Si-B alloy at 1000–1300 °C: The effect of a multicomponent Mo-Si-B coating, Applied Surface Science 470 (2019) 289-295.
[222] J.H. Perepezko, High temperature environmental resistant Mo-Si-B based coatings, International Journal of Refractory Metals and Hard Materials 71 (2018) 246-254.
[223] P. Jain, K.S. Kumar, Tensile creep of Mo–Si–B alloys, Acta Materialia 58(6) (2010) 2124-2142.
[224] M. Heilmaier, H. Saage, M. Krüger, P. Jehanno, M. Böning, H. Kestler, Current Status of Mo-Si-B Silicide Alloys for Ultra-high Temperature Applications, MRS Proceedings 1128 (2008) 1128-U07-07.
[225] M.A. Azim, D. Schliephake, C. Hochmuth, B. Gorr, H.J. Christ, U. Glatzel, M. Heilmaier, Creep Resistance and Oxidation Behavior of Novel Mo-Si-B-Ti Alloys, JOM 67(11) (2015) 2621-2628.
[226] D. Schliephake, M. Azim, K. von Klinski-Wetzel, B. Gorr, H.-J. Christ, H. Bei, E.P. George, M. Heilmaier, High-Temperature Creep and Oxidation Behavior of Mo-Si-B Alloys with High Ti Contents, Metallurgical and Materials Transactions A 45(3) (2014) 1102-1111.
[227] S. Obert, A. Kauffmann, M. Heilmaier, Characterisation of the oxidation and creep behaviour of novel Mo-Si-Ti alloys, Acta Materialia 184 (2020) 132-142.
[228] J.A. Lemberg, R.O. Ritchie, Mo-Si-B Alloys for Ultrahigh-Temperature Structural Applications, Advanced Materials 24(26) (2012) 3445-3480.
[229] J.H. Perepezko, M. Krüger, M. Heilmaier, Mo-Silicide Alloys for High-Temperature Structural Applications, Materials Performance and Characterization 10(2) (2021) 122-145.
[230] S.K. Makineni, A.R. Kini, E.A. Jägle, H. Springer, D. Raabe, B. Gault, Synthesis and stabilization of a new phase regime in a Mo-Si-B based alloy by laser-based additive manufacturing, Acta Materialia 151 (2018) 31-40.
[231] L. Liu, L.F. Wood, P. Nelaturu, T. Duan, C. Zhang, F. Zhang, D.J. Thoma, J.H. Perepezko, Oxidation of a Mo-Si-B-Ti Alloy, High Temperature Corrosion of Materials 101(6) (2024) 1383-1393.
[232] C. Frey, H. You, S. Kube, G.H. Balbus, K. Mullin, S. Oppenheimer, C.S. Holgate, T.M. Pollock, High Temperature B2 Precipitation in Ru-Containing Refractory Multi-principal Element Alloys, Metallurgical and Materials Transactions A 55(6) (2024) 1739-1764.

[233] K.M. Mullin, C. Frey, S.I.A. Jalali, M.S. Patullo, C.S. Holgate, K.J. Hemker, T.M. Pollock, Solidification behavior and cracking mechanisms of Ru-containing BCC-B2 superalloys, Scripta Materialia 267 (2025) 116731.
[234] N. David, T. Benlaharche, J.M. Fiorani, M. Vilasi, Thermodynamic modeling of Ru–Zr and Hf–Ru systems, Intermetallics 15(12) (2007) 1632-1637.
[235] B.J. Crossman, J. Wang, L. Perrière, S.A. Chen, J.-P. Couzinié, M. Ghazisaeidi, M.J. Mills, Multi-modal characterization of the B2 phase in the Ta-Re binary system, Acta Materialia 293 (2025) 121097.